\documentclass[letterpaper]{article}
\usepackage[USenglish]{babel}

\usepackage{pdfpages,graphicx}
\usepackage[caption=false]{subfig}
\usepackage{rerunfilecheck} 
\usepackage{tabularx}
\usepackage{booktabs}
\usepackage{cite}
\usepackage[utf8]{inputenc}
\usepackage{amsmath}
\usepackage[T1]{fontenc}
\usepackage{csquotes}
\usepackage[titletoc]{appendix}
\usepackage[hidelinks]{hyperref}
\DeclareUnicodeCharacter{202F}{\,}
\begin{document}
\selectlanguage{USenglish}
\title{Parallelism in Neurodegenerative Biomarker Tests: Hidden Errors
  and the Risk of Misconduct}
\author{Axel Petzold\thanks{University College London, Queen Square
    Institute of Neurology, London, WC1N 3BG, United Kingdom.\\ Email:
    \href{mailto:a.petzold@ucl.ac.uk}{a.petzold@ucl.ac.uk}} and
  Joachim Pum\thanks{LABanalytics GmbH, Anton-Bruckner-Weg 22, Jena,
    Germany} and David P. Crabb\thanks{City St.\ George's, University
    of London, Northampton Square, London, EC1V 0HB, United Kingdom}}
\date{November 19, 2025}
\maketitle
\begin{abstract}
  Biomarkers are critical tools in the diagnosis and monitoring of
  neurodegenerative diseases.  Reliable quantification depends on
  assay validity, especially the demonstration of parallelism between
  diluted biological samples and the assay's standard
  curve. Inadequate parallelism can lead to biased concentration
  estimates, jeopardizing both clinical and research applications.
  Here we systematically review the evidence of analytical parallelism
  in body fluid (serum, plasma, cerebrospinal fluid) biomarker assays
  for neurodegeneration and evaluate the extent, reproducibility, and
  reporting quality of partial parallelism.

  This systematic review was registered on PROSPERO
  (\href{https://www.crd.york.ac.uk/PROSPERO}{CRD42024568766}) and
  conducted in accordance with PRISMA guidelines. We included studies
  published between December 2010 to July 2024 without language
  restrictions. Eligible studies included original research assessing
  biomarker concentrations in body fluids with data suitable for
  evaluating serial dilution and standard curve parallelism. The data
  extraction for interrogating parallelism included dilution steps,
  measured concentrations, and sample types. For each study we
  generated
  \href{https://discovery.ucl.ac.uk/id/eprint/10185313/1/parallel-v9-accepted.pdf}{parallelism
    plots} in a uniform and comparable way. These graphs were used to
  come to a balanced decision on whether parallelism or
  \href{https://discovery.ucl.ac.uk/id/eprint/10185313/1/parallel-v9-accepted.pdf}{partial
    parallelism} were present. The risk of bias was assessed based on
  sample preparation, buffer consistency, and methodological
  transparency.

  Of 44 eligible studies, 19 provided sufficient data for generating
  49
  \href{https://discovery.ucl.ac.uk/id/eprint/10185313/1/parallel-v9-accepted.pdf}{partial
    parallelism plots}. Of these plots, only 7 (14\%) demonstrated
  clear partial parallelism. Partial parallelism was typically
  achieved over a narrow dilution range of about three doubling
  steps. Most assays deviated from parallelism, risking over- or
  underestimation of biomarker levels if determined at different
  dilution steps. A high risk of bias was identified in 9 studies
  using spiked or artificial samples, inconsistent dilution buffers,
  or incomplete reporting. Several studies assessed sample-to-sample
  parallelism rather than sample-to-standard, contrary to guidelines
  by regulatory authorities.

  In conclusion,
  \href{https://discovery.ucl.ac.uk/id/eprint/10185313/1/parallel-v9-accepted.pdf}{partial
    parallelism} was infrequently observed and inconsistently reported
  in most biomarker assays for neurodegeneration. Narrow dilution
  ranges and variable methodologies limit
  generalizability. Transparent reporting of dilution protocols and
  adherence to established analytical validation guidelines is
  needed. This systematic review has practical implications for
  clinical trial design, regulatory approval processes, and the
  reliability of biomarker-based diagnostics.
\end{abstract}
 \tableofcontents
 \section{Introduction}
 Biomarker assays have become central to advancing research and
 clinical applications in neurodegeneration. For instance, the
 astrocytic biomarker glial fibrillary acidic protein (GFAP) has been
 approved by the U.S. Food and Drug Administration (FDA) as part of a
 panel to guide decisions on brain imaging after
 trauma~\cite{Abd2022_158}. Similarly, the FDA granted rapid approval
 for novel treatments in multiple sclerosis (MS)\cite{Hau2020_546} and
 amyotrophic lateral sclerosis (ALS)~\cite{Mil2022_1099} on the basis
 of neurofilament (Nf) biomarkers~\cite{Kha2024_269}. The accurate
 quantification of such biomarkers relies on robust assay
 performance. To ensure this, the FDA and other regulatory authorities
 have established comprehensive frameworks for biomarker validation,
 covering all stages from sample collection and processing to
 analytical validation, clinical application, quality control, and
 accreditation. However, the breadth of this framework can also create
 ambiguity. For example, the terms ``sensitivity'' and ``specificity''
 are used differently in analytical versus clinical contexts
 (Table~\ref{t_validation}). This distinction is illustrated by
 neurofilament light chain (NfL), which has a reported analytical
 specificity of 99.3\%\cite{Lee2022_} due to minimal cross-reactivity
 with other Nf isoforms, yet demonstrates low clinical specificity
 because blood NfL levels rise across a wide range of neurological
 diseases~\cite{Hau2020_546,Mil2022_1099} and in many conditions that
 compromise the integrity and function of neurons and their
 connections~\cite{Pet2022_179,Kha2024_269}.
 
\begin{table*}\centering
  \caption{\textbf{Framework for laboratory test validation as
      provided by regulatory authorities.} Abbreviations: U.S. Food
    and Drug Administration = FDA, Conformit\'{e} Europ\'{e}enne = CE,
    College of American Pathologists = CAP, Clinical and Laboratory
    Standards Institute = CLSI, International Organization for
    Standardization = ISO, analytical measurement range = AMR, limit
    of the blank = LoB, limit of detection = LoD, limit of
    quantitation = LoQ. Table adapted from
    reference~\cite{Pum2019_215} with a focus on quantitative
    laboratory tests.~\label{t_validation}}
  \scriptsize
  \begin{tabular}{p{3.5cm}p{2.5cm}p{2.5cm}p{2.5cm}p{2.5cm}}
    \toprule
\textbf{Laboratory test}&\multicolumn{2}{c}{\textbf{ISO 15189 / 17025}}&\multicolumn{2}{c}{\textbf{CAP}}\\
&FDA approved / CE certified&In-house \& modified FDA approved / CE certified&FDA approved / CE certified&In-house \& modified FDA approved / CE certified\\
  \midrule
  Precision \& Bias&Verify&Establish&Verify&Establish\\
&CLSI EP15-A3&CLSI EP15-A3&CLSI EP15-A3&CLSI EP15-A3\\
Method Comparison Experiment&Compare with current method&Compare with reference method&Compare with current method&Compare with reference method\\
&CLSI EP09-A3 Regression Analysis, Difference Plot&CLSI EP09-A3 Regression Analysis, Difference Plot&CLSI EP09-A3 Regression Analysis, Difference Plot&CLSI EP09-A3 Regression Analysis, Difference Plot\\
Analytical Sensitivity&---&Establish&Verify&Establish\\
&---&CLSI EP17-A2 LoB, LoD, LoQ, Precision Profile Approach, Probit Analysis&Documentation from manufacturer or literature, CLSI EP17-A2 Verification of LoB, LoD, LoQ&CLSI EP17-A2 LoB, LoD, LoQ, Precision Profile Approach, Probit Analysis\\
Analytical Specificity&---&Establish&Verify &Establish\\
&---&CLSI EP07-A Test for hemolysis, icterus and lipemia, potential cross-reactivities&Documentation from manufacturer or literature&CLSI EP07-A Test for hemolysis, icterus and lipemia, potential cross-reactivities\\
Diagnostic Sensitivity&---&---&---&---\\
&---&---&---&---\\
Diagnostic Specificity&---&---&---&---\\
&&---&---&---\\
Linearity&Verify&Establish&Verify&Establish\\
&Evaluation of linearity, Calibration Verification, Verification of AMR&Evaluation of linearity, Calibration Verification, Establish AMR&Evaluation of linearity, Calibration Verification, Verification of AMR&Evaluation of linearity, Calibration Verification, Establish AMR\\
Carryover&Verify&Establish&Verify&Establish\\
&Standard protocol or Short protocol&Standard protocol&Standard protocol or Short protocol&Standard protocol\\
Measurement Uncertainty&Establish&Establish&Establish&Establish\\
&ISO 11352 / NORDTEST Method&ISO 11352 / NORDTEST Method&ISO 11352 / NORDTEST Method&ISO 11352 / NORDTEST Method\\
Reference Range / Cut-off&Verify&Establish&Verify&Establish\\
&Documentation from manufacturer or literature&CLSI EP28-A3c Use direct or indirect method for data collection&CLSI EP28-A3c (Use direct or indirect method for data collection) OR Documentation from manufacturer or literature&CLSI EP28-A3c (Use direct or indirect method for data collection)\\
    \bottomrule
  \end{tabular}
    \normalsize 
\end{table*}

The framework for laboratory test evaluation presented in
Table~\ref{t_validation} requires additional clarification because
certain terms are frequently misinterpreted in the
literature~\cite{And2015_}. A notable example concerns the distinction
between ``parallelism'' and ``dilution linearity'' Guidance documents
such as ISO/IEC 17025 and EURACHEM refer to parallelism when
illustrating method validation, but the concept largely originates
from applied laboratory practice. In practical use, parallelism and
dilution linearity are sometimes conflated because both involve serial
dilutions. The key distinction lies in the sample type: dilution
linearity experiments use samples spiked with analyte at
concentrations designed to minimize matrix effects, whereas spiked
samples are not permissible for assessing
parallelism~\cite{And2015_,Bur1994_}. Although this difference may
appear subtle, as will be argued in the present critical review, it
has important implications for assay validation and interpretation.

Beyond these terminological distinctions, core analytical parameters
such as precision, accuracy, linearity, and the evaluation of matrix
effects remain central to assay
validation~\cite{Pum2019_215}. Precision and accuracy ensure that
results are both reproducible and correct, while linearity confirms
that measured values remain proportional to analyte concentrations
across the assay's dynamic range~\cite{Pum2019_215}. These parameters
can be compromised by matrix effects, which introduce systematic
bias~\cite{And2015_,Zha2024_1}. This challenge is particularly
pronounced when analyzing complex biological samples such as serum and
plasma~\cite{Pum2019_215,And2015_,Zha2024_1}, especially in mass
spectrometry-based
assays~\cite{Ric2016_2475,Pic2019_2207,Fer2022_66}. To mitigate such
issues, harmonization across laboratories and analytical platforms has
become a major focus in biomarker validation
studies~\cite{Pet2010_23,Oec2016_404,Wil2024_322,Boo2025_120447}. The
goal is to ensure comparability of results and reliability of clinical
decision-making~\cite{And2015_}. As emphasized earlier, one of the
hidden sources of error is the lack of
parallelism~\cite{Lu2011_143,Mus2024_119679}. For this reason,
regulatory authorities require demonstration of parallelism between
serially diluted samples and the assay calibration
curve~\cite{ICH2023_M10,Her2022_627}. Demonstrating parallelism
ensures that measured concentrations remain consistent and accurate
across the relevant dilution range. Yet, the extent to which
guidelines~\cite{ICH2023_M10,Her2022_627} for testing parallelism are
systematically applied in biomarker validation studies remains
unknown.

Therefore, we systematically reviewed the biomarker literature on
neurodegeneration for evidence of parallelism testing, beginning with
the first report of its absence in the neurofilament heavy chain (NfH)
ELISA, which was attributed to protein aggregate
formation~\cite{Lu2011_143}. Protein aggregation is a hallmark of many
neurodegenerative diseases~\cite{Juc2013_45}, meaning that the lack of
parallelism observed in NfH assays has implications far beyond a
single biomarker~\cite{Lu2011_143}. Building on this
observation~\cite{Lu2011_143}, we examined subsequent studies to
determine whether other biomarkers are similarly affected. Deviations
from parallelism~\cite{Pet2024_602} can lead to systematic over- or
underestimation of biomarker concentrations, creating risks for
misinterpretation that are particularly consequential in the context
of clinical trials~\cite{Hau2020_546,Mil2022_1099} and regulatory
submissions~\cite{Abd2022_158,ICH2023_M10,Her2022_627}. Finally, our
critical review underscores the limitations of current reporting
practices in biomarker validation studies, which often include
inconsistent dilution protocols, incomplete reporting of dilution
ranges, and reliance on non-representative spiked or artificial
samples.

With regard to terminology, we begin by clarifying the concept of
parallelism, a term originally rooted in geometry. Although
statistical approaches to testing parallelism are available, detailed
discussion of these methods is provided in the supplementary
materials. In the main text, we instead focus on visual methods, as
they are more accessible to readers without a strong mathematical
background. This approach is supported by illustrative examples to
ensure clarity. We then present the methodology of our systematic
review, conducted in strict accordance with the PRISMA 2020
guidelines. The resulting findings are interpreted in the context of
clinical laboratory science and extended to address broader
methodological and regulatory issues. Finally, we highlight practical
recommendations for laboratories, manufacturers, and regulators, with
the aim of ensuring applicability beyond the research community.

\section{Terminology and concept}
The formal concept of parallelism can be traced to the Greek
mathematician, geometer, and logician Euclid who, around 300 BCE,
authored the seminal treatise Elements. In Book I, Definition 23,
Euclid defined parallel lines as ``straight lines which, being in the
same plane and being produced indefinitely in both directions, do not
meet one another in either direction.'' However, it is Euclid's fifth
postulate, known as the parallel postulate, that has historically
attracted the most scrutiny. Over the centuries, mathematicians sought
to prove the fifth postulate using Euclid's other axioms. These
efforts persisted for over two millennia and produced many flawed or
incomplete proofs, reflecting a significant historical
misconception~\cite{Win1925_356}. It was not until the 19th century
that mathematicians like Carl Friedrich Gauss (working on parallelism
1779--1844), Lobachevsky (1829--30), and Bolyai (1832) demonstrated
that entirely consistent non-Euclidean geometries could be constructed
by replacing the fifth postulate with alternative
versions~\cite{Jen2023_707,Win1925_356}.  In a letter to Bolayi
(17-DEC-1799) Gauss wrote: ``It is true that I have come upon much
which by most people would be held to constitute a proof: but in my
eyes it proves as good as \emph{nothing}.''

Similarly, in the context of present, critical, systematic review,
parallelism, though seemingly straightforward at first glance, demands
careful scrutiny. Just as for Euclid's parallel postulate, analytical
parallelism in quantitative assays must not be assumed but must be
rigorously tested, substantiated, and importantly \emph{lack of
  parallelism must be understood.}

\subsection{Parallelism and standard curves}
The accurate determination of analyte concentrations using a standard
curve requires that the dilution series of the test sample exhibits
parallelism with the standard curve~\cite{Pli1994_2441}. This means
the two curves must be similar functions, differing only by a scaling
factor along the dose axis (in present systematic review this is
always the y-axis), so that interpolation yields valid and
reproducible results.  Without demonstrated parallelism, calculations
derived from a standard curve may lead to inaccurate
quantification~\cite{Pli1994_2441}.  In its most simple form
\emph{interpolation} is linear. Hence the to verify/establish
linearity for a laboratory test in Table~\ref{t_validation}.  Linear
interpolation is a method of curve fitting using linear polynomials to
construct new data points within the range of a discrete set of known
data points~\cite{Mei2002_319}. It is absolutely crucial to recognise
that for biomarker concentrations from samples calculations are only
permitted for data points within the range of the points of the
standard curve. Expanding from linear interpolation, quadratic, cubic,
four-parameter logistic (4PL)~\cite{Rod1974_165,Rod1978_469}, 4PL with
logarithmic scaling of the dose (y-axis)~\cite{Yan2016_}, and
five-parameter logistic
(5PL)~\cite{Pre1976_761,Rod1978_469,Dud1985_1264,Got2005_204} standard
curves have entered contemporary laboratory routine.  Extrapolation is
not allowed~\cite{Pum2019_215,Yan2016_,Mei2002_319}. It is mandatory
to demonstrate parallelism for the selected type of a standard curve
and real-world samples~\cite{ICH2023_M10}.

Despite its importance, the validation of parallelism presents a
practical challenge for many laboratory
workflows~\cite{Pum2019_215}. The concept, as reviewed here and long
known in analytical chemistry, has only more recently gained sustained
attention in the context of regulatory method validation for biomarker
assays~\cite{Her2022_627,Zha2024_1}. Variability in internal standards
(ISV) has been identified as a core factor contributing to measurement
error, further underscoring the need for robust validation.

Regulatory agencies have begun to address the issue of
parallelism. The FDA, through its 2022 M10 Bioanalytical Method
Validation (BMV) guidance, and the European Medicines Agency (EMA)
both highlight parallelism as a critical validation
criterion~\cite{us2023m10,ICH2023_M10}. According to the EMA,
parallelism is defined as follows: ``Parallelism demonstrates that the
serially diluted incurred sample response curve is parallel to the
calibration curve''~\cite{ICH2023_M10}.

\paragraph{Spiked samples in parallelism}
The use of spiked samples in bioassay validation is an interesting
strategy to assess analytical performance, particularly in relation to
matrix effects and assay specificity~\cite{Zha2024_1}. The use of
spiked samples is frequently employed in the context of mass
spectrometry-based methods, where matrix-induced variability is a
major concern to the accuracy of
quantification~\cite{Cul2012_8,Ric2016_2475}. Current guidelines
suggest that variations of less than 20\% are generally acceptable
when comparing spiked samples with calibrators or quality
controls~\cite{Fu2020_545,JW2019_1679,Zha2024_1}. However, this
approach typically focuses on performance at fixed concentration
levels and does not explicitly evaluate assay behaviour across wider
dilution ranges or in the presence of other factors that apply to
real-world samples~\cite{JW2019_1679}. 

Therefore one of the key limitations of spiked samples is the
requirement for high-concentration material to enable serial dilution
and minimize matrix effects~\cite{JW2019_1679}. Although spiking can
be informative under controlled conditions, it introduces several
methodological concerns. First, the reference material used for
spiking may differ structurally or functionally from the endogenous
biomarker. This includes the use of truncated peptide sequences or
proteins derived from non-human
species~\cite{Law1982_1,Rob2014_211}. Second, the stability of spiked
material can be suboptimal, particularly over long-term storage, where
protein aggregation may occur~\cite{Rob2014_211}. Aggregation, a known
source of error in parallelism assessment~\cite{Lu2011_143}, cannot be
adequately simulated using spiked samples.

Moreover, spiked samples do not replicate the complexity of
physiological matrices. As noted in prior reviews, their utility is
limited to evaluating internal standard variability introduced by
physiological differences between control and patient
matrices~\cite{JW2019_1679}. Even when the spiked analyte behaves
additively with the endogenous biomarker, this assumption may not hold
under all conditions, and further concerns regarding non-linear
interactions persist~\cite{Tu2017_1107}.

For these reasons, real-word, patient-derived samples remain the gold
standard for assessing parallelism~\cite{And2015_,Bur1994_}. Their use
better reflects the biochemical and biophysical properties relevant to
clinical measurement~\cite{Loh2023_502}. In the context of this
systematic review, the use of spiked samples is discussed as a
potential source of bias.

\subsection{Determination of parallelism}
Visual assessment offers a complementary and often more intuitive
approach to evaluating parallelism in
bioassays~\cite{Pli1994_2441}. Unlike formal statistical testing (see
supplementary materials), graphical methods allow analysts to inspect
the behaviour of dilution curves directly, which is particularly
valuable in the presence of noisy or incomplete datasets. Statistical
expertise required to properly interpret hypothesis testing
methods~\cite{Fay2020_721,Got2005_437,Guy2023_229,Jon2009_818} is not
commonly included in the core training of laboratory scientists,
clinicians, or regulatory reviewers. This gap may lead to an
overreliance on statistical outcomes, potentially overlooking
meaningful deviations in curve behaviour across extended dilution
ranges~\cite{Pet2024_602}. In real-world applications, complete
parallelism is rare, while
\href{https://discovery.ucl.ac.uk/id/eprint/10185313/1/parallel-v9-accepted.pdf}{partial
  parallelism} is often observed within limited dilution
ranges~\cite{Pet2024_602}. This practical observation has led to the
operational definition of \emph{partial parallelism}, acknowledging
that assays may behave acceptably within specific, predefined ranges
without meeting idealized criteria across the entire
curve~\cite{Smi1998_509}.

\begin{figure*}  \centering
  \includegraphics[width=1\textwidth]{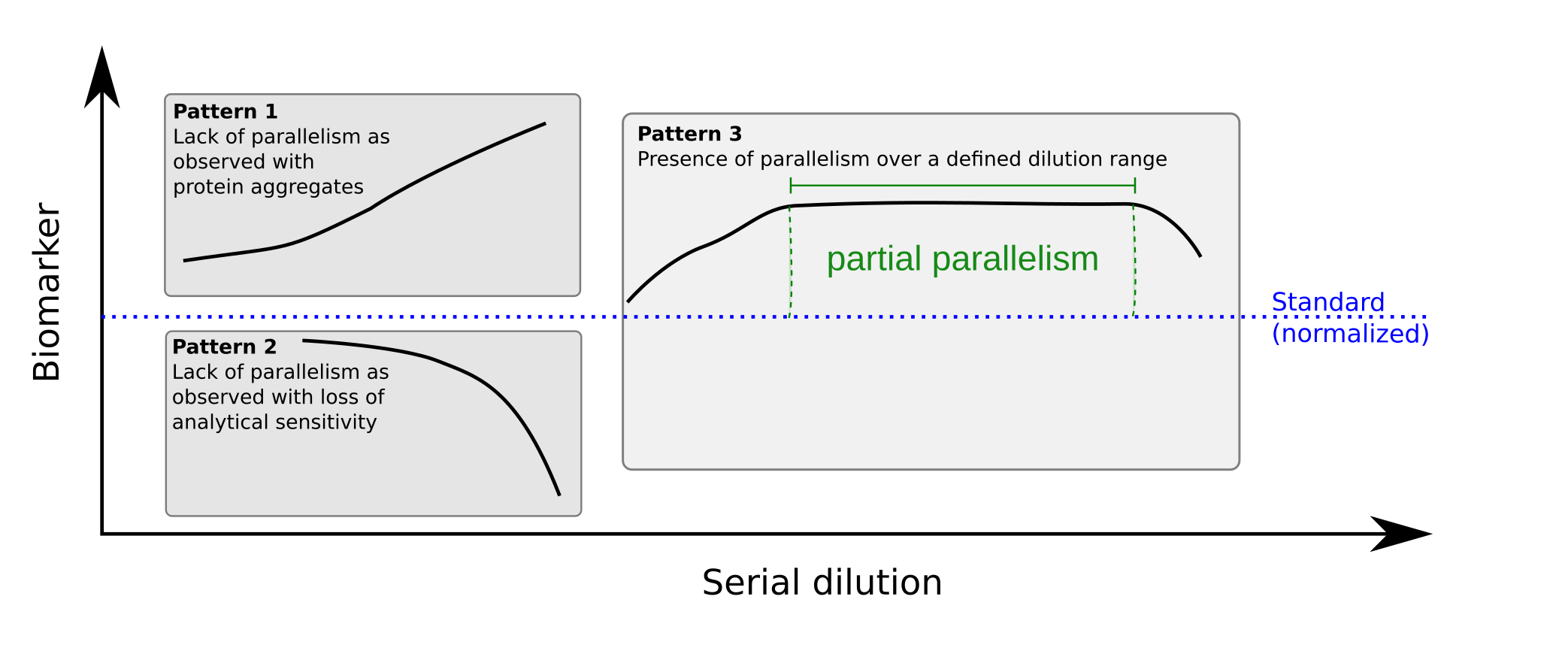}
  \caption{Visual assessment of parallelism. The figure illustrates
    characteristic patterns used to compare dilution ranges (x-axis)
    of the assay calibration standard (blue dotted line) with
    theoretical sample responses (black lines). Such visual patterns
    provide an accessible means of determining whether parallelism is
    maintained or lost and form the basis of our approach for
    evaluating parallelism across the assays included in this critical
    and systematic review. \textbf{Pattern 1} shows an apparent
    increase in biomarker concentration with greater sample dilution,
    a phenomenon typically associated with the release of biomarker
    molecules from protein
    aggregates~\cite{Lu2011_143}. \textbf{Pattern 2} shows a decrease
    in measured biomarker concentration with successive dilution
    steps, commonly observed when concentrations approach the assay's
    lower limit of detection (LoD\@; see
    Table~\ref{t_validation}). This underscores why regulatory
    authorities require validation of analytical
    sensitivity~\cite{Pum2019_215}. \textbf{Pattern 3} demonstrates
    that parallelism may be achieved within a limited dilution range
    but lost again at higher dilutions. This pattern is termed
    \href{https://discovery.ucl.ac.uk/id/eprint/10185313/1/parallel-v9-accepted.pdf}{partial
      parallelism}~\cite{Pet2024_602}. As highlighted throughout this
    review, observed parallelism is most often
    partial.~\label{f_example}}
\end{figure*}

The value of visual methods for detecting lack of parallelism has been
recognized in the literature since the landmark paper by
Plikaytis~\cite{Pli1994_2441}, and has been expanded by subsequent
studies~\cite{Gil2002_47,Kle1999_35,Gai2002_35,Pet2024_602}. Figure~\ref{f_example}
illustrates the most common patterns observed.

The interpretation of the partial parallelism plots goes back to
Euclid's geometry. The two lines in the
\href{https://discovery.ucl.ac.uk/id/eprint/10185313/1/parallel-v9-accepted.pdf}{partial
  parallelism plot} need to be parallel. To be more precise they need
to be parallel to each other, on the y-axis, for at least part of the
graph (Pattern 3 in Figure~\ref{f_example}). Hence the term
\href{https://discovery.ucl.ac.uk/id/eprint/10185313/1/parallel-v9-accepted.pdf}{\emph{partial
    parallelism}}. A vertical offset is permitted, which is also
entirely consistent with the requirements for statistical
testing~\cite{Got2005_437}. Interpretation of partial parallelism
plots is best if data for the standard curves are present. For the
purpose of this systematic review partial parallelism plots will be
created for each study included. Based on the graphical
interpretation, which is presented for each study, a binary decision
will be made: presence (see Figure~\ref{f_example} pattern 3) or
absence (Figure~\ref{f_example} patterns 1\&2, and other patterns of
deviation from parallel alignment) of
\href{https://discovery.ucl.ac.uk/id/eprint/10185313/1/parallel-v9-accepted.pdf}{partial
  parallelism}.

\section{Clinical context}
The clinical implications of reporting artificially elevated biomarker
concentrations due to lack of parallelism (for example Pattern 1 in
Figure~\ref{f_example}) can be illustrated using neurofilaments. Both
NfL~\cite{Mar2020_70,Axe2018_143} and
NfH~\cite{Pet2006_1071,Pet2009_42} have been established as valuable
prognostic biomarkers in patients with Guillain-Barr\'{e} syndrome
(GBS)~\cite{Yuk2012_2294,Gui1916_1462}.

GBS is characterized by evolving paralysis, which in severe cases can
compromise respiratory function and necessitate intensive care unit
(ICU) admission with mechanical ventilation. While most patients
experience transient paralysis followed by substantial recovery, a
subset develop long-term disabilities, including the permanent loss of
ambulation. Early identification of high-risk patients would therefore
be of great clinical value, enabling personalized management
strategies ranging from ICU admission decisions to treatment
escalation.

Targeting high-efficacy, but often more costly, treatments to patients
at greatest risk offers the dual benefit of optimizing clinical
outcomes and improving resource allocation. Enhanced functional
recovery in this subgroup would not only improve patient quality of
life but also reduce downstream healthcare costs, including those
associated with long-term disability, loss of work capacity, increased
care needs, and broader economic deadweight losses. The importance of
such strategies is reflected in World Health Organization (WHO)
guidance developed in collaboration with the Universal Health Coverage
Partnership \href{www.uhcpartnership.net}{(UHC Partnership)}. Their
report, covering 115 countries and representing more than three
billion people, explicitly recognizes the essential role of clinical
laboratory services in healthcare
delivery~\cite{WHO2022_framework}. While the report does not
specifically address technical parameters such as parallelism, this
level of detail is beyond its scope, it implies that such issues may
be incorporated under the remit of National Quality Control
Laboratories~\cite{WHO2022_framework}.

To illustrate how the lack of demonstrated parallelism can have
wide-ranging consequences, we highlight one recent example from the
literature of NfL in GBS~\cite{Til2024_105072}. In this study, NfL
concentrations were measured using a commercially available assay. The
manufacturer recommends quantification at a dilution of 1:4. However,
for a subset of samples, measurements were performed at a much higher
dilution (1:400) without demonstrating parallelism between the diluted
samples and the calibration standard across this extended range. The
inferred implications are substantial:

\begin{itemize}
\item Inflated biomarker concentrations: Reported values ranged from
approximately 1 pg/mL to nearly 10,000 pg/mL, which is several orders
of magnitude higher than expected under either physiological variation
or disease-related pathology.
\item Potential patient misclassification: Elevated NfL levels were
  associated with more severe phenotypes defined by electrodiagnostic
  criteria~\cite{Yuk2012_2294,Had1998_780}. The data imply that
  artificially inflated concentrations may have been interpreted as
  markers of irreversible axonal degeneration.
\item Bias in predictive modeling: Predictive models are invaluable
  for risk stratification, but if their statistical significance
  arises from artificially elevated biomarker concentrations in a
  subgroup of severely affected patients, these models are unlikely to
  replicate in independent cohorts, clinical trials, or clinical
  practice.
\item Impact on treatment evaluation: The study reported no treatment
  effect on NfL levels. One clinical interpretation might be to
  withhold a potentially effective therapy. However, the extreme
  variability in reported concentrations likely increased noise and
  obscured real effects, deviating from prior treatment trials that
  adhered to validated dilution
  ranges~\cite{Hau2020_546,Mil2022_1099}.
\item Risk of deliberate misuse: A particularly concerning possibility
  arises in the context of treatment trials. If placebo samples were
  analyzed at a higher dilution range than treatment samples under
  Pattern 1 conditions (Figure~\ref{f_example}), or at a lower range
  under Pattern 2 conditions, this would introduce systematic bias
  favoring a positive drug effect. Such practices would constitute
  scientific misconduct.
\end{itemize}

This example underscores, on multiple levels, the necessity of
ensuring rigorous and reliable biomarker quantification. Inaccuracies
not only risk patient misclassification but also undermine predictive
modeling, treatment evaluation, and ultimately regulatory
confidence. To evaluate how consistently this principle has been
upheld in the field, we now proceed with a systematic review of the
literature.

\section{Methods}\label{methods}
The protocol for this systematic review was submitted to the PROSPERO
registry and published on the PROSPERO website. The study protocol can
be searched under the registration number CRD42024568766. The 2020
PRISMA guidelines for reporting systematic reviews are
followed~\cite{Pag2021_71}. The 27-item PRISMA checklist is uploaded
(see section~\ref{prisma_list}).

\subsection{Eligibility Criteria}\label{Eligibility_Criteria}
\textbf{Inclusion Criteria}: All studies involving the analysis of
parallelism of biomarkers in body fluids were considered for
inclusion. No exclusion criteria were applied based on participant
demographics or specific disease conditions. Body fluids included
cerebrospinal fluid (CSF), urine, saliva, and other relevant bodily
fluids. Various analytical techniques were considered, including
immunoassays, mass spectrometry, ELISA, and other relevant methods for
quantitative biomarker
detection~\cite{Her2022_627,Ric2016_2475,Pic2019_2207,Fer2022_66}. Data
was permitted to be derived from purely analytical studies,
experimental studies or clinical settings.

\textbf{Exclusion Criteria}: Studies not involving biomarker sampling
from body fluids.  Research focused solely on tissue biomarkers or
biomarkers derived from non-fluid sources.  Articles lacking detailed
methodology or insufficient data on biomarker sampling
techniques~\cite{Pum2019_215}.  Case reports, reviews, and opinion
articles without original research data.

\subsection{Information Sources}\label{Information_Sources}
Two databases were searched, Medline and Google Scholar.

\subsection{Search Strategy}\label{Search_Strategy}
A search of the MEDLINE database was conducted covering the period
after publication of lack of parallelism for
NfH~\cite{Lu2011_143}. The dates entered to the search strategy were
between 9-Dec-2010 and 12-July-2024. There were no language
restrictions. The Entrez Programming Utilities (E-utilities), provided
by the National Center for Biotechnology Information (NCBI), were used
in a
\href{https://www.crd.york.ac.uk/PROSPEROFILES/568766_STRATEGY_20240713.pdf}{Python
  script} which can be downloaded from the PROSPERO register.

\textbf{Search Strategy Details}: The search terms for the biomarkers
(first search term) were: ``neurofilament'', ``neurofilaments'', ``tau
protein'', ``T-tau'', ``P-tau'', ``glial fibrillary acidic protein'',
``amyloid beta'', ``ubiquitin C-terminal hydrolase 1'',
``neurogranin'', and ``YKL-40''. The search terms for the analytical
methods (second search term) were: ``method'', ``development'',
``linearity'', ``parallelism'', and ``doubling dilution''. The first
search term and the second search term were combined
individually. These combinations were coded in python. The complete
literature search code can be downloaded from the study protocol on
the \\
\href{https://www.crd.york.ac.uk/PROSPERO}{PROSPERO} Registry under
item number 17. The literature search was performed on 02-SEP-2024.

\subsection{Study Selection}\label{Study_Selection}
Neurodegeneration is a prevalent feature in various human diseases,
and biomarkers play a crucial role in its indirect
assessment. Therefore, no specific disease-related restrictions were
imposed. Limitations were focused on the analytical development of
biomarker tests~\cite{Pum2019_215}. Studies were selected based on the
availability of quantitative body fluid samples.

\textbf{Selection Process}: All studies identified underwent a
full-text review and review of supplementary data where available.

\textbf{Tools Used}: A spreadsheet was kept using the PubMed
Identifier (PMID) for each study. Studies included and excluded after
review were clearly marked as such, including the reason for that
decision.

\begin{figure}\centering
  \includegraphics[width=0.45\textwidth]{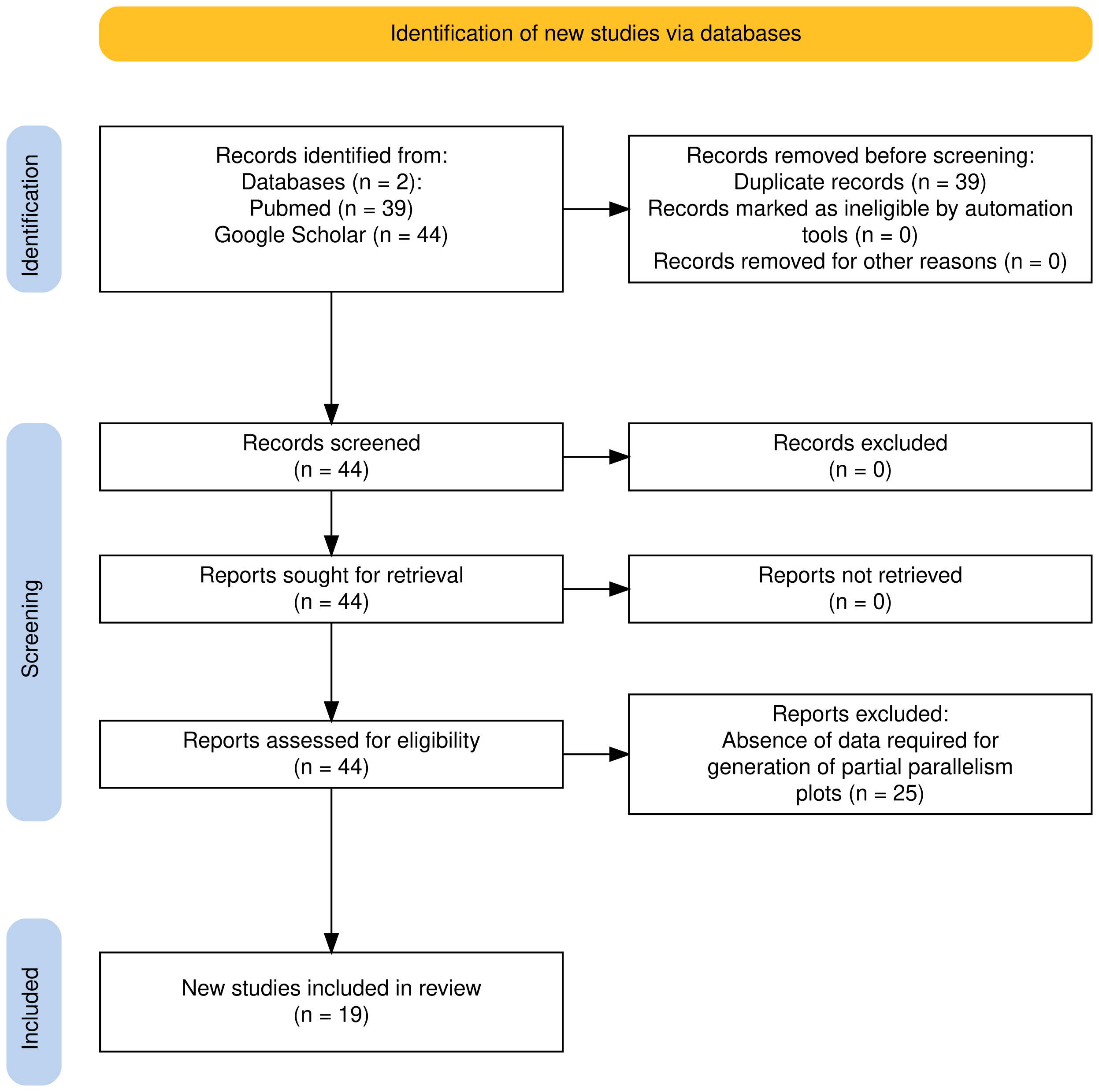}
  \caption{PRISMA flow diagram for the literature search, based on the
    PRISM template
    \href{https://estech.shinyapps.io/prisma_flowdiagram/}{code}~\cite{Had2022_}.}\label{f_flowchart}
\end{figure}

\subsection{Data Collection Process and Data items}\label{data_collection}
\textbf{Data Extraction}: The data were extracted by hand from the
values provided for samples at each dilution step into a
spreadsheet. If data were only available in a figure, the
corresponding author was contacted by email, containing the PROSPERO
study number, asking to share the raw data.

\textbf{Author contact}: Two email requests for sharing the data items
required were sent to all corresponding authors. The first email was
sent on 12-SEP-2024. The second email was sent to the non-responders
on 12-OCT-2024.


\textbf{Data Items}: The three specific variables for which data were
collected included: 1.\ dilution step, 2.\ measured biomarker
concentration, 3.\ sample type (standard, body fluid,
artificial/spiked).

\subsection{Risk of Bias in Individual Studies}\label{risk_bias}
The risk of bias in individual studies was evaluated by carefully
reviewing the methodologies used in sample selection and
preparation. Specifically, we assessed whether the samples were native
or spiked with protein standards and whether both the samples and
standards were diluted using the same buffer, ensuring that it was
indeed the correct dilution buffer~\cite{Pum2019_215}. Additionally,
we recorded the selection of appropriate disease and control samples,
noted whether the laboratory analyst was blinded during the
experiments, and considered the reproducibility of the experiments
through repeat assessments.

\subsection{Summary Measures}\label{summary_measures}
The summary measures used in this review include the deviation from
the line of unity, which is represented by the horizontal line at a
y-value of one in the
\href{https://discovery.ucl.ac.uk/id/eprint/10185313/1/parallel-v9-accepted.pdf}{partial
  parallelism plots}~\cite{Pet2024_602}. The data are categorical,
indicating the presence or absence of partial parallelism
(yes/no). For studies where parallelism is observed, continuous data
were presented, specifically the dilution range within which partial
parallelism was demonstrated. These two effect measures will be
analysed for each sample individually and, when meaningful (n$>$5),
also as a group mean.

\subsection{Synthesis of Results}\label{synthesis_results}
The synthesis of results is primarily visual, utilising
\href{https://discovery.ucl.ac.uk/id/eprint/10185313/1/parallel-v9-accepted.pdf}{partial
  parallelism plots}~\cite{Pet2024_602} derived from the included
studies. These visual representations are then summarised in table
format and complemented by a narrative synthesis.

\subsection{Risk of Bias Across Studies}\label{risk_bias_across}
The primary focus is on determining whether partial parallelism exists
between a sample from an individual with a specific disease and the
protein standard used in the test. Practically this requires to
provide data for a dilution series of the sample into the assay buffer
(the same dilution buffer that is used for the standard
curve)~\cite{Pum2019_215}.  Therefore, a high risk of bias was
assigned if no such sample was available and instead, an artificial
sample was created by spiking the body fluid collected with the
protein. Similarly, if different buffers were used for dilution
between the sample and the standard, the bias was also rated as
high. A moderate risk of bias was assigned when an appropriate disease
sample was not used, or if the laboratory analyst was not blinded, or
if data on reproducibility were missing. In all other cases, the risk
of bias was rated as low.

\section{Results}

\subsection{Study Selection}\label{r_study_selection}
The flow diagram in Figure~\ref{f_flowchart} outlines the process of
study selection, detailing the number of records identified, included,
and excluded~\cite{Had2022_}. A comprehensive literature search
initially yielded 39 articles for further
evaluation~\cite{But2021_58,Cul2012_8,Her2017_244,Huy2023_734,Kan2014_17,
  Kir2021_267,Koe2014_43,Kra2017_465,Kru2016_153564,Lac2013_897,
  Lac2015_42,Lif2019_30,Son2016_58,Thi2021_9736,
  Tri2020_1417,Waa2016_21,Waa2017_310,Yan2017_9304,Che2019_5277,Cio2023_115174,
  De2017_159,Del2016_12,Far2020_165,Gal2016_85,Ge2019_3294,
  Hup2018_123,Kha2022_122946,Kha2022_695,Kov2014_90,Lam2011_9,
  Liu2022_1186,Pei2020_4775,Per2017_421,Pil2023_,Sha2021_744,
  Van2014_44,Wan2024_325,Wil2024_322,Zha2023_19304}.  Additionally, a
search on Google Scholar, using the same terms, identified five more
references~\cite{Yam2021_22,Woj2023_,Lee2022_,Wil2022_761,Bay2021_}. All
44
articles~\cite{But2021_58,Cul2012_8,Her2017_244,Huy2023_734,Kan2014_17,
  Kir2021_267,Koe2014_43,Kra2017_465,Kru2016_153564,Lac2013_897,
  Lac2015_42,Lif2019_30,Son2016_58,Thi2021_9736,
  Tri2020_1417,Waa2016_21,Waa2017_310,Yan2017_9304,Che2019_5277,Cio2023_115174,
  De2017_159,Del2016_12,Far2020_165,Gal2016_85,Ge2019_3294,
  Hup2018_123,Kha2022_122946,Kha2022_695,Kov2014_90,Lam2011_9,
  Liu2022_1186,Pei2020_4775,Per2017_421,Pil2023_,Sha2021_744,
  Van2014_44,Wan2024_325,Wil2024_322,Zha2023_19304,Yam2021_22,Woj2023_,Lee2022_,Wil2022_761,Bay2021_},
including supplementary analyses where available, were thoroughly
reviewed. The corresponding author was contacted twice per email for
sharing data if this was necessary for creating
\href{https://discovery.ucl.ac.uk/id/eprint/10185313/1/parallel-v9-accepted.pdf}{partial
  parallelism plots}.  For four articles, no presently valid author
contact details could be
found~\cite{Lac2013_897,Lac2015_42,Thi2021_9736,Waa2016_21}.  For the
remaining papers, a careful examination of all data available resulted
in the exclusion of 25
articles~\cite{Che2019_5277,Cio2023_115174,De2017_159,
  Del2016_12,Far2020_165,Gal2016_85,Ge2019_3294,Hup2018_123,
  Kha2022_122946,Kha2022_695,Kov2014_90,Lac2015_42,Lam2011_9,
  Liu2022_1186,Pei2020_4775,Per2017_421,Pil2023_,Sha2021_744,
  Thi2021_9736,Van2014_44,Waa2016_21,Waa2017_310,Wan2024_325,
  Wil2024_322,Zha2023_19304} due to the absence of data necessary for
generating
\href{https://discovery.ucl.ac.uk/id/eprint/10185313/1/parallel-v9-accepted.pdf}{partial
  parallelism plots} as per our predefined
\href{https://www.crd.york.ac.uk/PROSPEROFILES/568766_STRATEGY_20240713.pdf}{protocol}. Consequently,
19 studies were selected for further
evaluation~\cite{Bay2021_,But2021_58,Cul2012_8,Her2017_244,Huy2023_734,Kan2014_17,Kir2021_267,Koe2014_43,Kra2017_465,Kru2016_153564,Lac2013_897,Lee2022_,Lif2019_30,Son2016_58,Tri2020_1417,Wil2022_761,Woj2023_,Yam2021_22,Yan2017_9304}.


\subsection{Study Characteristics}\label{r_study_character}
The characteristics of the included studies are summarised in
Table~\ref{t_studies}. The majority of studies focused on human
samples, primarily using plasma or CSF, with only a few investigating
serum. Four studies examined brain tissue homogenates, while two used
rodent tissue samples. All studies included control samples, and most
also incorporated samples from a variety of disease conditions, with
Alzheimer's disease (AD) being the most frequently studied. The
biomarkers analyzed across the studies included the Amyloid $\beta$
fragments, A$\beta_{1-40}$, A$\beta_{1-42}$, A$\beta$ oligomers,
$\alpha$-synuclein, DJ-1, total tau protein (Tau), phospho tau
proteins (pTau181, pTau217, pTau231), Apolipoprotein E (ApoE) isoforms
E2, E3, E4, neurofilament light chain (NfL), neurofilament heavy chain
(NfH), GFAP, and poly(GP).

\begin{table*}\centering
  \caption{Study characteristics of the included
    studies~\cite{But2021_58,Kru2016_153564,Wil2022_761,Bay2021_,Cul2012_8,Huy2023_734,Kan2014_17,Koe2014_43,Lee2022_,Tri2020_1417,Yam2021_22,Her2017_244,Kir2021_267,Yan2017_9304,Lac2013_897,Lif2019_30,Son2016_58,Woj2023_,Kra2017_465},
    sorted for biomarkers tested.\label{t_studies} Column
    abbreviations: P=Plamsa, S=Serum, C=CSF, T=Tissue, H=Human,
    R=Rodent. Remaining abbreviations in alphabetical order:
    A$\beta$=Amyloid beta, AD=Alzheimer disease, ALS=Amyotrophic
    Lateral Sclerosis, ApoE=Apolipoprotein E, Ctrl=control,
    ELISA=enzyme linked immonosorbent assay,
    Elecsys=ElectroChemiLuminescence (ECL) technology for immunoassay
    analysis (Roche), FTL/FTLD=fronto temporal lobar dementia,
    GFAP=glial fibrillary acidic protein, IMR=Immunomagnetic
    reduction, LC-MS/MS=Liquid chromatography-mass spectrometry/mass
    spectrometry, MALDI-TOF=matrix-assisted laser
    desorption/ionisation time of flight, MCI=Minimal Cognitive
    Impairement, MS=Multiple Sclerosis, MSD=Meso Scale Discovery,
    NfH=Neurofilament heavy chain, NfL=Neurofilament light chain,
    PD=Parkinson disease, Ptau=phospho-tau, Simoa=Single Molecule
    Arrays, sAPP=soluble amyloid-$\beta$ precursor protein,
    sFIDA=surface-based fluorescence intensity distribution analysis,
    VD=Vascular Dementia.  Studies are sorted alphabetically according
    to the biomarker analysed.}
  \footnotesize
  \begin{tabular}{lllllllp{2.5cm}p{3cm}p{3cm}}
    \toprule
    Ref&P&S&C&T&H&R&Disease(s)&Assay&Biomarker\\
    \midrule
    \cite{Yam2021_22}&x&-&-&-&x&-&Ctrl&HISCL&A$\beta_{1-40}$, A$\beta_{1-42}$\\
    \cite{Kir2021_267}&x&-&-&-&x&-&Ctrl&LC-MS/MS&A$\beta$40, A$\beta_{1-42}$\\
   \cite{Lac2013_897}&x&-&x&-&x&-&Ctrl, AD&ELISA&A$\beta_{1-40}$, A$\beta_{1-42}$\\
    \cite{Son2016_58}&x&-&-&-&x&-&Ctrl, AD&Simoa \& ELISA&A$\beta_{1-42}$\\
    \cite{Cul2012_8}&-&-&x&-&x&-&Ctrl, AD&ELISA&A$\beta_{1-42}$\\
    \cite{Kan2014_17}&x&-&-&-&x&-&Ctrl&MALDI-TOF MS&A$\beta$ peptides\\
    \cite{Her2017_244}&x&-&x&-&x&-&Ctrl&sFIDA&A$\beta$ aggregates\\
    \cite{Kra2017_465}&x&-&-&-&x&-&Ctrl, VD, PD, MCI, FTL, AD&sFIDA&A$\beta$ oligomers\\
    \cite{Kru2016_153564}&-&-&x&-&x&-&Ctrl&MSD multi-array&$\alpha$-synuclein, A$\beta_{1-42}$, DJ-1, total tau\\
    \cite{Huy2023_734}&-&-&x&-&x&-&ApoE genotypes&LC-MS/MS&ApoE (E2, E3, E4)\\
    \cite{But2021_58}&x&x&-&x&-&x&Ctrl, TBI, AD&MSD &GFAP\\
    \cite{Lee2022_}&x&x&x&-&x&-&Ctrl, MS, ALS&ADVIA\&Atellica&NfL\\
    \cite{Koe2014_43}&-&-&x&x&x&-&Ctrl&ELISA, Luminex&NfL, NfH\\
    \cite{Wil2022_761}&-&-&x&-&x&-&Ctrl, C9or7f2&Simoa&poly(GP)\\
    \cite{Yan2017_9304}&x&-&x&x&x&x&Ctrl, AD&MSD, IMR&tau\\
    \cite{Lif2019_30}&-&-&x&-&x&-&Ctrl, AD&Elecsys&tau, Tau-p181\\
    \cite{Woj2023_}&x&-&x&-&x&-&Ctrl, MCI, AD, FTLD&Simoa&Tau-p181-p231\\
    \cite{Tri2020_1417}&-&-&x&x&x&-&Ctrl, AD&Simoa&Tau-p217\\
   \cite{Bay2021_}&x&-&-&-&x&-&Ctrl, AD&Simoa&tau-p217, tau-p231\\
    \bottomrule
  \end{tabular}
  \normalsize
\end{table*}

\subsection{Risk of Bias Within Studies}\label{r_risk_bias} The risk
of bias across the included studies is summarised in
Table~\ref{t_bias}. Only one study was assessed as having a low risk
of bias~\cite{But2021_58}. This study provided clear documentation on
key factors, including blinding and sample handling.

Eight studies were rated as having a moderate risk of
bias~\cite{Cul2012_8,Tri2020_1417,Lee2022_,Yam2021_22,Kan2014_17,Huy2023_734,Koe2014_43,Wil2022_761},
mainly due to incomplete information regarding blinding procedures,
particularly in relation to the blinding of the analyst.

Another, eight studies were categorised as having a high risk of
bias~\cite{Kru2016_153564,Son2016_58,Kra2017_465,Yan2017_9304,Lif2019_30,Her2017_244,Woj2023_,Kir2021_267}. This
rating was due either to the use of spiked
samples~\cite{Kru2016_153564,Son2016_58,Kra2017_465,Yan2017_9304,Her2017_244,Woj2023_,Kir2021_267},
or to incomplete documentation on whether the samples were spiked and
how they were processed~\cite{Lif2019_30}.

\begin{table*}\centering
  \caption{Risk of bias assessment of the studies included. Studies
    are sorted according to the level of bias risk.\label{t_bias}}
  \footnotesize
  \begin{tabular}{llp{9cm}}
    \toprule
    Ref & Bias risk & Explanation for risk of bias assessment \\
    \midrule
    \cite{But2021_58}&Low &Native \& spiked samples used \& investigator blinded\\
    \midrule        
    \cite{Bay2021_}&Moderate &Native samples used, but not stated if the two analysts blinded\\
    \cite{Kru2016_153564}&Moderate &Native \& spiked CSF samples used, but not stated if investigator blinded\\
    \cite{Cul2012_8}&Moderate &Native \& spiked samples used, but not stated if analyst blinded\\
    \cite{Huy2023_734}&Moderate &Native samples used, but not stated if the analyst blinded\\
    \cite{Kan2014_17}&Moderate &Native samples used, but not stated if the analyst blinded\\
    \cite{Koe2014_43}&Moderate &Native \& spiked samples used, but not stated if analyst blinded\\
    \cite{Lee2022_}&Moderate &Native \& spiked samples used, but not stated if analyst blinded\\
    \cite{Tri2020_1417}&Moderate &Native \& spiked samples used, but not stated if analyst blinded\\
    \cite{Yam2021_22}&Moderate &Native samples used, but not stated if the analyst blinded\\
    \midrule
    \cite{Her2017_244}&High &Spiked samples used\\
    \cite{Kir2021_267}&High &Spiked samples used for high concentrations\\
    \cite{Yan2017_9304}&High &Spiked samples of health donor plasma purchased from vendors, unknown how they were processed. Parallelism was done with spiked buffer only. \\
    \cite{Lac2013_897}&High &Spiked samples  used\\
    \cite{Lif2019_30}&High &Left over samples purchased from vendors, unknown whether or not they had been spiked or how they were processed. \\
    \cite{Son2016_58}&High &Pooled plasma samples first immunodepleted \& then artificially spiked with amyloid-beta\\
    \cite{Wil2022_761}&High &Native \& spiked samples used \& analyst blinded, but dilution not done in buffer\\
    \cite{Woj2023_}&High &Spiked samples  used\\
    \cite{Kra2017_465}&High &Spiked samples of human plasma \& spiked PBS solution used\\
    \bottomrule
  \end{tabular}
  \normalsize
\end{table*}

\subsection{Results of Individual Studies}\label{r_results_indiv_studies}
Detailed results for each included study are summarised in
Table~\ref{t_ppp}. 

\begin{table*}\centering
  \caption{\href{https://discovery.ucl.ac.uk/id/eprint/10185313/1/parallel-v9-accepted.pdf}{Partial
      parallelism plots} (PP-plots) from studies with available raw
    data, obtained either through publication or direct email request
    (details provided in the Methods section). Comprehensive PP-plots
    for all included studies are presented in
    Figures~\ref{f_But2021_58} to~\ref{f_Kra2017_465}. The specific
    dilution ranges demonstrating partial parallelism are explicitly
    indicated.  Studies are sorted as in
    Table~\ref{t_bias}.\label{t_ppp}}
  \footnotesize
  \begin{tabular}{llp{4.5cm}p{2.5cm}}
    \toprule
    Ref & PP-plot & Partial parallelism achieved & Dilution range \\
    \midrule
    \cite{But2021_58}& Figure~\ref{f_But2021_58}&Yes& 1:4--1:16\\
    \cite{Kru2016_153564}& Figure~\ref{f_Kru2016_153564}&No&---\\
    \cite{Bay2021_}& Figure~\ref{f_Bay2021_}& Yes (Lilly pTau217 assay)& 1:8--1:32\\
        &&Not for the pTau181 assays&---\\
        &&Not for the pTau231 assay&---\\
    \cite{Cul2012_8}& Figure~\ref{f_Cul2012_8}&No&---\\
    \cite{Huy2023_734}& Figure~\ref{f_Huy2023_734}&No&---\\
    \cite{Kan2014_17}& Figure~\ref{f_Kan2014_17}&No&---\\
    \cite{Koe2014_43}& Figure~\ref{f_Koe2014_43}&No&---\\
    \cite{Lee2022_}& Figure~\ref{f_Lee2022_}&Yes&1:2--1:8\\
    \cite{Tri2020_1417}& Figure~\ref{f_Tri2020_1417}&No&---\\
    \cite{Yam2021_22}& Figure~\ref{f_Yam2021_22}&No&---\\
    \cite{Her2017_244}& Figure~\ref{f_Her2017_244}&No&---\\
    \cite{Kir2021_267}& Figure~\ref{f_Kir2021_267}&No&---\\
    \cite{Yan2017_9304}& Figure~\ref{f_Yan2017_9304}&No&---\\
    \cite{Lac2013_897}& Figure~\ref{f_Lac2013_897}&No&---\\
    \cite{Lif2019_30}& Figure~\ref{f_Lif2019_30}&Yes (pTau181)&1:1.43--1:2.5\\
        & &Not for the total Tau assay&---\\
    \cite{Son2016_58}& Figure~\ref{f_Son2016_58}&Yes&1:4--1:16\\
    \cite{Wil2022_761}& Figure~\ref{f_Wil2022_761}&Yes& 1:4--1:16\\
    \cite{Woj2023_}& Figure~\ref{f_Woj2023_}&No&---\\
    \cite{Kra2017_465}& Figure~\ref{f_Kra2017_465}&No (EDTA)&---\\
        & &Yes (Citrate, Heparin)&1:100--1:10,000\\
    \bottomrule
  \end{tabular}
  \normalsize 
\end{table*}

\subsubsection{Amyloid \boldmath$\beta$}
The amyloid cascade hypothesis, proposed by Hardy and
Higgins~\cite{Har1992_184}, has positioned amyloid $\beta$ (A$\beta$)
as a central pathological driver in AD, following proteolytic cleavage
of the amyloid precursor protein (APP). Quantification of A$\beta$
peptides, particularly A$\beta_{1-42}$ and A$\beta_{1-40}$, and their
ratios has since become integral to subsequent revisions of diagnostic
criteria for AD~\cite{Dub2007_46,Dub2014_614,Dub2024_1304}. However,
aggregation-prone properties of
A$\beta$~\cite{Pit1998_832,Rez2023_1427} complicate immunoassay
quantification, as epitope masking and altered conformations impair
antibody recognition and disrupt dilutional
parallelism~\cite{Lu2011_143}.

On review of individual
\href{https://discovery.ucl.ac.uk/id/eprint/10185313/1/parallel-v9-accepted.pdf}{partial
  parallelism plots} this emerges as a consistent and persistent
analytical problem:

\begin{figure}\centering
  \includegraphics[width=0.4\textwidth]{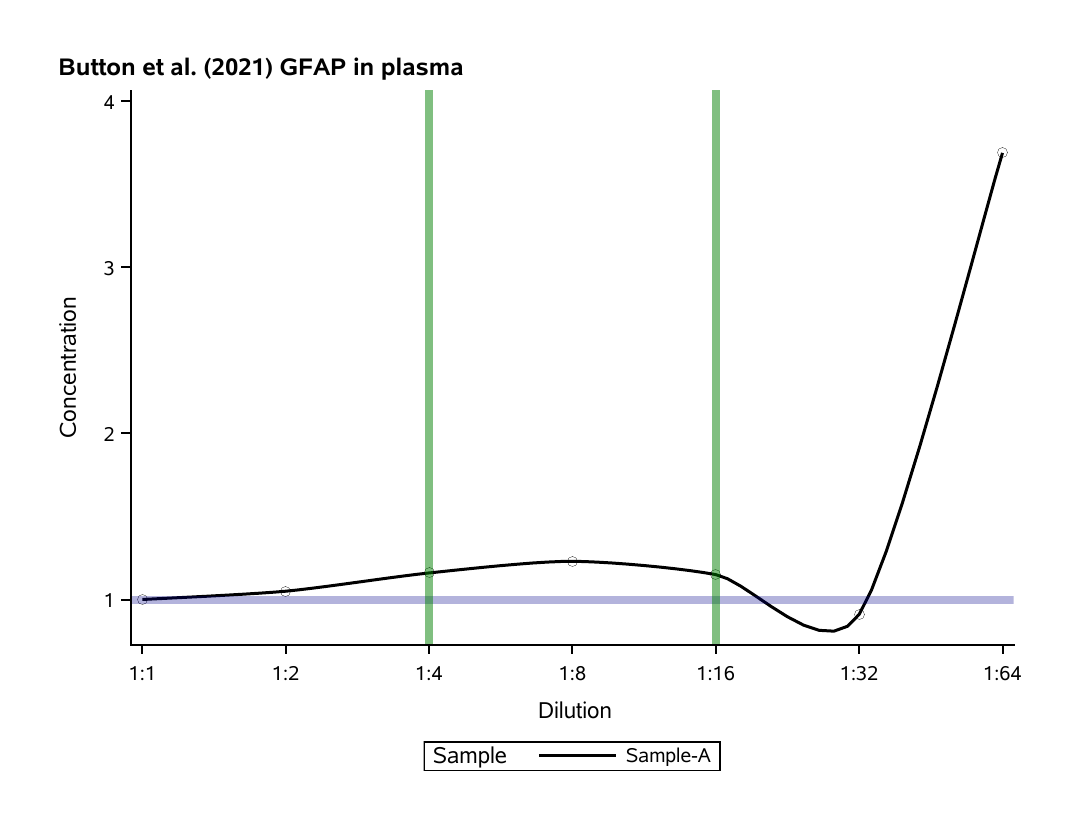}
  \caption{Partial Parallelism Plot for endogenous GFAP from plasma
    samples.  The partial parallelism plot illustrates the average
    GFAP concentrations from three native plasma samples, based on the
    data summarised in Table 7 of reference~\cite{But2021_58}. The
    individual sample data are shown in Figure 4D of the same
    reference~\cite{But2021_58}. Partial parallelism is evident for
    dilution steps ranging from 1:4 to 1:16. However, at the final
    dilution step of 1:64, while GFAP remains detectable by the assay,
    the partial parallelism plot clearly indicates a loss of
    parallelism.}\label{f_But2021_58}
\end{figure}

\begin{figure}\centering
  \includegraphics[width=0.34\textwidth]{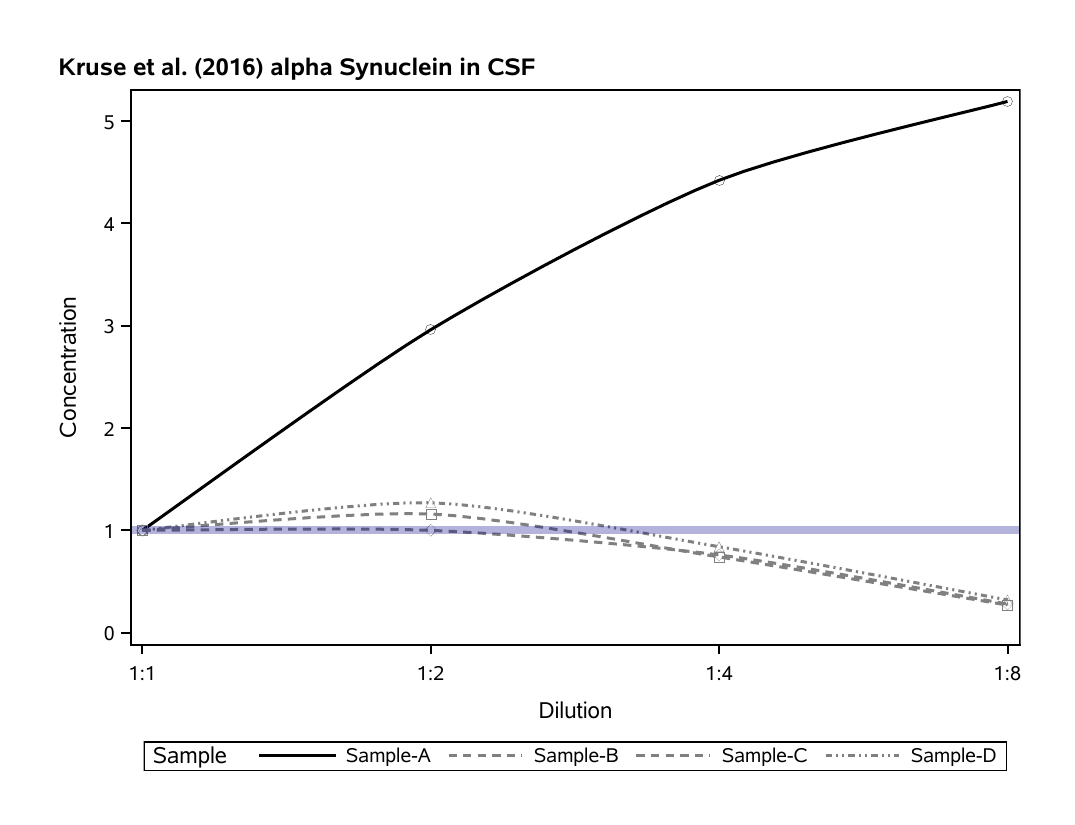}
  \includegraphics[width=0.34\textwidth]{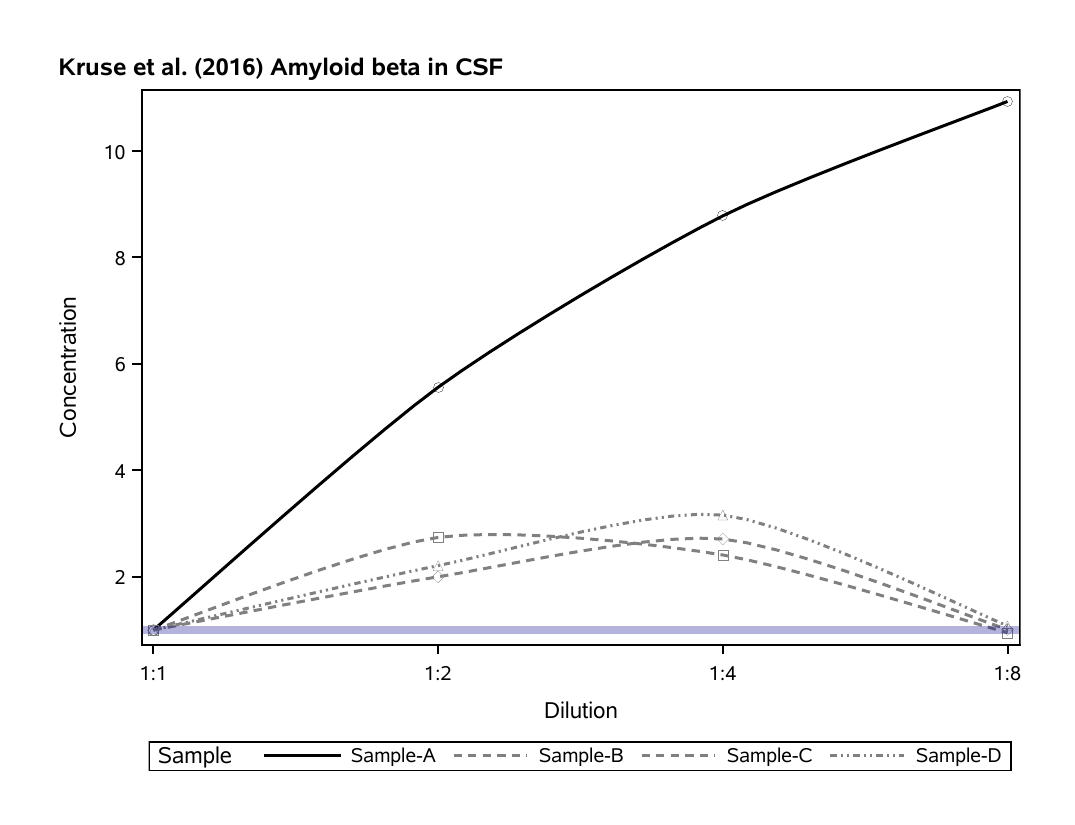}
  \includegraphics[width=0.34\textwidth]{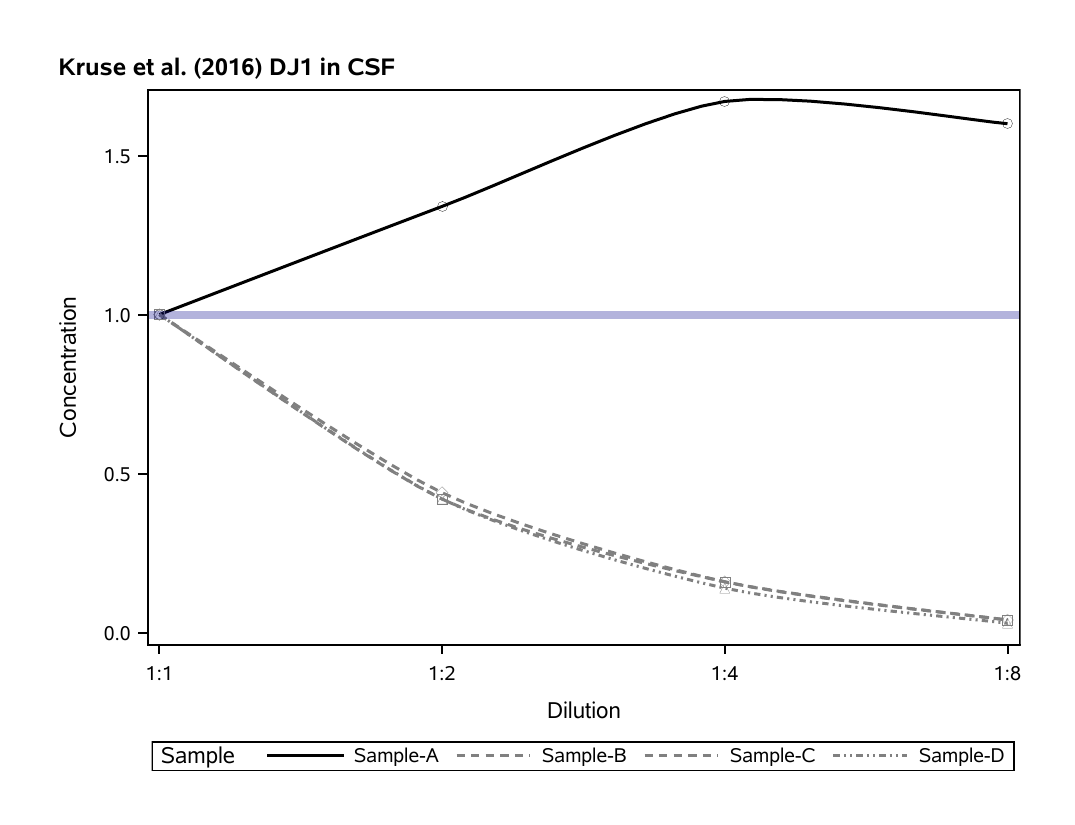}
  \includegraphics[width=0.34\textwidth]{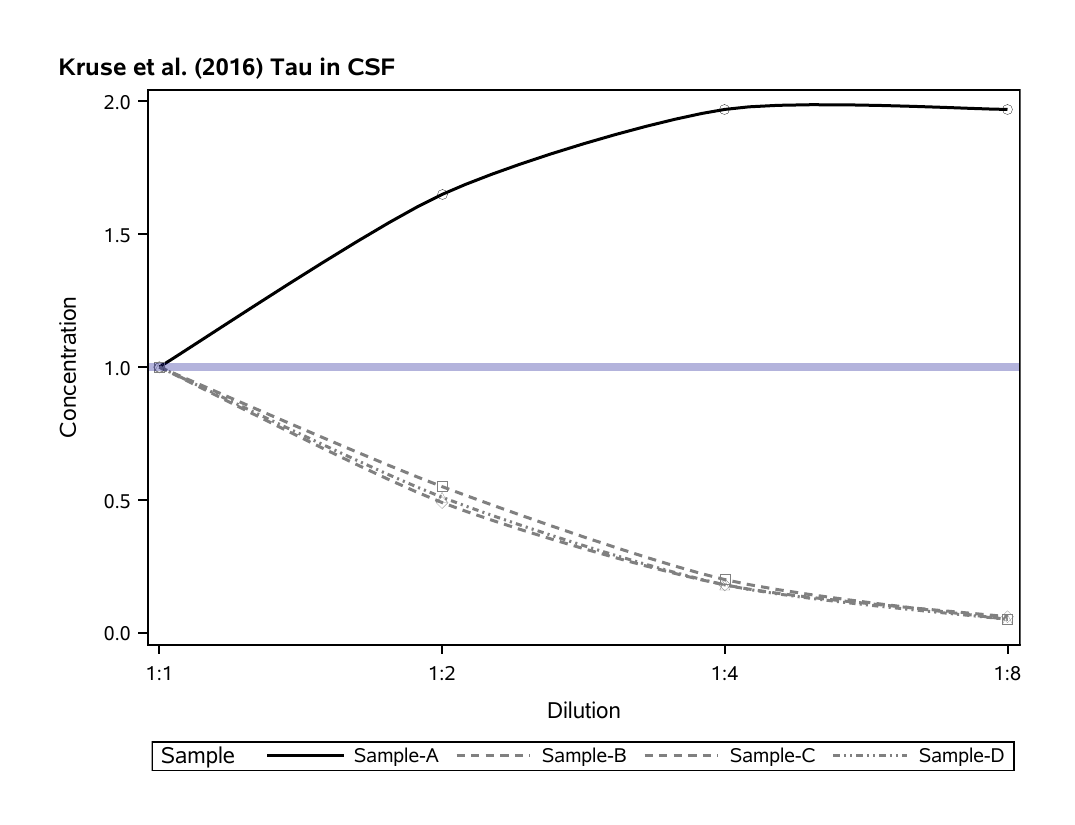}
  \caption{Partial parallelism plots for four proteins:
    $\alpha$-synuclein, A$\beta_{1-42}$, DJ-1, total tau. Raw data
    were obtained from Supplementary Table 5 in
    reference~\cite{Kru2016_153564}. The plots demonstrate a lack of
    partial parallelism across the tested dilution
    ranges.}\label{f_Kru2016_153564}
\end{figure}

\begin{itemize}
\item \textbf{A \boldmath$\beta_{1-40}$}: None of the assessed immunoassays
  achieved partial parallelism across three independent
  studies~\cite{Yam2021_22,Kir2021_267,Lac2013_897}. For each study
  deviations from expected parallelism was shown in the partial
  parallelism plots (see Figures~\ref{f_Yam2021_22},
  \ref{f_Kir2021_267}, \ref{f_Lac2013_897}).

\item \textbf{A \boldmath$\beta_{1-42}$}: Likewise, most studies
  failed to demonstrate partial
  parallelism~\cite{Yam2021_22,Kir2021_267,Lac2013_897,Kru2016_153564,Cul2012_8}. A
  single study reported successful parallelism~\cite{Son2016_58};
  however, this study was flagged as high risk for bias
  (Table~\ref{t_bias}), as it involved spiking immunodepleted pooled
  plasma with synthetic A$\beta$, a method that may not replicate the
  conformational complexity of endogenous peptides (see
  Figures~\ref{f_Cul2012_8}, \ref{f_Kru2016_153564},
  \ref{f_Yam2021_22}, \ref{f_Kir2021_267}, \ref{f_Lac2013_897},
  \ref{f_Son2016_58}).

\item \textbf{Other A \boldmath$\beta$ species}: No evidence of
  partial parallelism was found for truncated A$\beta$
  peptides~\cite{Kan2014_17}, aggregated forms~\cite{Her2017_244}, or
  oligomeric A$\beta$ in EDTA plasma
  samples~\cite{Kra2017_465}. However, in the latter study, the use of
  citrate or heparin as anticoagulants enabled partial parallelism,
  highlighting the importance of pre-analytical variables for future
  immunoassay development (see Figure~\ref{f_Kra2017_465}).
\end{itemize}

Although current diagnostic strategies emphasize A$\beta$ cleavage
products, APP itself remains a biomarker of
interest~\cite{Obr2011_185}. Recent attention has turned to its
soluble fragments (sAPP$\alpha$, sAPP$\beta$)~\cite{Ant2025_63},
though the only identified assay for these analytes~\cite{Waa2016_21}
could not be included in this systematic review.  Together, these
findings underscore ongoing analytical challenges in A$\beta$
biomarker development. Despite three decades of intensive
research~\cite{Har1992_184}, robust quantification of circulating
A$\beta$, especially A$\beta_{1-42}$, remains elusive.

\subsubsection{\boldmath Alpha-Synuclein}
$\alpha$-Synuclein remains a central candidate biomarker in the
differential diagnosis of movement disorders, particularly PD and
other
synucleinopathies~\cite{Dan2009_192,Hon2010_713,Gro2019_593,Mag2022_}. However,
its intrinsic biochemical properties pose substantial challenges for
quantitative assay development. Notably, $\alpha$-synuclein exhibits a
pronounced tendency to aggregate~\cite{Uve2007_17,Dan2009_192}, a
characteristic that directly impairs immunoassay performance by
compromising parallelism~\cite{Lu2011_143}. In line with these known
biophysical limitations, the present review identified a failure to
achieve partial parallelism in the only study that evaluated this
criterion using standard dilution series~\cite{Kru2016_153564} (see
Figure~\ref{f_Kru2016_153564}).

\begin{figure*}\centering
  \includegraphics[width=0.38\textwidth]{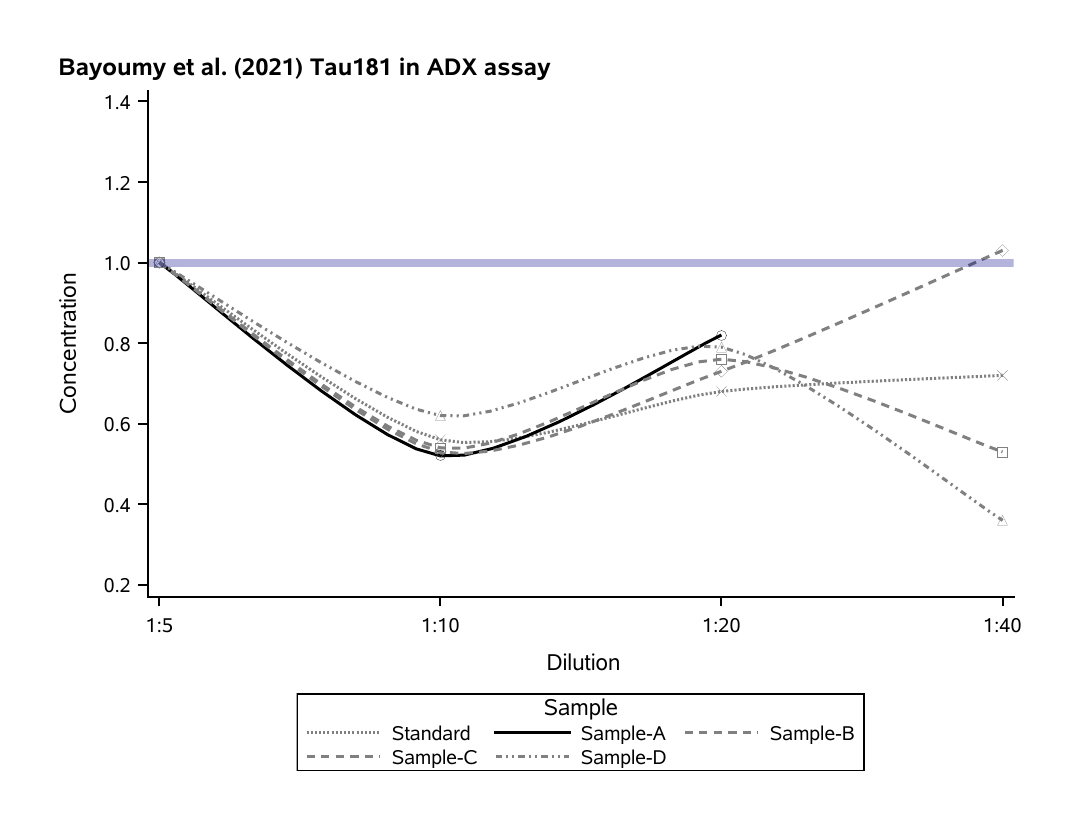}
  \includegraphics[width=0.38\textwidth]{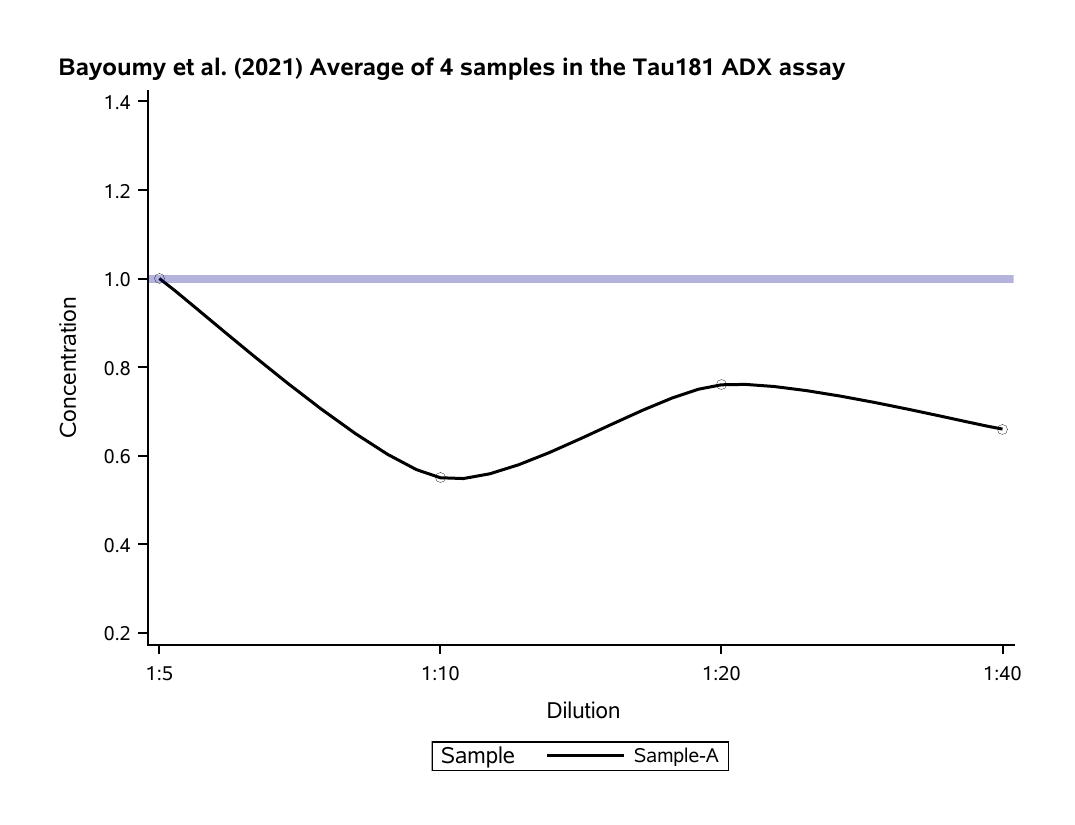}
  \includegraphics[width=0.38\textwidth]{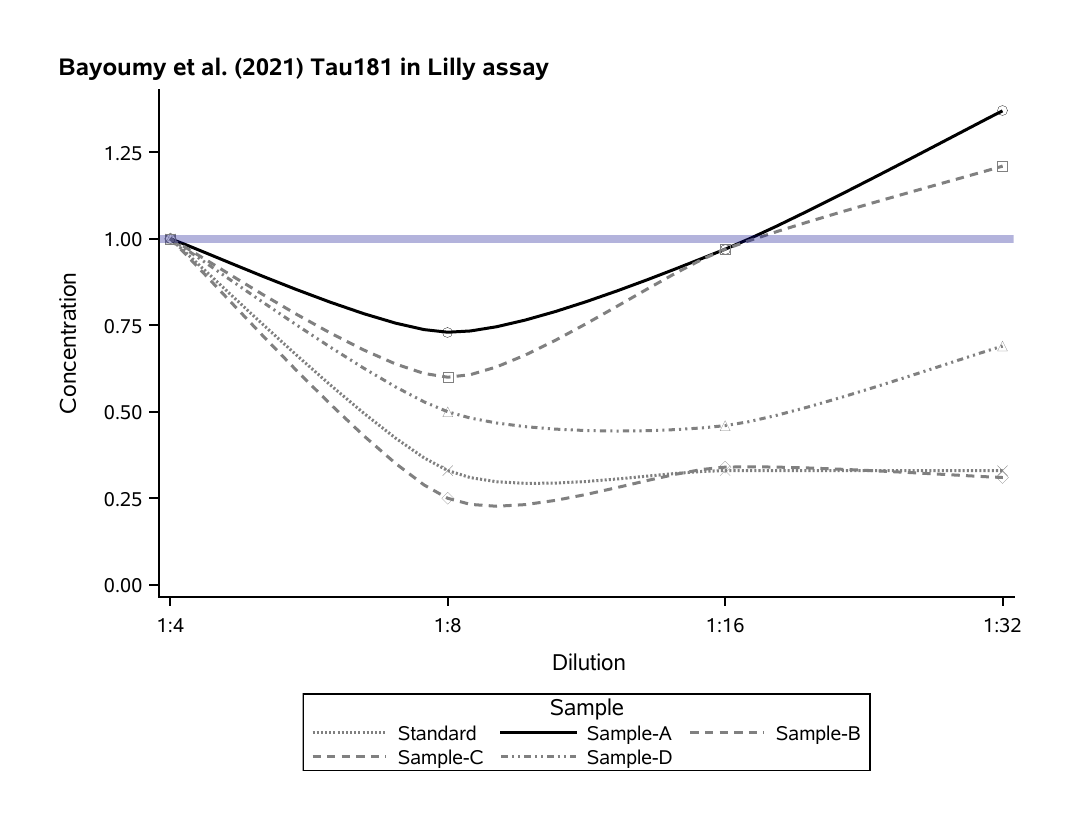}
  \includegraphics[width=0.38\textwidth]{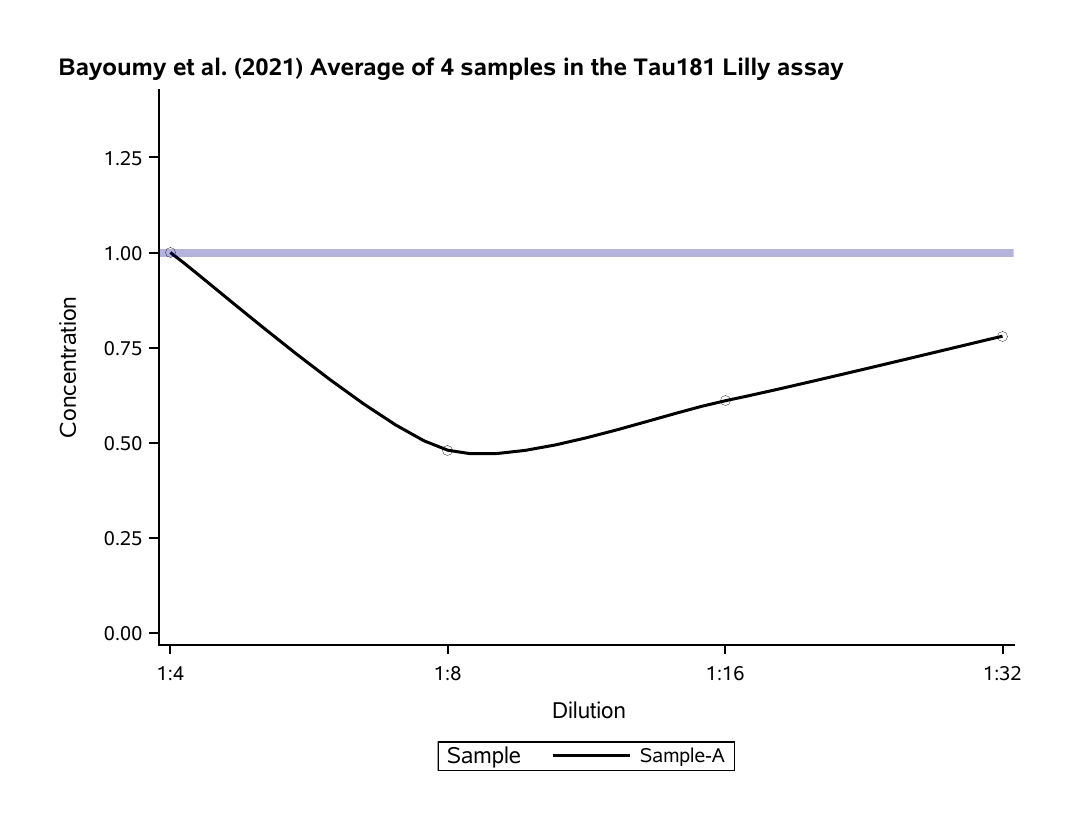}
  \includegraphics[width=0.38\textwidth]{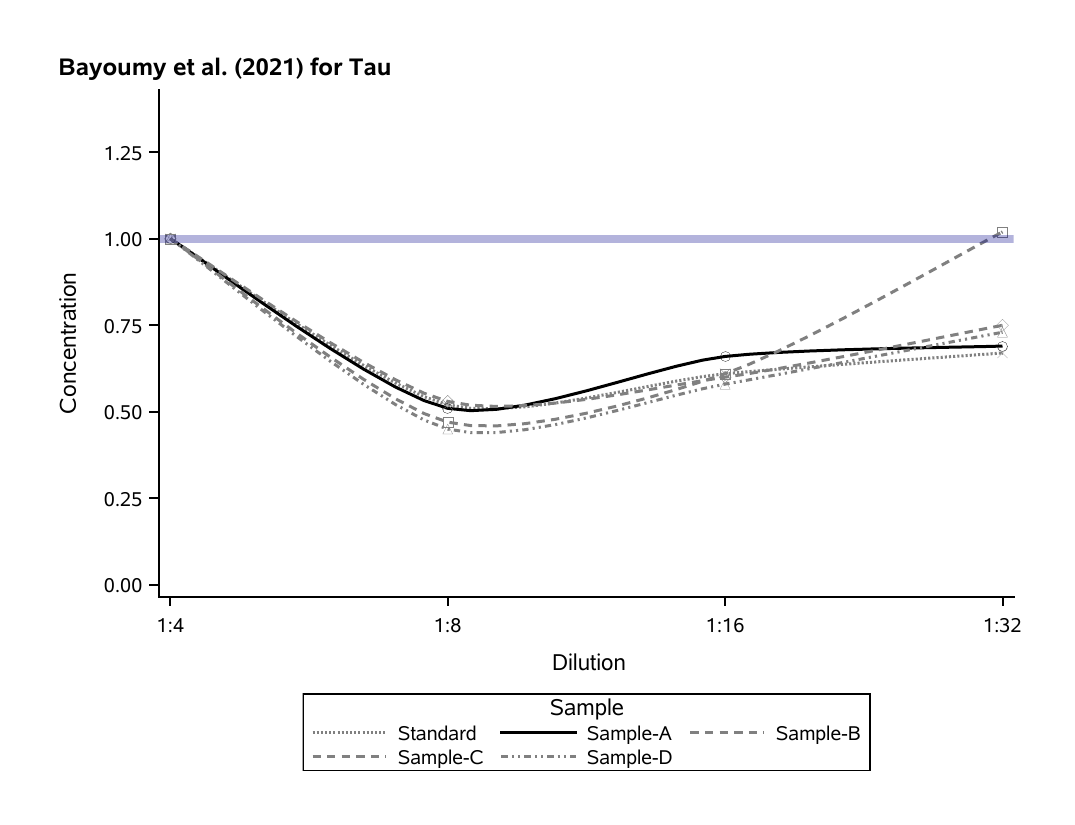}
  \includegraphics[width=0.38\textwidth]{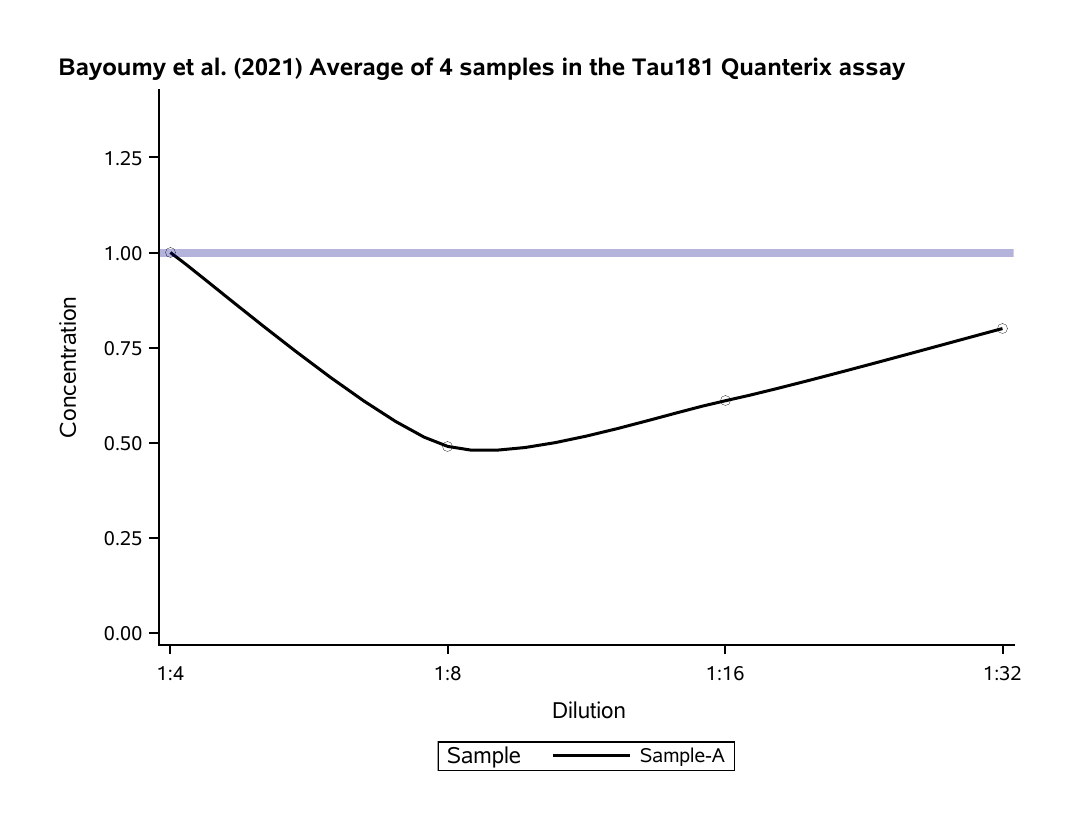}
  \includegraphics[width=0.38\textwidth]{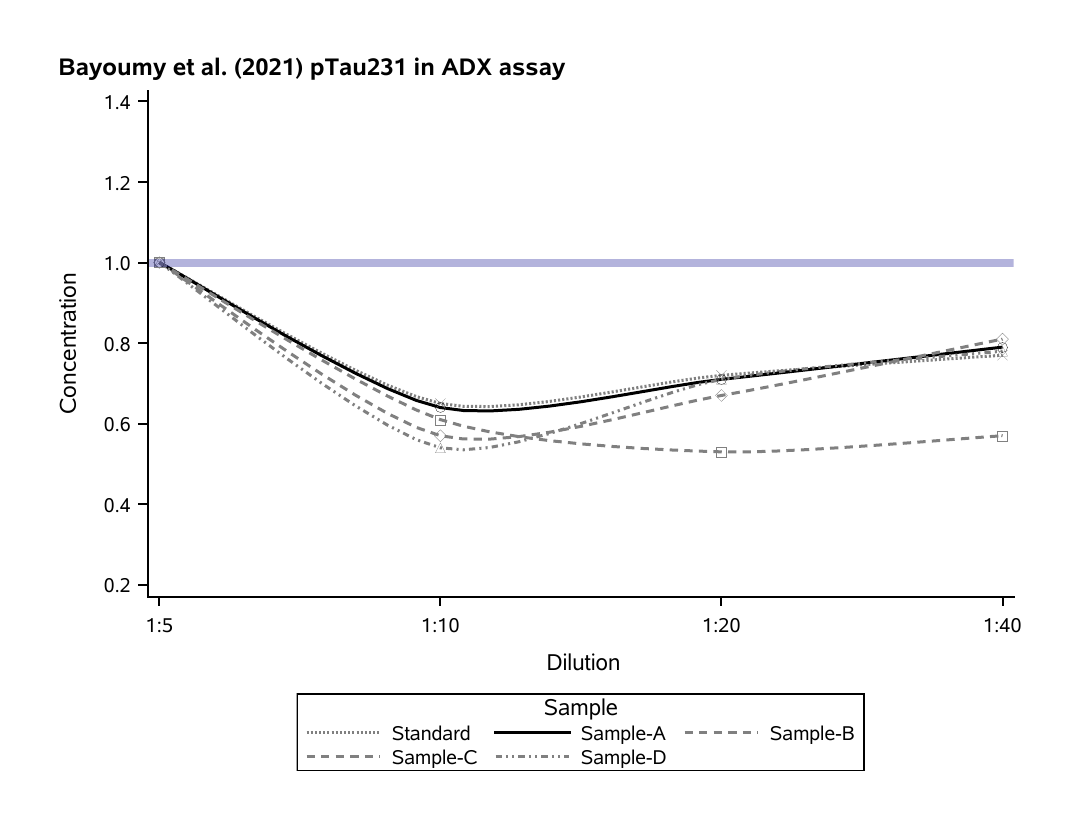}
  \includegraphics[width=0.38\textwidth]{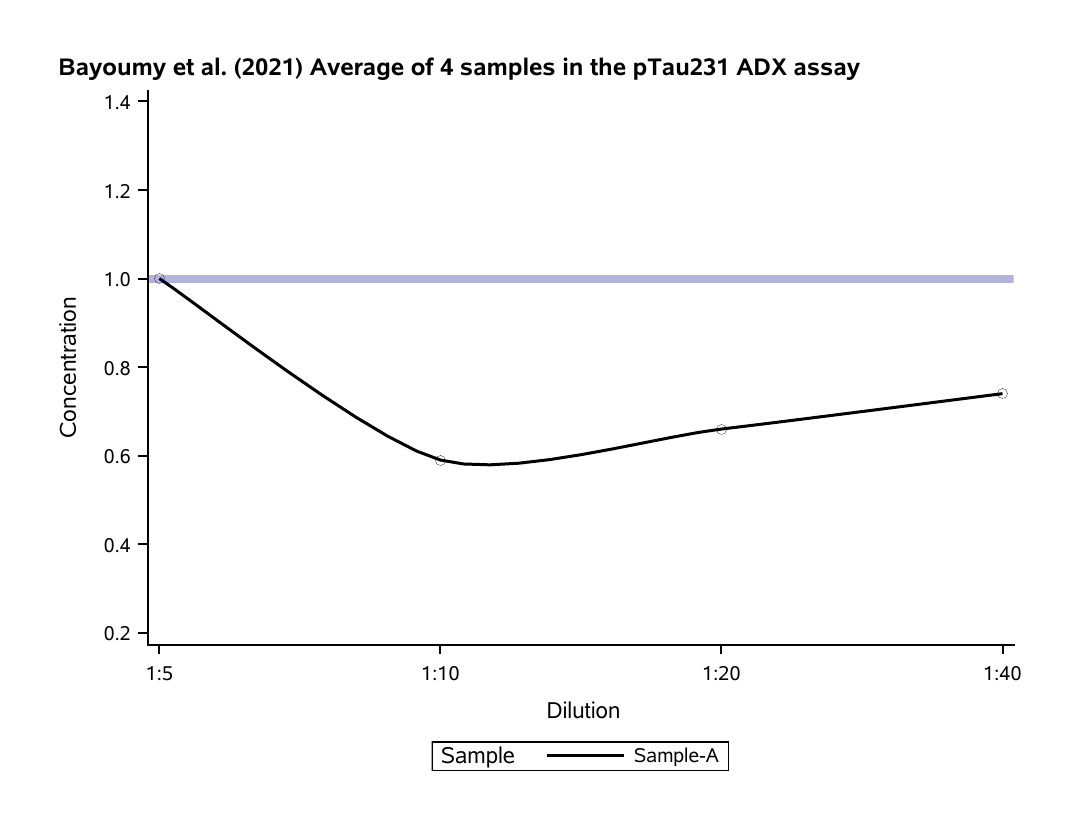}
  \includegraphics[width=0.38\textwidth]{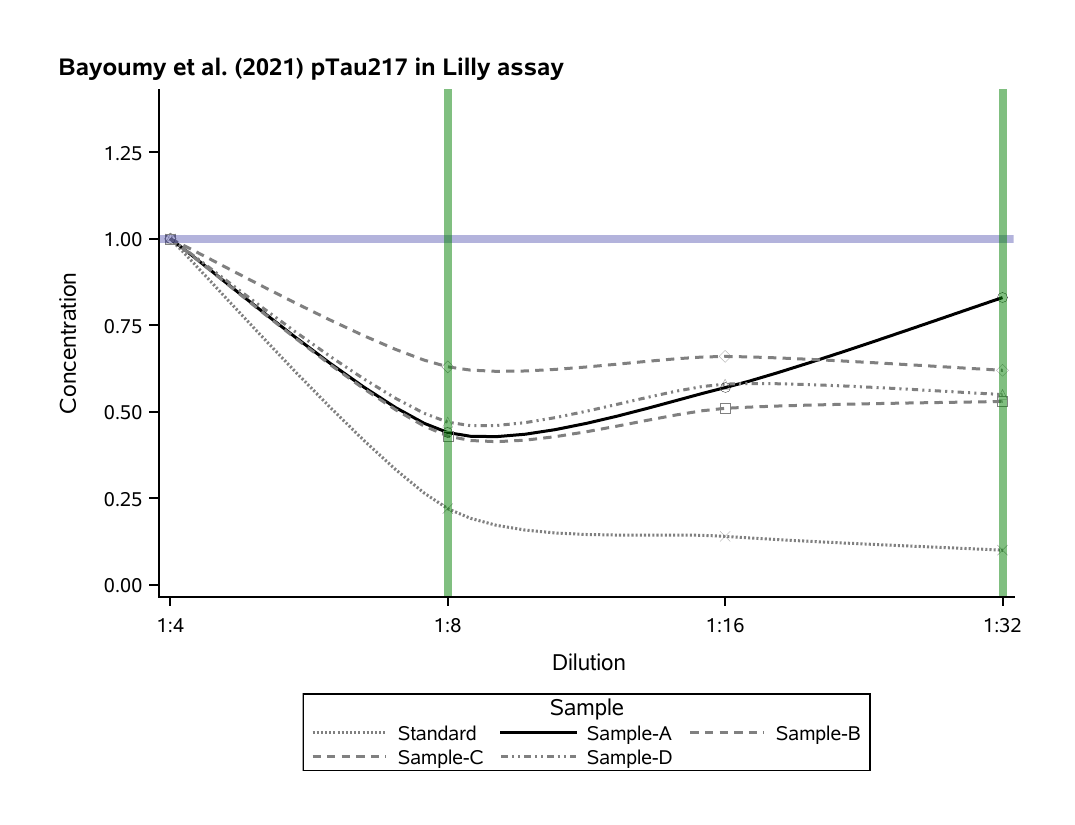}
  \includegraphics[width=0.38\textwidth]{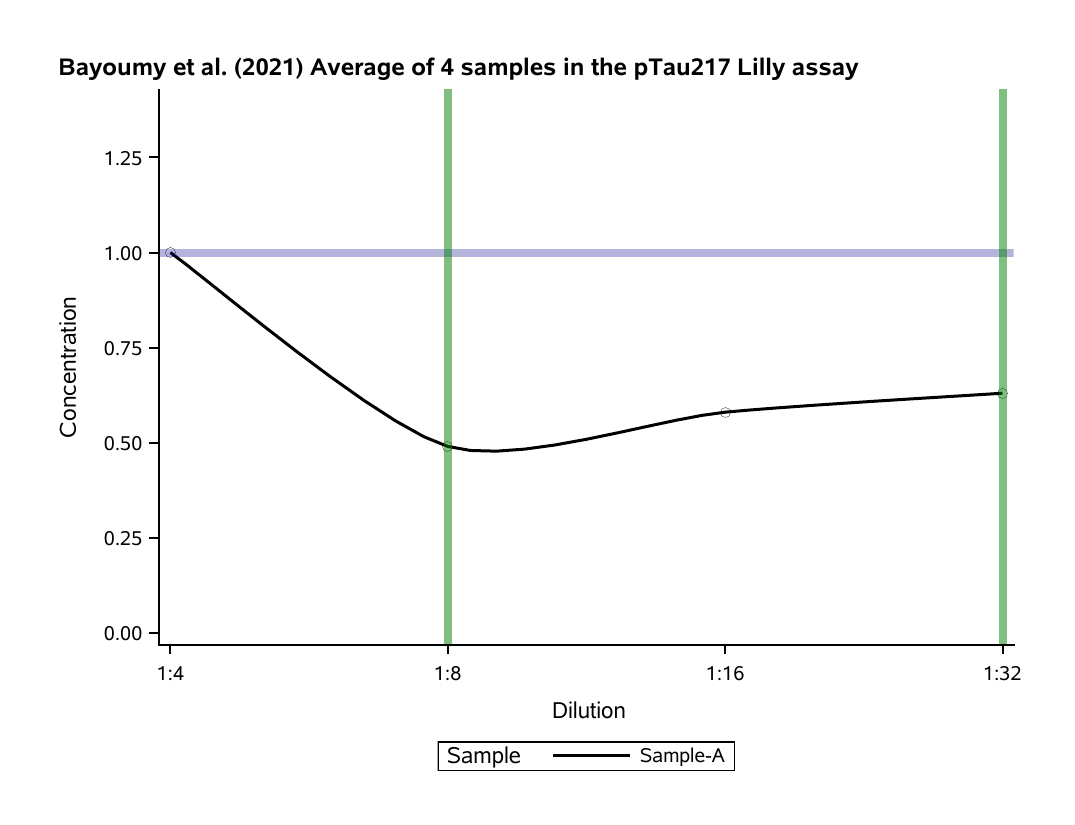}
  \caption{Partial parallelism plot for tau from native plasma
    samples~\cite{Bay2021_}. Raw data were kindly provided by the
    authors. The plots show a lack of partial parallelism for all
    pTau181 and pTau231 assays across the tested dilution ranges. In contrast,
    partial parallelism is observed for the Lilly pTau217 assay
    between dilutions of 1:8 to 1:32.}\label{f_Bay2021_}
\end{figure*}

\subsubsection{DJ-1}
DJ-1 (PARK7) has been implicated in the pathophysiology of both
familial and sporadic Parkinson's disease, with several studies
reporting elevated levels in patient
cohorts~\cite{Hon2010_713,War2006_967}. Despite its biomarker
potential, the analytical validation of DJ-1 assays remains
insufficient. Notably, published work utilizing in-house Luminex-based
immunoassays did not report formal testing for dilutional
parallelism~\cite{Hon2010_713}. Within the current review, the only
available dataset assessing this parameter demonstrated a lack of
partial parallelism~\cite{Kru2016_153564} (see
Figure~\ref{f_Kru2016_153564}).

\begin{figure}\centering
  \includegraphics[width=0.4\textwidth]{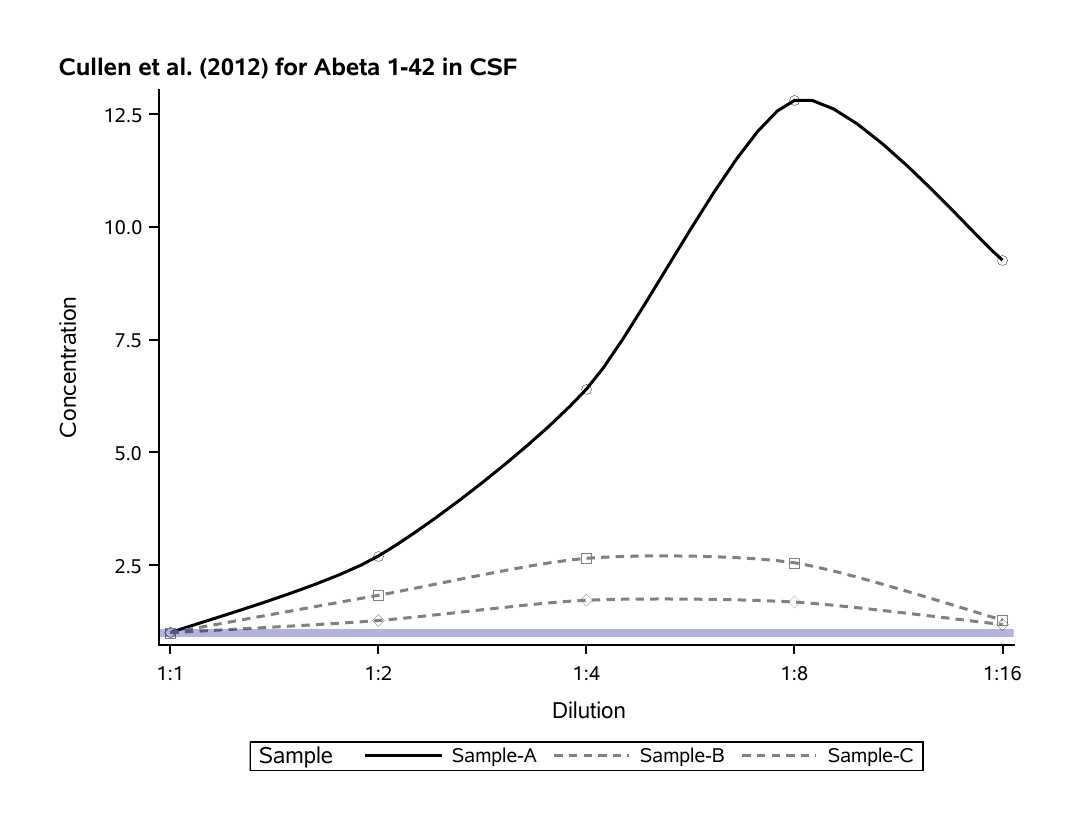}
  \caption{Partial Parallelism plot for A$\beta_{1-42}$ in CSF\@. Data
    were taken from Figure 5 of reference~\cite{Cul2012_8}. Datapoints
    were measured in pixel distances from that Figure. The plots
    demonstrate a lack of partial parallelism across the tested
    dilution ranges.}\label{f_Cul2012_8}
\end{figure}

\begin{figure*}\centering
  \includegraphics[width=0.4\textwidth]{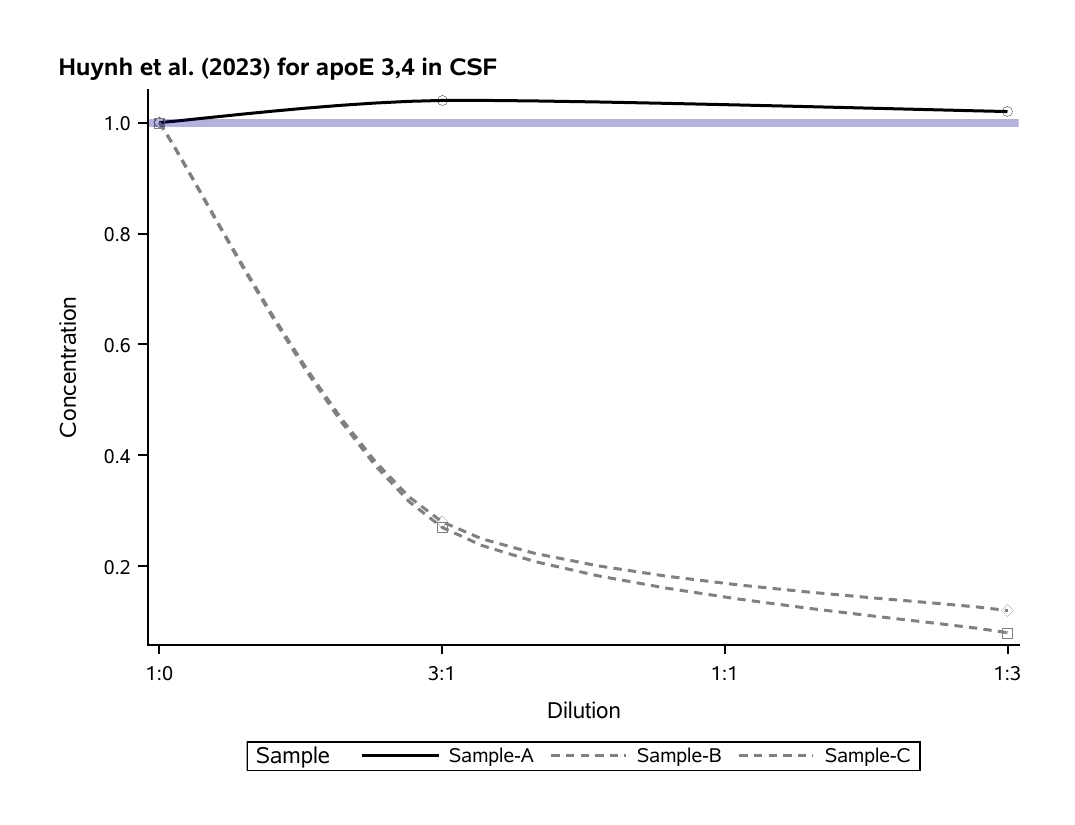}
  \includegraphics[width=0.4\textwidth]{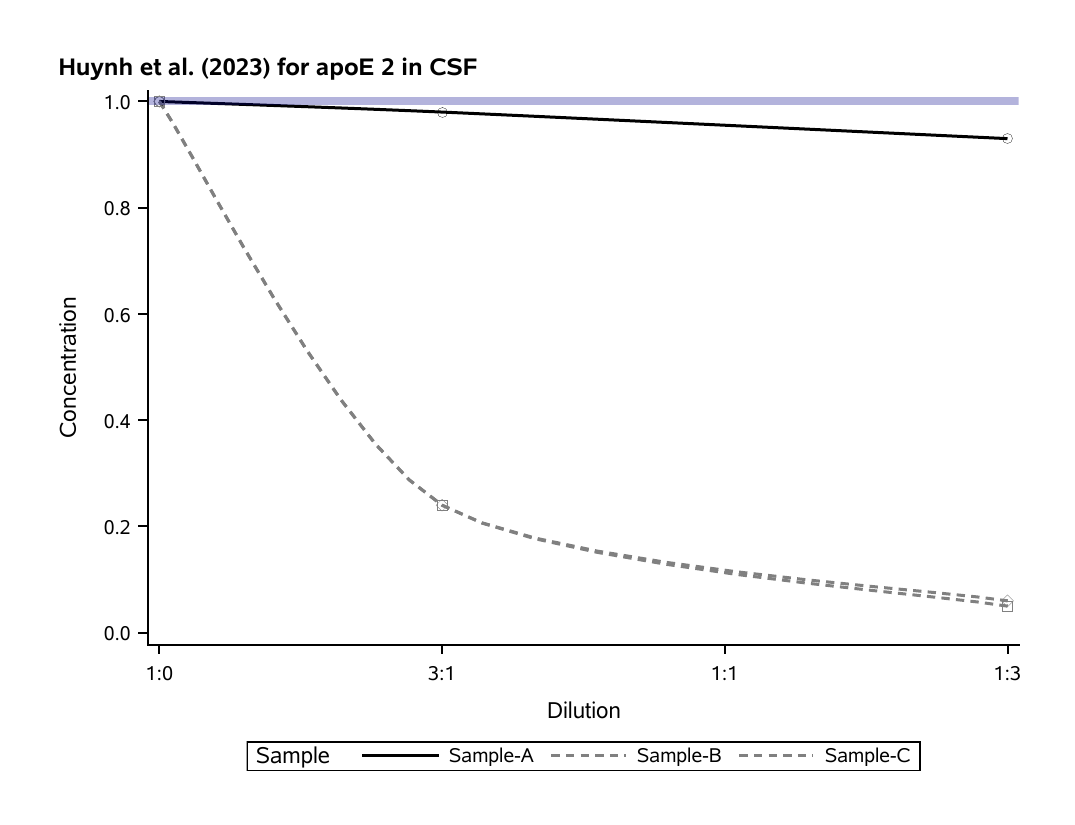}
  \includegraphics[width=0.4\textwidth]{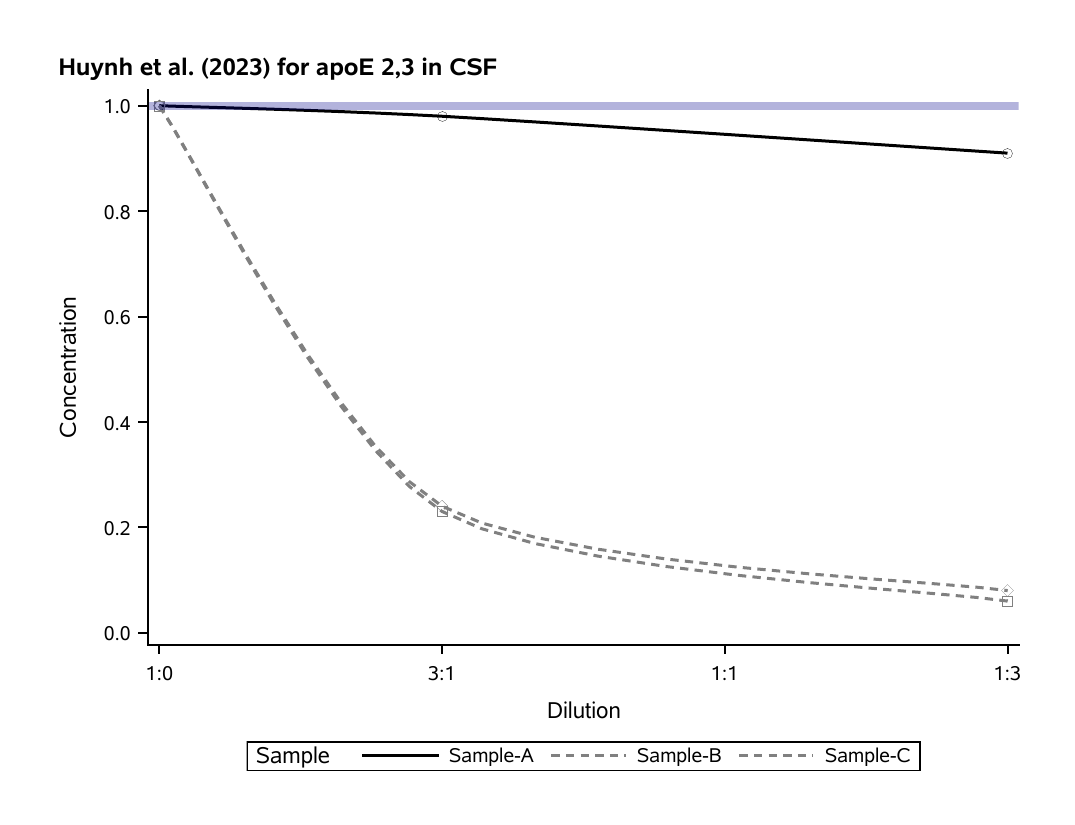}
  \includegraphics[width=0.4\textwidth]{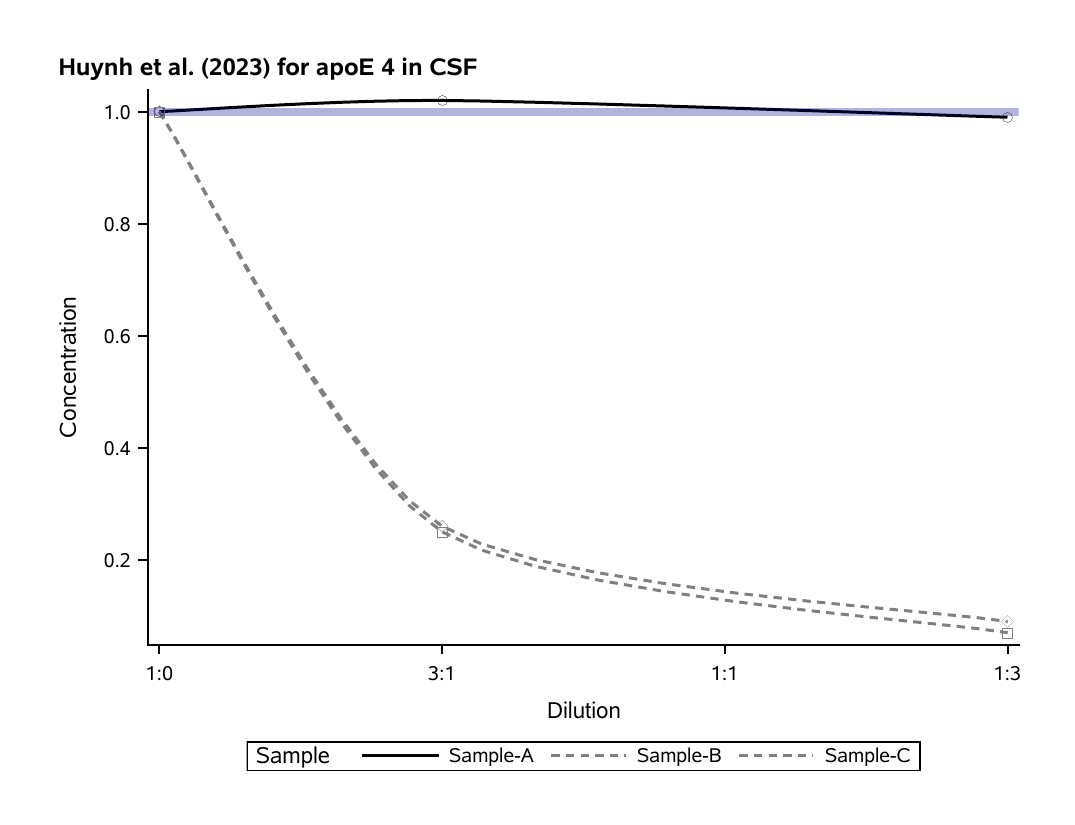}
  \caption{Partial Parallelism Plot for ApoE isoforms in CSF samples.
    This plot depicts the partial parallelism of apoE isoforms diluted
    in lumbar CSF, ventricular CSF, and bovine CSF\@. The data are
    derived from Figure 4 of reference~\cite{Huy2023_734}. The
    measured peptides include LAVYQAGAR ($\epsilon$3\&$\epsilon$4, SNP
    Arg158), CLAVYQAGAR ($\epsilon$2, SNP Cys158), LGADMEDVCGR
    ($\epsilon$2\&$\epsilon$3, SNP Cys112) and LGADMEDVR ($\epsilon$4,
    Arg112). The plots demonstrate a lack of partial parallelism
    across the tested dilution ranges.}\label{f_Huy2023_734}
\end{figure*}

\begin{figure*}\centering
  \includegraphics[width=0.4\textwidth]{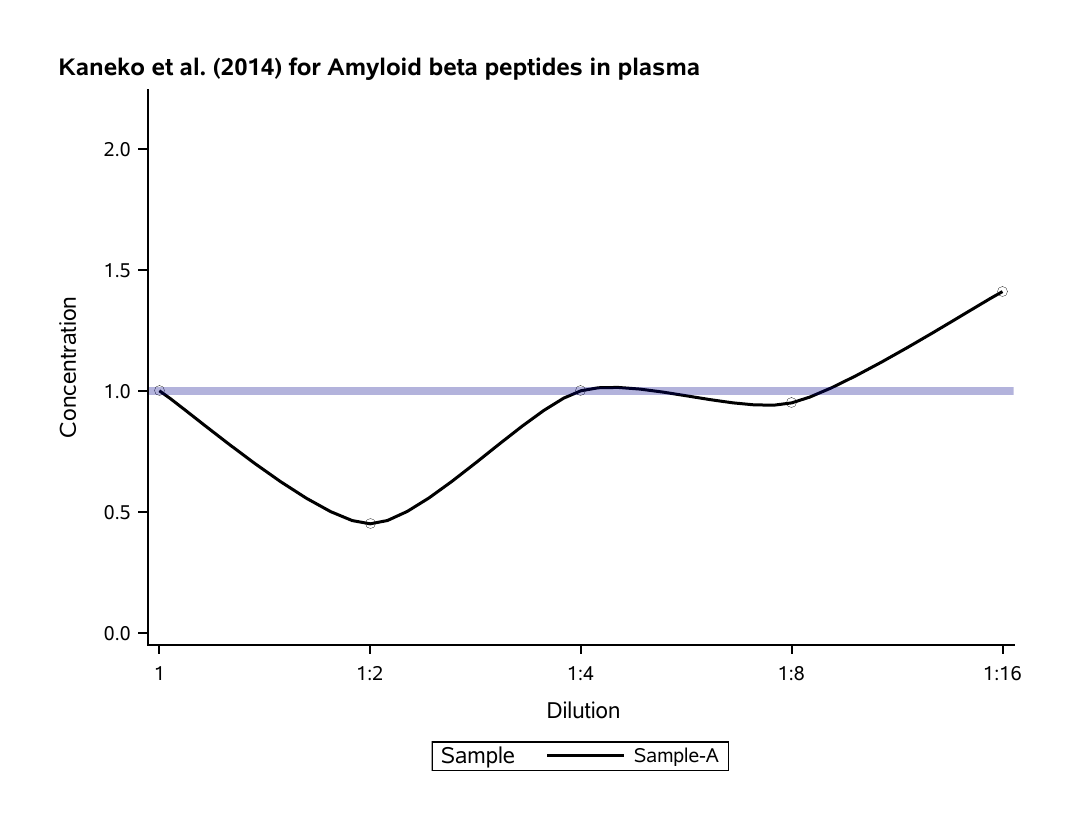}
  \includegraphics[width=0.4\textwidth]{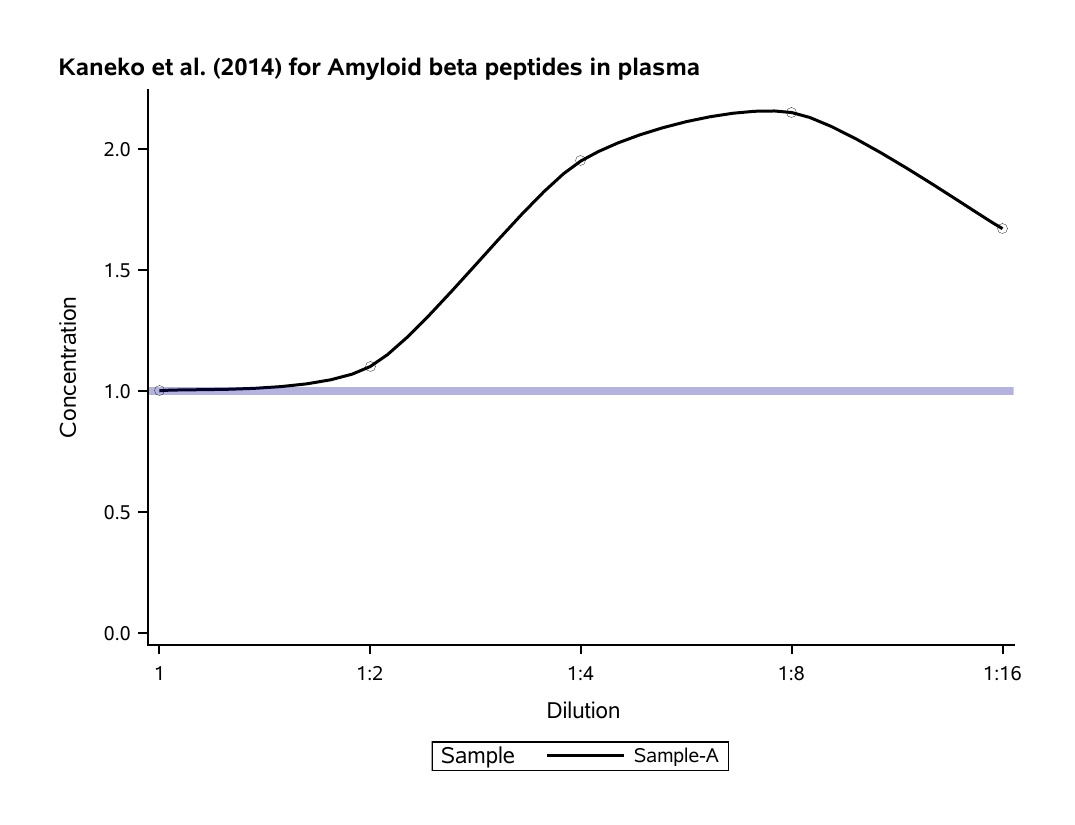}
  \includegraphics[width=0.4\textwidth]{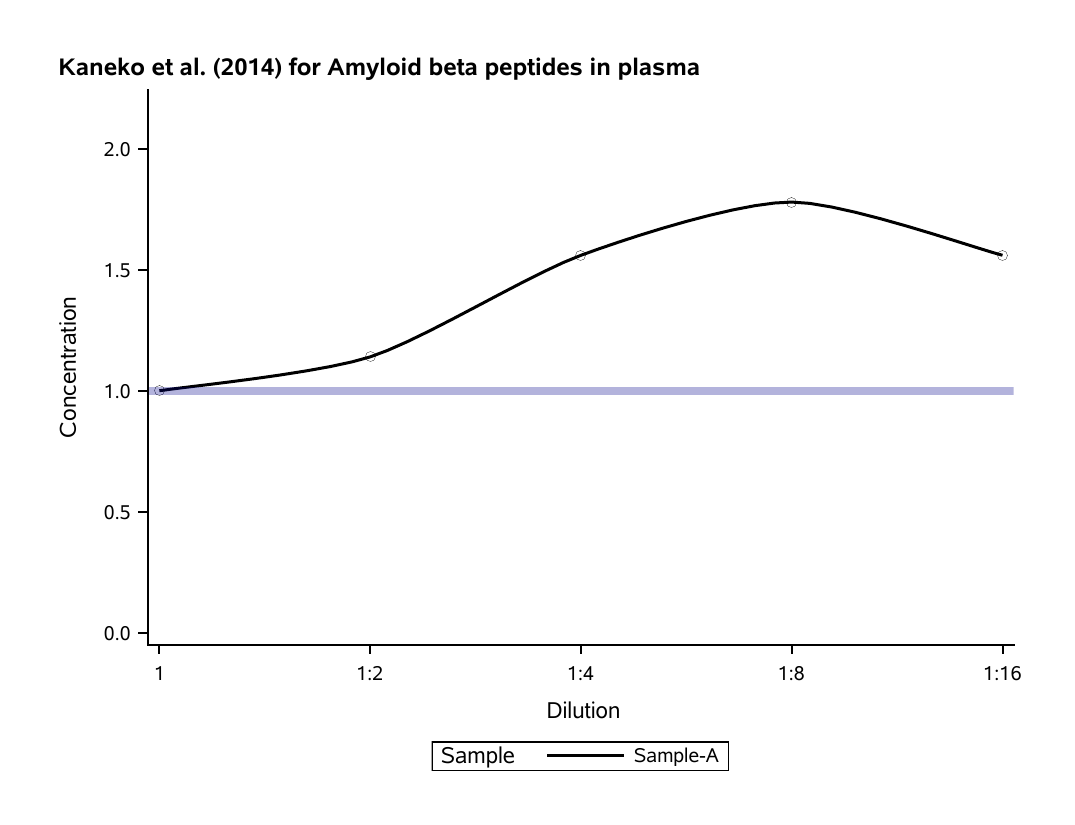}
  \includegraphics[width=0.4\textwidth]{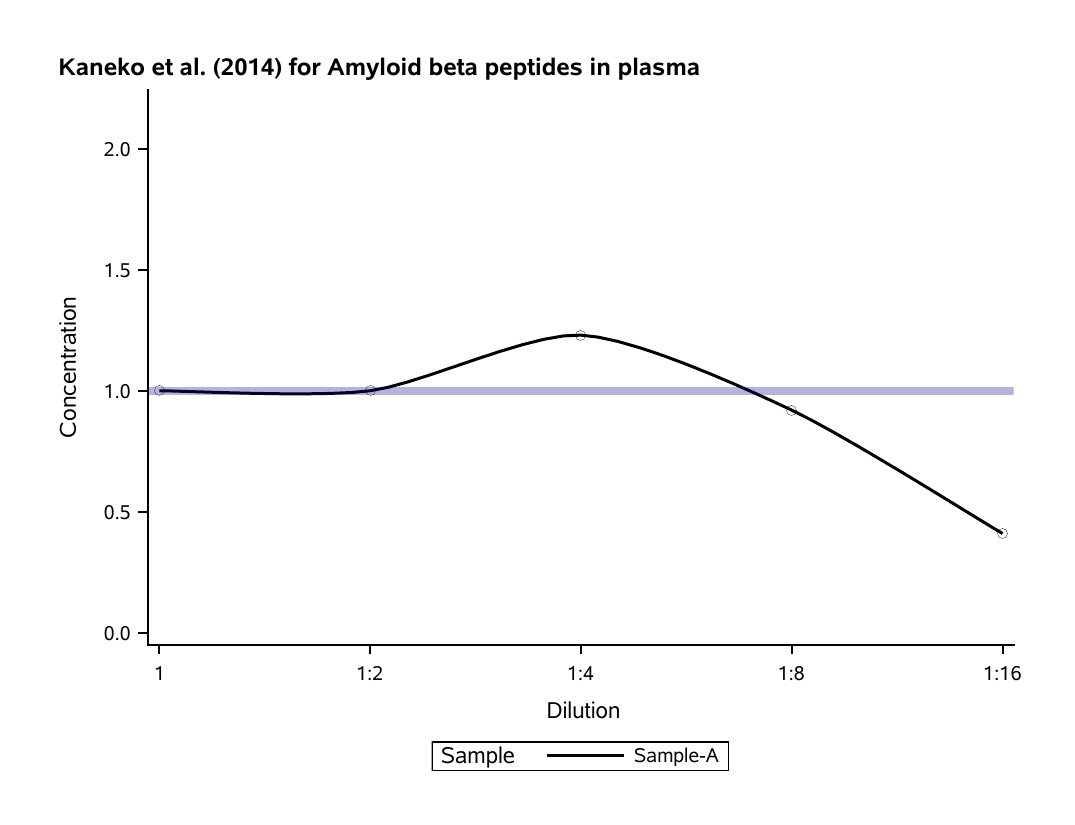}
  \caption{Partial Parallelism Plot for serial dilution of
    Amyloid$\beta$ peptides in plasma.  This plot illustrates the
    partial parallelism for serial dilutions of various
    amyloid-$\beta$ peptides in plasma. Data were analyzed across
    complete dilution ranges (five data points) for comparison in the
    partial parallelism plots. Such data were available for 4 out of
    18 (22\%) plots presented in Figure 4 of
    reference~\cite{Kan2014_17}. The peptides included
    A$\beta_{6-40}$, A$\beta_{5-40}$, A$\beta_{1-38}$,
    A$\beta_{1-40}$.  The plots demonstrate a lack of partial
    parallelism across the tested dilution
    ranges.}\label{f_Kan2014_17}
\end{figure*}

\subsubsection{Tau Protein}
The incorporation of CSF tau into the diagnostic criteria for AD
represented a critical milestone in the development of
neurodegenerative biomarkers~\cite{Dub2007_46}. Early diagnostic
efforts predominantly focused on total tau, supported by robust
clinical data~\cite{Ble2010_131}. However, inter-laboratory
variability in cutoff values ranged from 4fmol/mL to
1140pg/mL~\cite{Ish1999_91,Sjo2001_624,Ara1995_649,Ble1995_231,Van1998_773}. This
highlighted challenges for assay
validation~\cite{Pet2010_432,Dub2007_46}.

\begin{figure}\centering
  \includegraphics[width=0.4\textwidth]{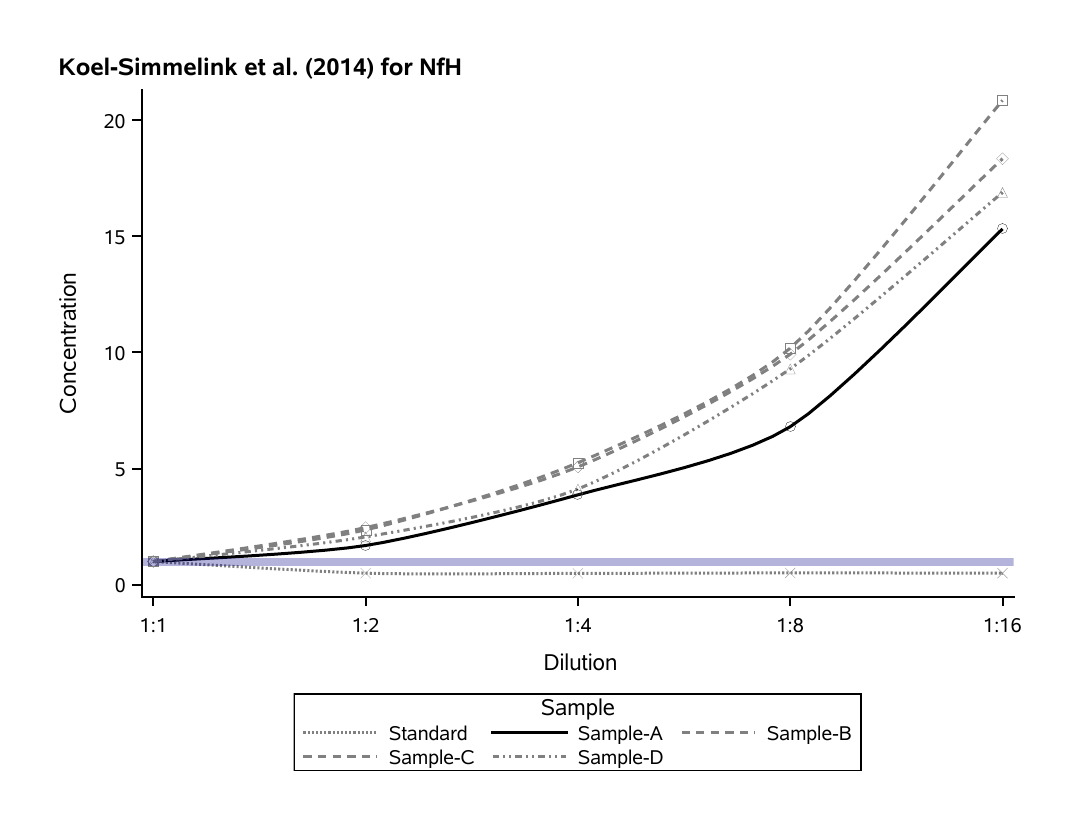}
  \caption{Partial parallelism plot for NfH quantification across
    different sample types. Sample A represents buffer spiked with
    NfH, while Samples B-D are derived from native
    CSF~\cite{Koe2014_43}. The raw data were provided by the
    corresponding author. The plot shows increasing NfH concentrations
    with increasing dilution steps, while the standard curve
    demonstrates partial parallelism only within the 1:1 to 1:16
    dilution range. Overall, the results indicate a lack of partial
    parallelism across the full dilution range
    tested.}\label{f_Koe2014_43}
\end{figure}

Subsequent advances in antibody technology targeting specific
phospho-epitopes of tau have facilitated the development of more
analytically rigorous immunoassays. Among these, assays for
phosphorylated tau at threonine-181 (p181), threonine-217 (p217), and
threonine-231 (p231) are now widely
utilized~\cite{Kar2022_400}. Notably, plasma p217-tau currently
demonstrates the highest diagnostic performance, with sensitivity and
specificity comparable to CSF-based tau
tests~\cite{Ash2024_255,Kar2022_400,Ble2010_131,Dub2007_46}.

The results for parallelism of these assays are summarized as follows:

\begin{itemize}
\item \textbf{Total Tau:} Partial parallelism was not achieved in any
  of the reviewed
  studies~\cite{Kru2016_153564,Yan2017_9304,Lif2019_30}, indicating
  persistent limitations in dilutional linearity across all tests
  available. This lack of parallelism raises concerns about the
  quantitative reliability of total tau assays for individualized or
  longitudinal use (see
  Figures~\ref{f_Kru2016_153564},\ref{f_Yan2017_9304},\ref{f_Lif2019_30}).

\item \textbf{Phospho-Tau p181:} Partial parallelism was demonstrated
  in one study~\cite{Lif2019_30}, but not in two
  others~\cite{Woj2023_,Bay2021_}, suggesting inter-assay variability
  or dependency on specific analytical conditions (see
  Figures~\ref{f_Woj2023_},~\ref{f_Bay2021_}).

\item \textbf{Phospho-Tau p217:} Achieved partial parallelism was
  reported in one study~\cite{Bay2021_}; however, conflicting results
  from another study\cite{Tri2020_1417} may reflect noise or technical
  artifacts. The overall evidence remains promising but requires
  further investigation (see
  Figures~\ref{f_Tri2020_1417},~\ref{f_Bay2021_}).

\item \textbf{Phospho-Tau p231:} No study to date has demonstrated
  satisfactory partial parallelism for p231-tau
  assays~\cite{Woj2023_,Bay2021_}, pointing to persistent limitations
  in current assay performance for this isoform (see
  Figures~\ref{f_Woj2023_},~\ref{f_Bay2021_}).
\end{itemize}

Taken together, these analytical findings support the emerging
consensus that phospho-tau p217 is currently the analytically most
robust and clinically best validated tau
biomarker~\cite{Bay2021_,Kar2022_400,Ash2024_255}.

\begin{figure}\centering
  \includegraphics[width=0.4\textwidth]{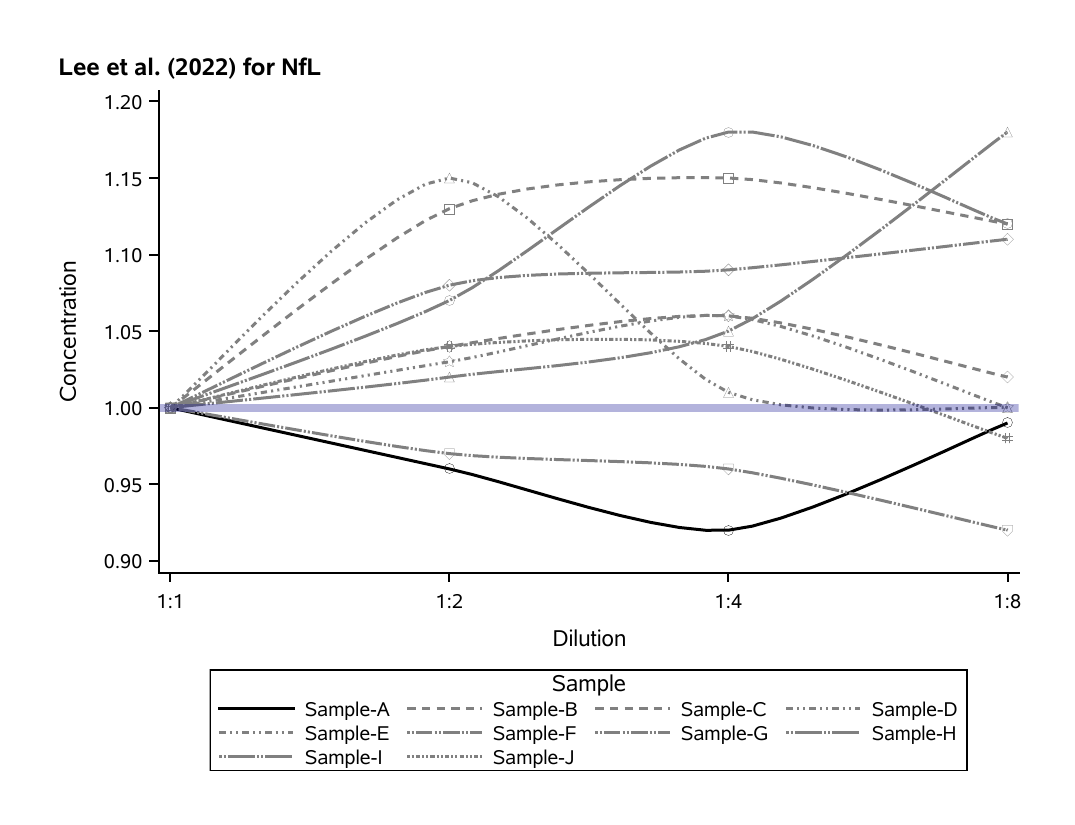}
  \includegraphics[width=0.4\textwidth]{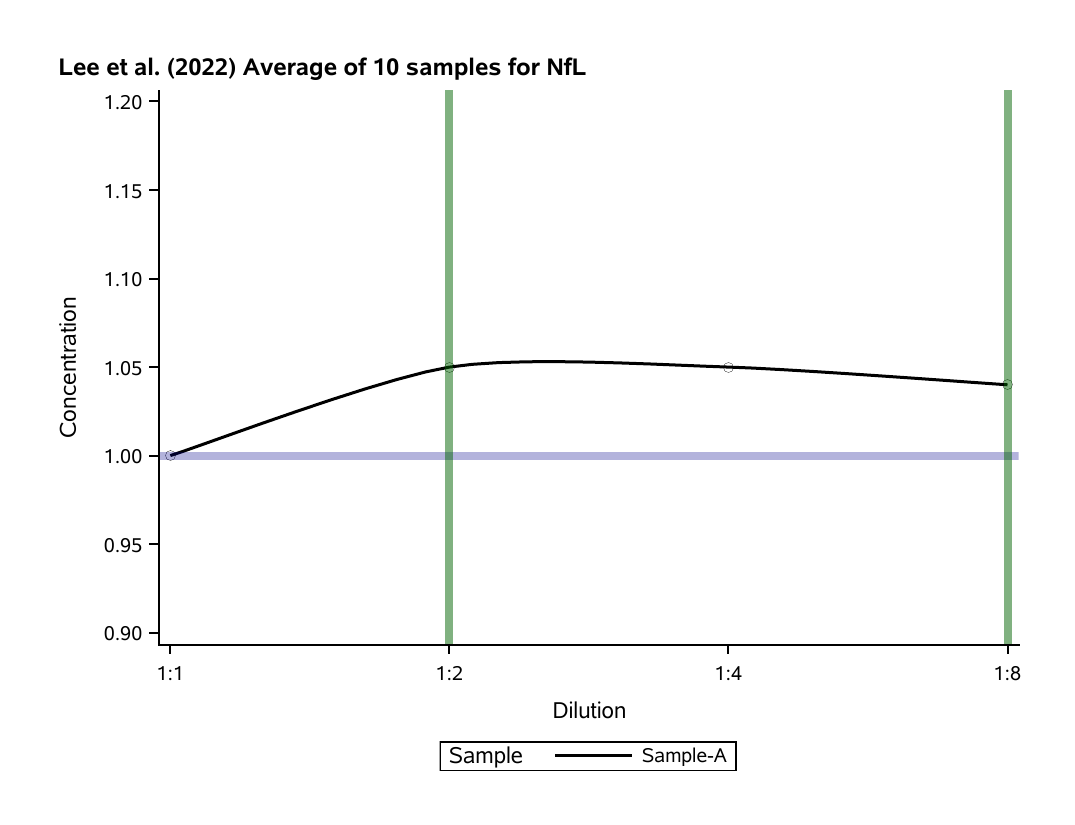}
  \caption{Partial Parallelism Plot for Neurofilament Light Chain
    (NfL) quantification.  The plot illustrates the quantification of
    NfL in serum samples spiked with 500 ng/mL of the peptide
    standard~\cite{Lee2022_}. The graph to the left displays the
    individual dilution curves, while the graph to the right shows the
    averaged data from 10 spiked serum samples. The raw data were
    provided by the corresponding author.  The graph demonstrates that
    parallelism is demonstrated, on a group level, for a dilution
    range of 1:2--1:8 for recombinant NfL in serum using the assay
    buffer.}\label{f_Lee2022_}
\end{figure}

\subsubsection{Apolipoprotein E}
Apolipoprotein E (ApoE) alleles (E2, E3, E4), are well established
genetic risk factors associated with various neurodegenerative
diseases~\cite{Huy2023_734,Wil1993_575,Hof1997_104,Nic1995_135}. In
the assay evaluated~\cite{Huy2023_734}, partial parallelism could not
be demonstrated for any of the ApoE isoforms (see
Figure\ref{f_Huy2023_734}). However, it is important to note that the
analysis was based on only three samples, limiting the robustness of
this conclusion. Notably, one sample (Sample A) did exhibit some
degree of partial parallelism for the E3/4 and E4 variants.  These
preliminary findings suggest potential for assay refinement, and
further developmental work~\cite{Huy2023_734} with a larger sample set
will be necessary to accurately assess dilutional behavior and
parallelism across ApoE isoforms.

\begin{figure}\centering
  \includegraphics[width=0.4\textwidth]{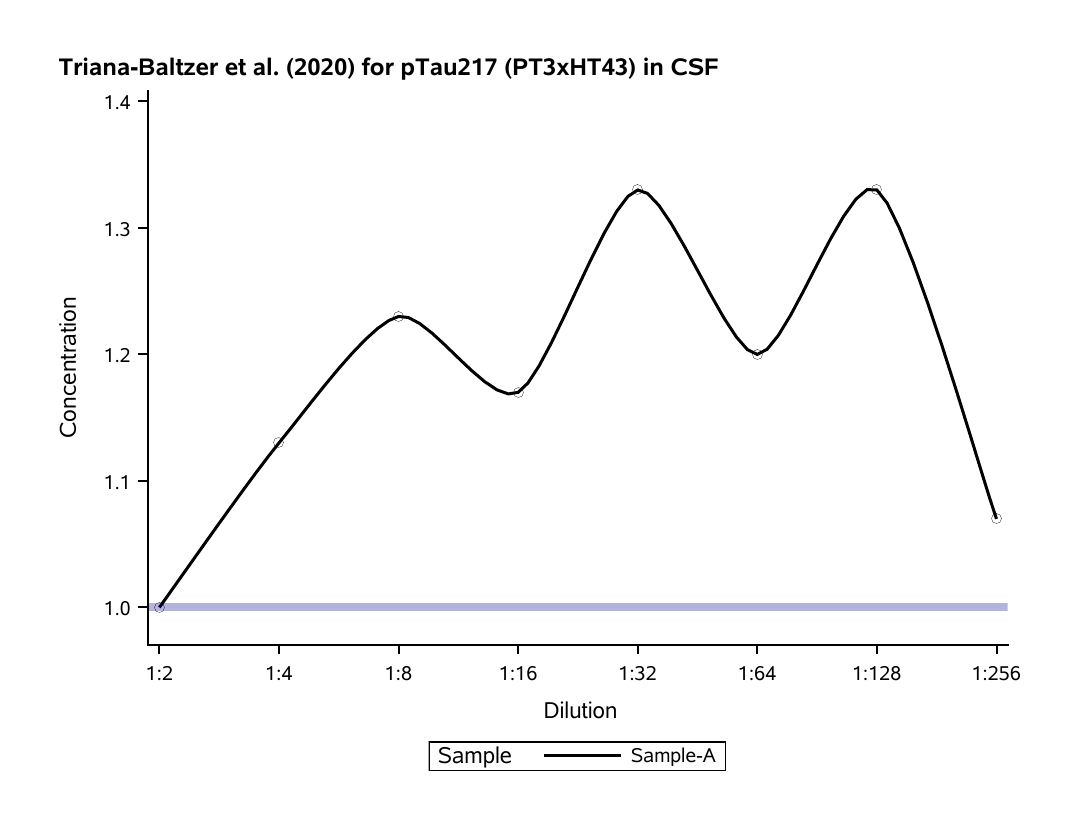}
  \includegraphics[width=0.4\textwidth]{Tri2020_1417.pdf}
  \caption{Partial parallelism plots for serial dilution of CSF
    measuring pTau217 using PT3xHT43 and PT3xPT82 assays. Data were
    extracted from Figure 2A in reference~\cite{Tri2020_1417}. The
    plots do not demonstrate partial parallelism; however, this may be
    influenced by data noise. Averaged data from additional samples
    tested within the dilution range of 1:8 to 1:64 could provide
    further clarity.}\label{f_Tri2020_1417}
\end{figure}

\subsubsection{Neurofilament proteins}
Among the five known neurofilament subunits~\cite{Kha2024_269},
neurofilament light (NfL) and heavy (NfH) chains have been
 assessed~\cite{Koe2014_43,Lee2022_}. For NfH, a
five-parameter logistic (5PL) model was employed to generate the
standard curve~\cite{Koe2014_43}. However, the authors' conclusion
that the assay demonstrated no evidence of non-parallelism is not
fully supported by the partial parallelism plot based on four samples
(see Figure~\ref{f_Koe2014_43}).

In contrast, the NfL assay showed group-level evidence of partial
parallelism across a dilution range of 1:2 to 1:8, as demonstrated in
a study using ten samples~\cite{Lee2022_} (see
Figure~\ref{f_Lee2022_}). Despite these findings at the group level,
deviations from partial parallelism remain evident at the level of
individual patient samples. These inconsistencies remain to be
addressed to enable reliable use of neurofilament measurements for
personalized clinical decision-making~\cite{Yal2025_}.

\begin{figure}\centering
  \includegraphics[width=0.4\textwidth]{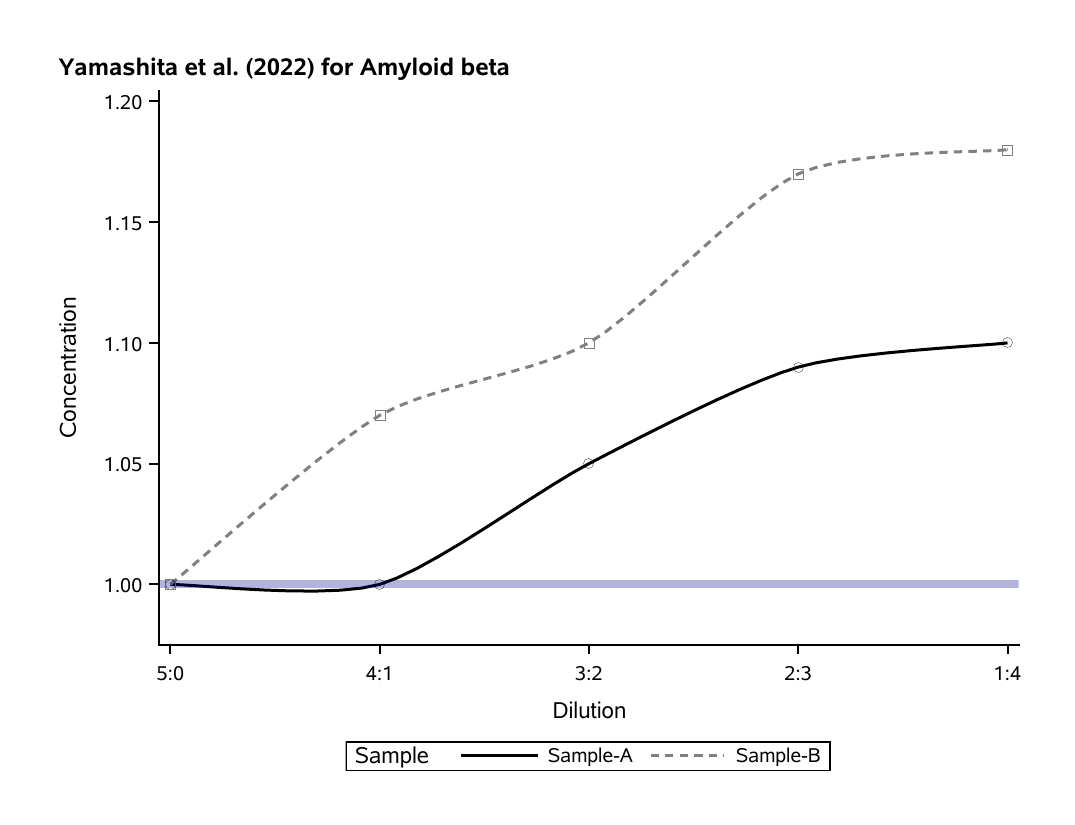}
  \caption{Partial Parallelism plot for A$\beta_{1-42}$ and
    A$\beta_{1-40}$ quantified from plasma samples spiked with the
    respective Amyloid$\beta$ peptides. The data were taken from
    supplementary Table 1 in reference~\cite{Yam2021_22}. The plot
    indicates a lack of partial parallelism across the dilution range
    (native sample to 1:4) tested. }\label{f_Yam2021_22}
\end{figure}

\subsubsection{Glial fibrillary acidic protein}
In the GFAP assay reported by Butcher et al.~\cite{But2021_58}, a
minor discrepancy was observed between the claimed dilution range for
parallelism (1:1 to 1:32) and the actual range showing partial
parallelism (1:4 to 1:16; see Figure~\ref{f_But2021_58}). The authors
also noted a deviation from the recommended protocol~\cite{And2015_}
(see Table in reference~\cite{But2021_58}), having used only 3 instead
of 5 plasma samples for parallelism assessment. Notably, all three
samples were derived from patients with acute traumatic brain injury
(TBI) and exhibited high GFAP concentrations, potentially introducing
a biomarker selection bias~\cite{Pet2014_1560}.  Previous studies have
reported a lack of dilutional parallelism for GFAP in CSF in chronic
conditions such as
ALS~\cite{Alb1985_301,Pet2015_17,Abd2022_158}. Additionally, the
pronounced deviation from the line of unity at a dilution of 1:64
observed in Figure~\ref{f_But2021_58} remains unexplained. However,
given that earlier findings were based on different ELISA platforms
and assessed GFAP in CSF rather than plasma, further investigation is
needed to determine whether partial parallelism can be reliably
demonstrated across neurodegenerative diseases in addition to TBI\@.

\begin{figure*}\centering
  \includegraphics[width=0.4\textwidth]{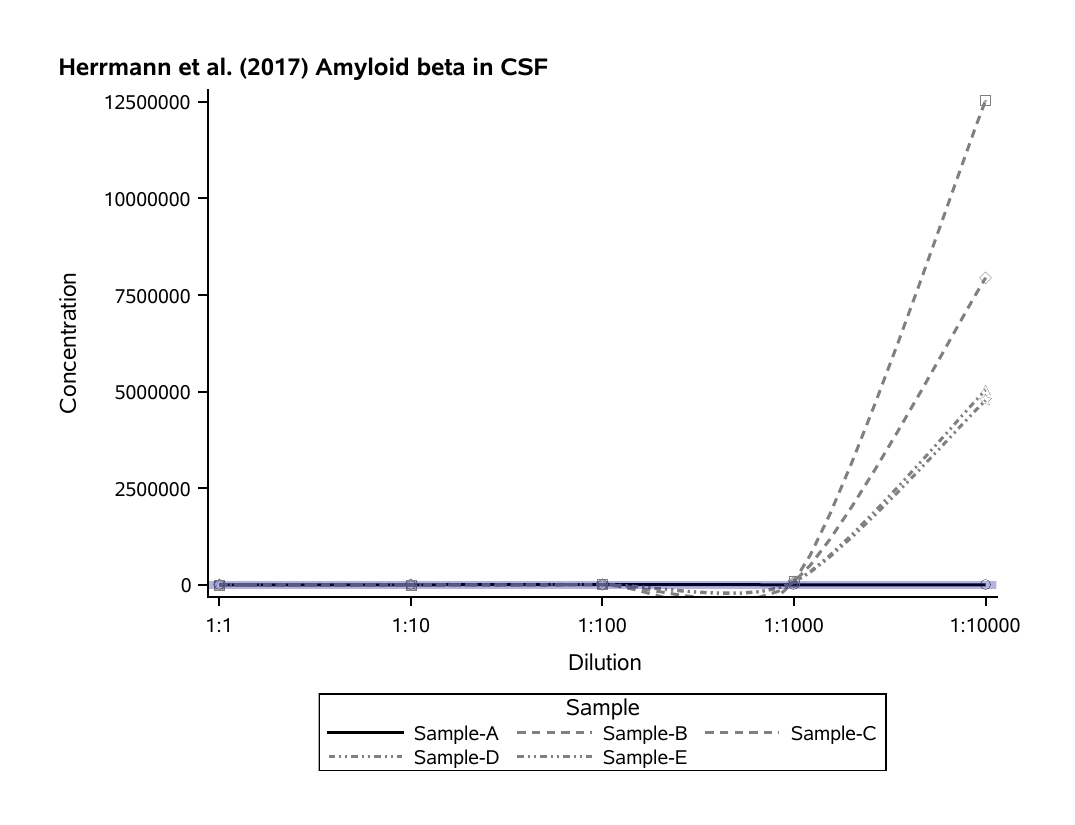}
  \includegraphics[width=0.4\textwidth]{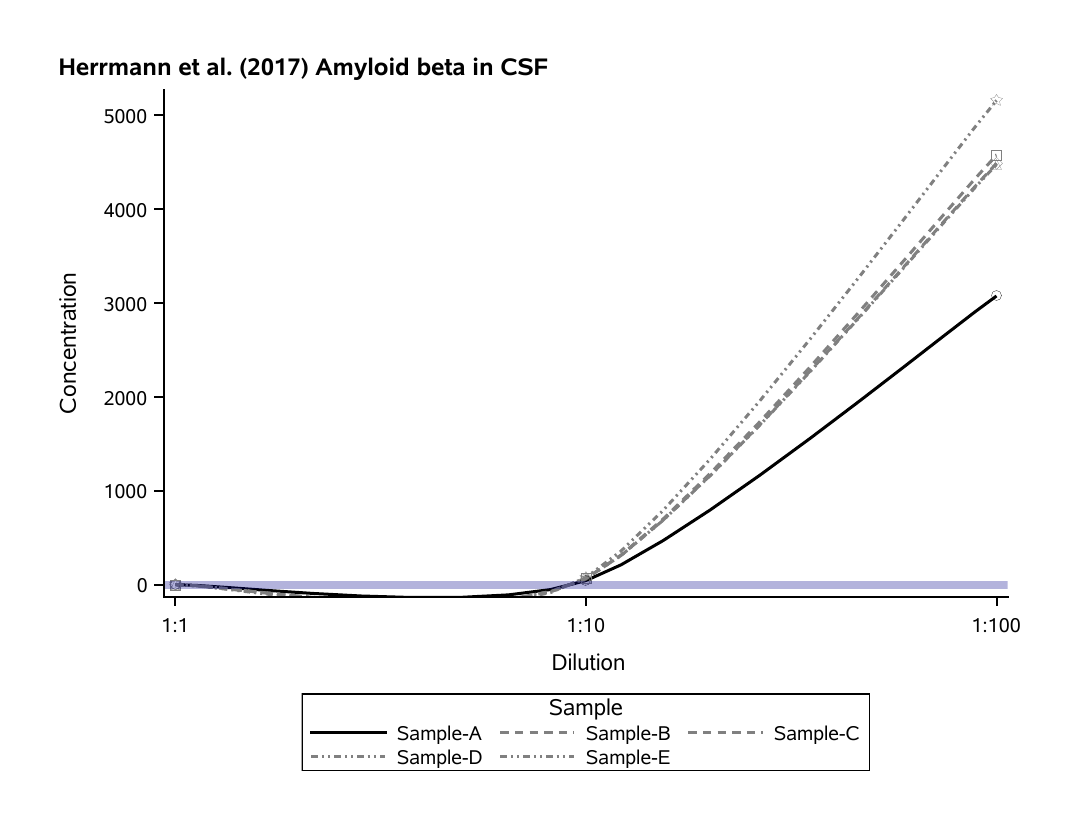}
  \includegraphics[width=0.4\textwidth]{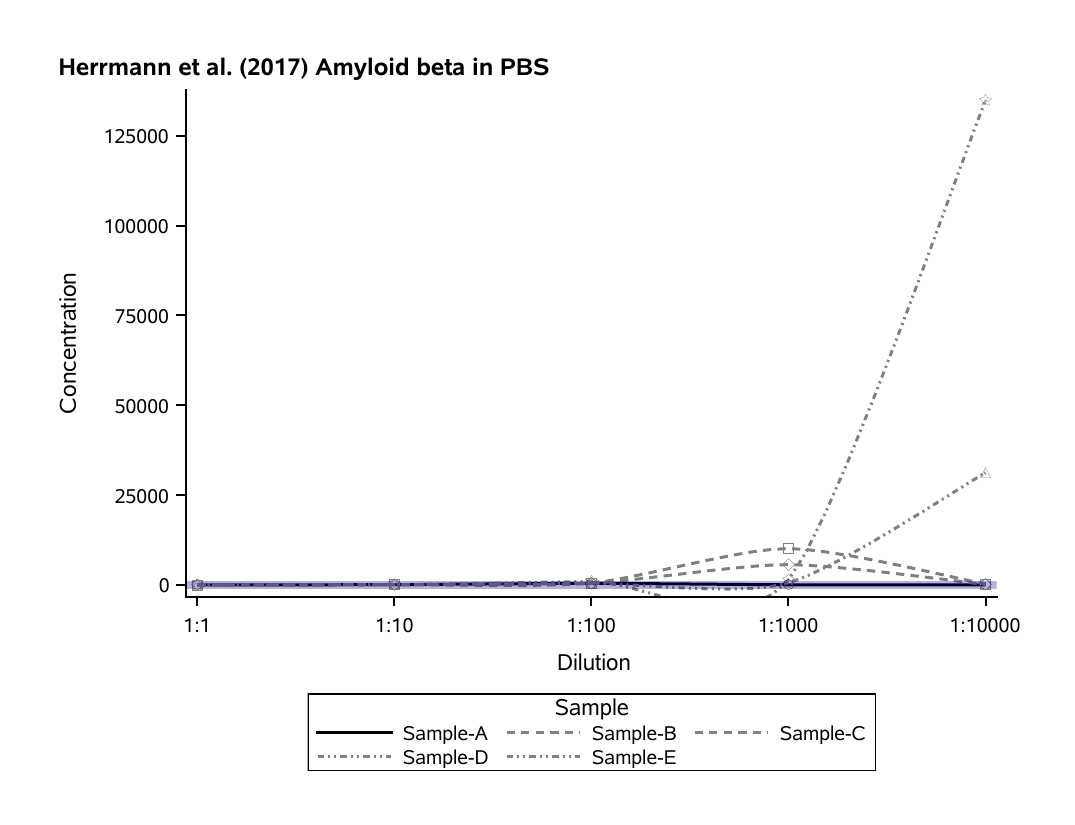}
  \includegraphics[width=0.4\textwidth]{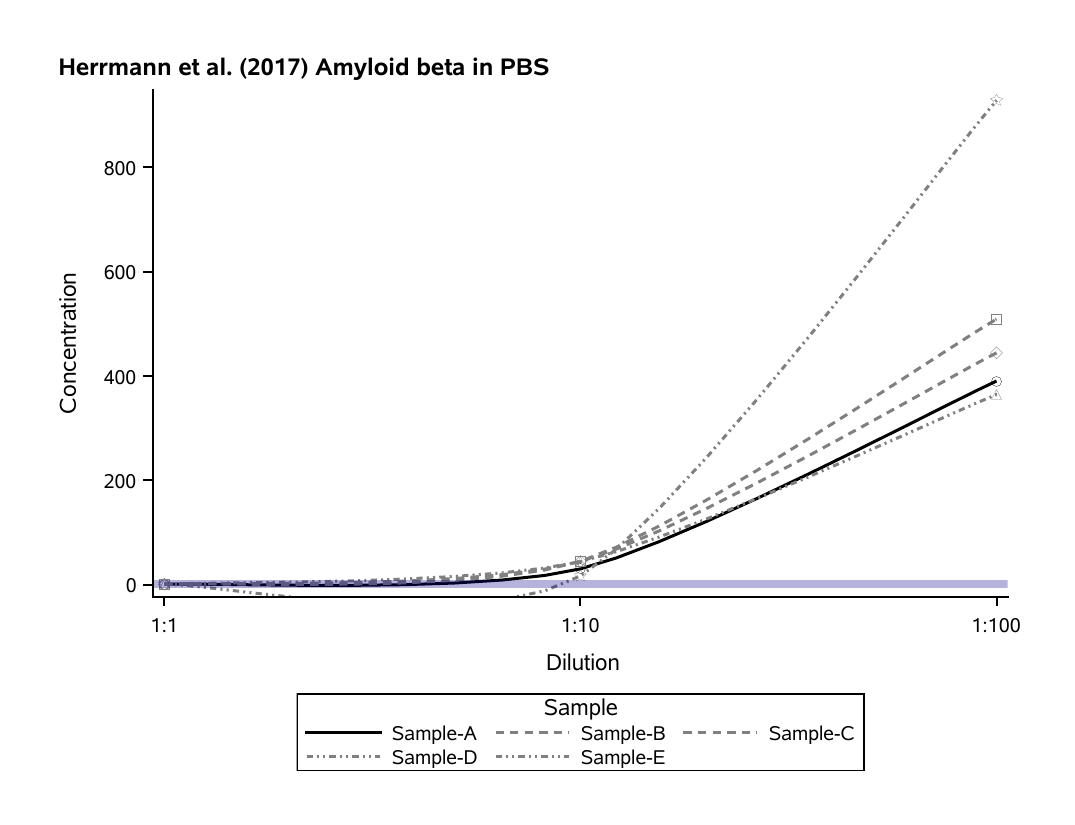}
  \caption{Partial Parallelism Plot for A$\beta$ oligomer
    quantification. This figure presents the partial parallelism plot
    for the quantification of A$\beta$ in CSF and
    PBS~\cite{Her2017_244}. Experiments were conducted with increasing
    iterations: Sample A had zero iterations, Sample B had one
    iteration, progressing to Sample E with four iterations. The raw
    data were kindly shared by the corresponding author. The plots
    demonstrate a lack of partial parallelism across the dilution
    range tested. The full dilution range is displayed in the plots on
    the left. The zoomed-in plots on the right highlight that A$\beta$
    oligomers are overestimated at dilutions beyond 1:10.
  }\label{f_Her2017_244}
\end{figure*}

\subsubsection{Dipeptide repeat proteins}
Among the five known dipeptide repeat proteins [poly(GP), poly(GA),
poly(GR), poly(PR), and poly(PA)], poly(GP) has been quantitatively
assessed~\cite{Wil2022_761}. The assay utilized a combination of
custom-made rabbit polyclonal antibodies and a mouse monoclonal
antibody (TALS 828.179). Due to heteroscedasticity observed in the
signal across the standard curve, the data required post-processing
prior to curve fitting. Both 4PL and 5PL models were initially
evaluated, with the 4PL model ultimately selected and weighted using
1/Y$^2$.  The test demonstrated a diagnostic sensitivity and
specificity of 100\% for distinguishing 40 individuals with
\emph{C9orf72} repeat expansions from 15 healthy
controls~\cite{Wil2022_761}. Notably, partial parallelism was observed
between dilutions of 1:4 and 1:16, a narrower dynamic range than the
less diluted 1:1 to 1:2 concentration proposed by the authors as
anchors for assessing dilutional parallelism (see
Figure~\ref{f_Wil2022_761}).

\begin{figure}\centering
  \includegraphics[width=0.4\textwidth]{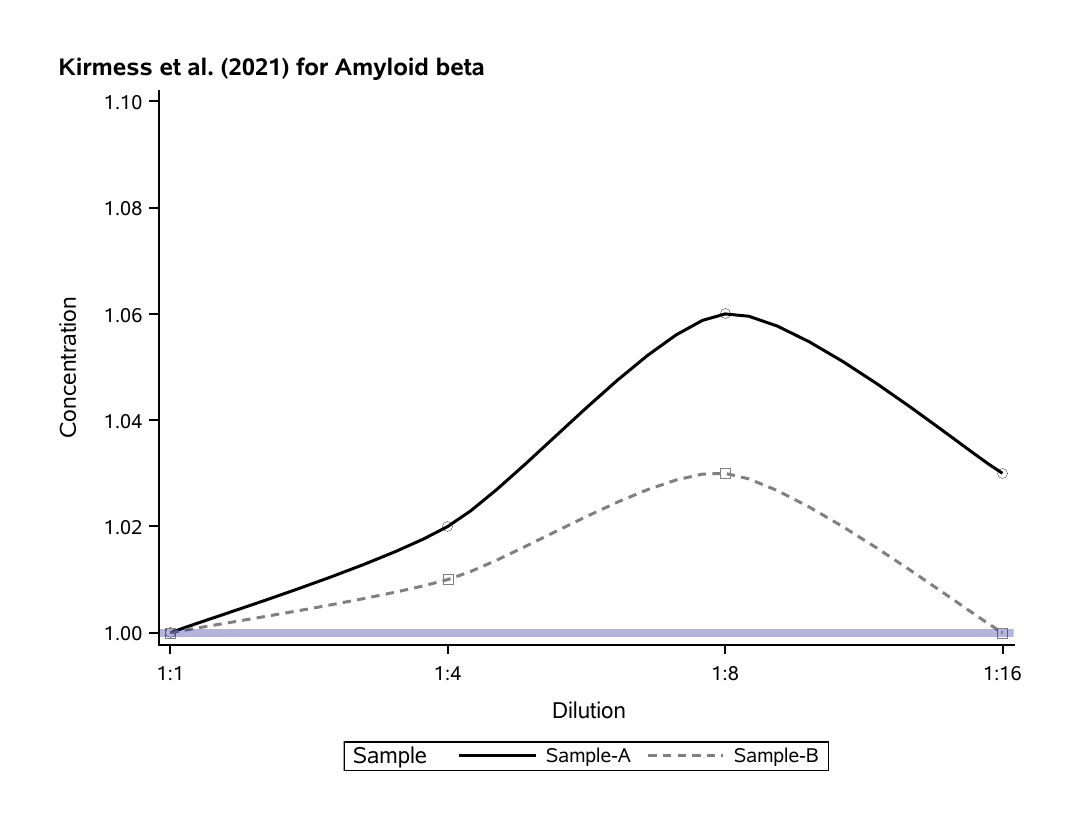}
  \caption{Partial Parallelism Plot for amyloid $\beta_{1-42}$ and
    amyloid $\beta_{1-40}$ quantified from artificial samples of
    recombinant human serum albumin which was spiked with full length
    recombinant amyloid $\beta_{1-42}$ and amyloid $\beta_{1-40}$
    proteins at concentrations about three times above the highest
    standard. The data were taken from Table 6 in
    reference~\cite{Kir2021_267}. The plots do not demonstrate partial
    parallelism.}\label{f_Kir2021_267}
\end{figure}

\subsection{Synthesis of Results}\label{r_synthesis}
\begin{figure}\centering
  \includegraphics[width=0.4\textwidth]{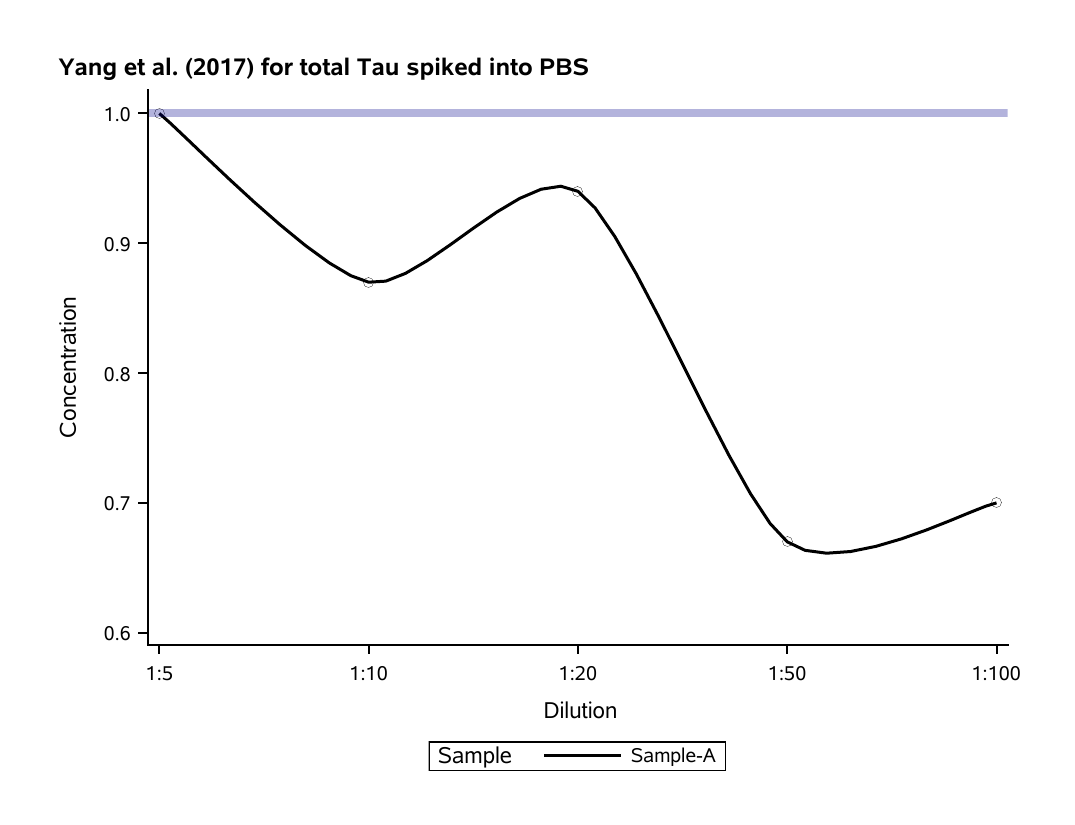}
  \caption{Partial Parallelism plot for total Tau spiked into
    PBS\@. The data were taken from table 5 in
    reference~\cite{Yan2017_9304}. There is lack of partial
    parallelism.}\label{f_Yan2017_9304}
\end{figure}

\begin{figure}\centering
  \includegraphics[width=0.4\textwidth]{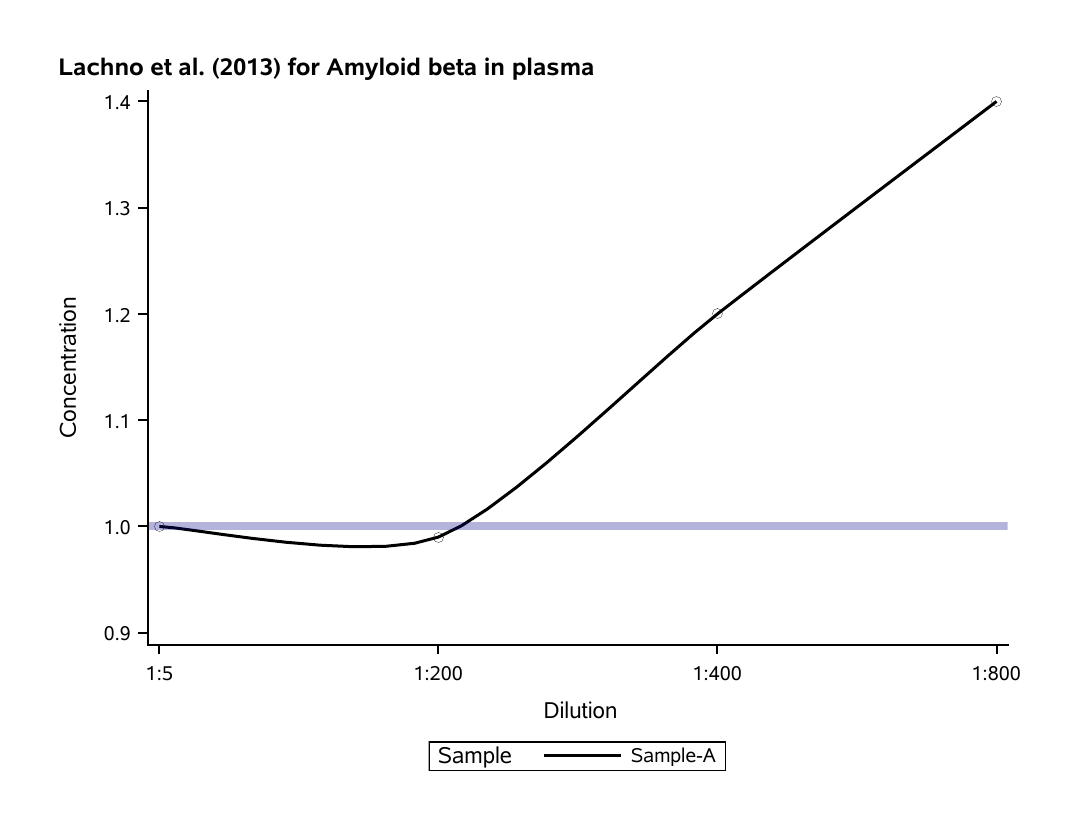}
  \caption{Partial Parallelism plot for spiked CSF A$\beta_{1-40}$
    (78,300 ng/L) diluted in an assay buffer. The data were taken from
    the top left graph in Figure 2 in reference~\cite{Lac2013_897}.
    The plots do not demonstrate partial
    parallelism.}\label{f_Lac2013_897}
\end{figure}

\begin{figure}\centering
  \includegraphics[width=0.4\textwidth]{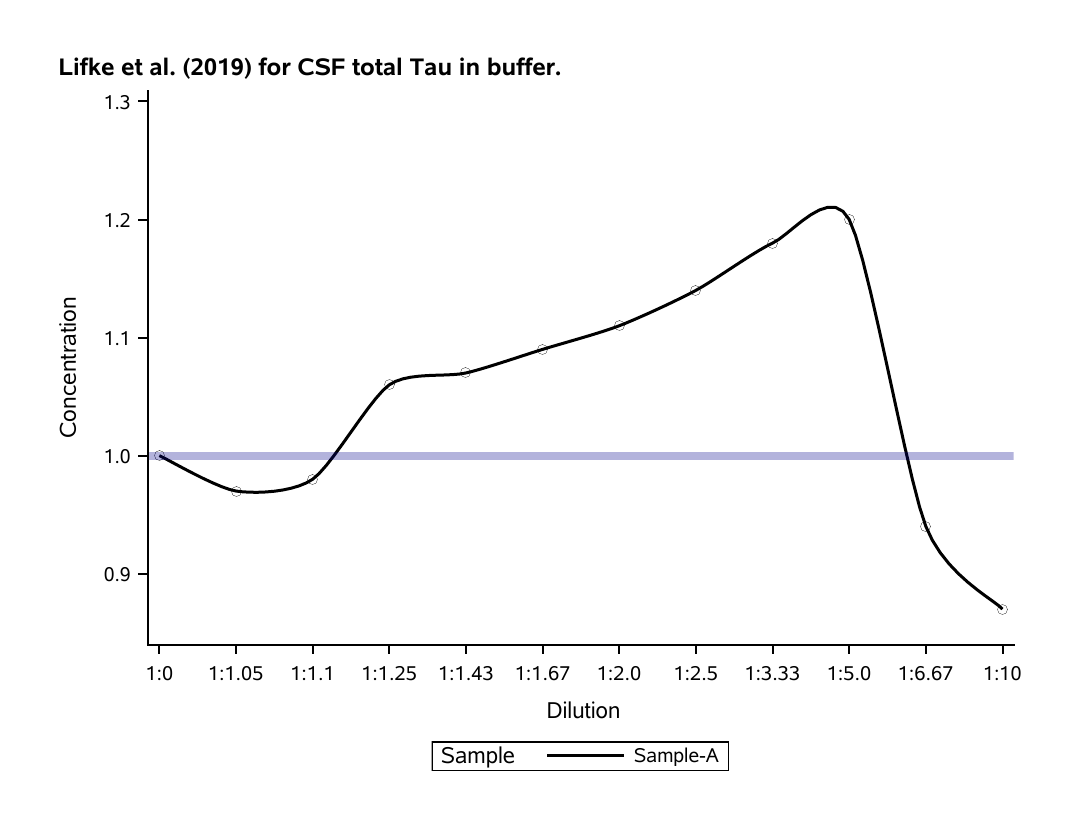}
  \includegraphics[width=0.4\textwidth]{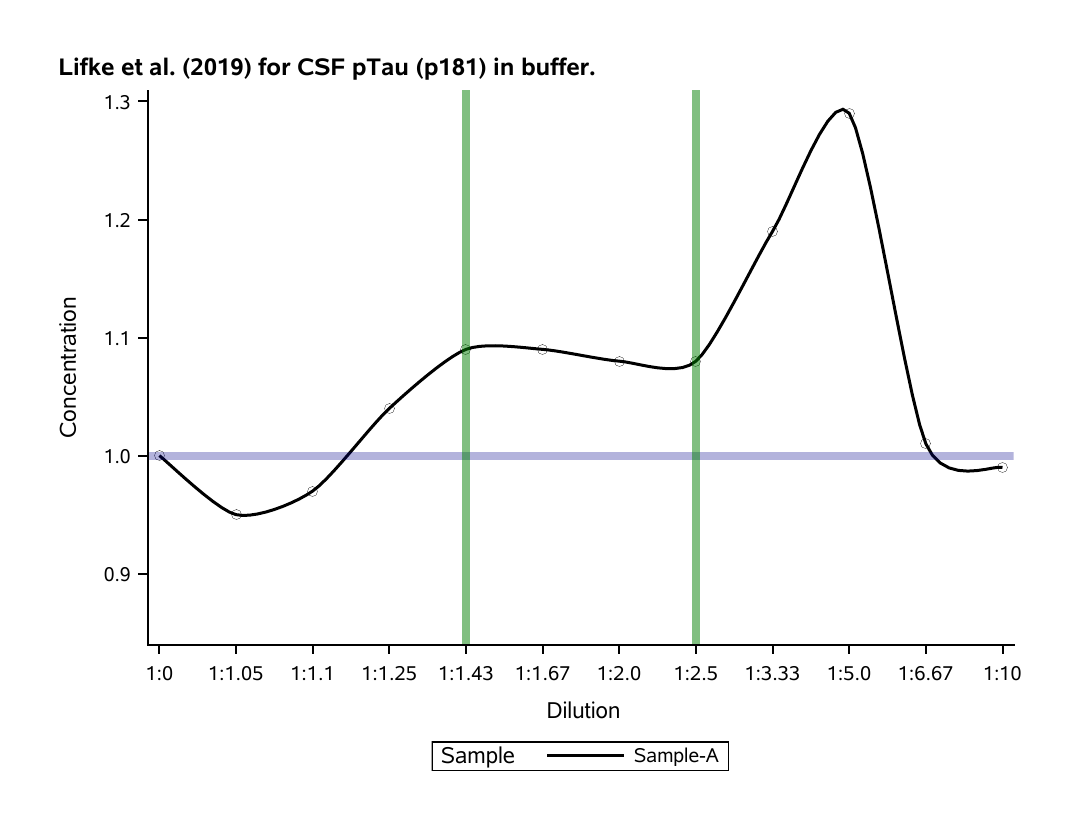}
  \caption{Partial parallelism plots for total Tau and pTau (181P)
    assays. Data were derived from ``representative results from one
    sample'' in Supplementary Figure 1 of
    reference~\cite{Lif2019_30}. The CSF sample dilution series in
    buffer was estimated from the x-axis (1, 0.95, 0.9, 0.8, 0.7, 0.6,
    0.5, 0.4, 0.3, 0.2, 0.15, 0.10), and measured concentrations were
    approximated from the y-axis. Partial parallelism was achieved
    within a dilution range of 1:1.43 to 1:2.5 for the pTau181 assay
    but not for the total Tau assay.}\label{f_Lif2019_30}
\end{figure}

\begin{figure}\centering
  \includegraphics[width=0.4\textwidth]{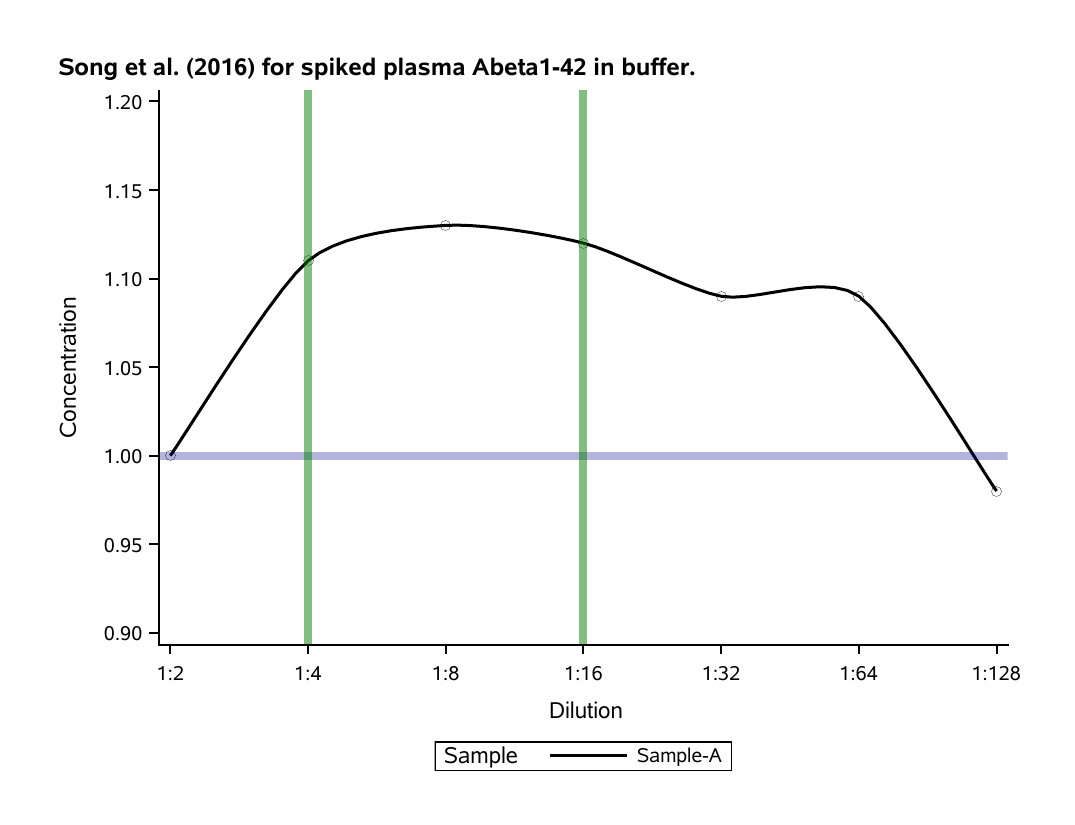}
  \caption{Partial Parallelism plot for A$\beta_{1-42}$ from plasma
    diluted into a sample buffer. Data were taken from the log-scaled
    (0--100) y-axis from Figure 3A in reference~\cite{Son2016_58}. The
    plot shows partial parallelism within a dilution range of
    1:4--1:16.}\label{f_Son2016_58}
\end{figure}

\begin{figure}\centering
  \includegraphics[width=0.4\textwidth]{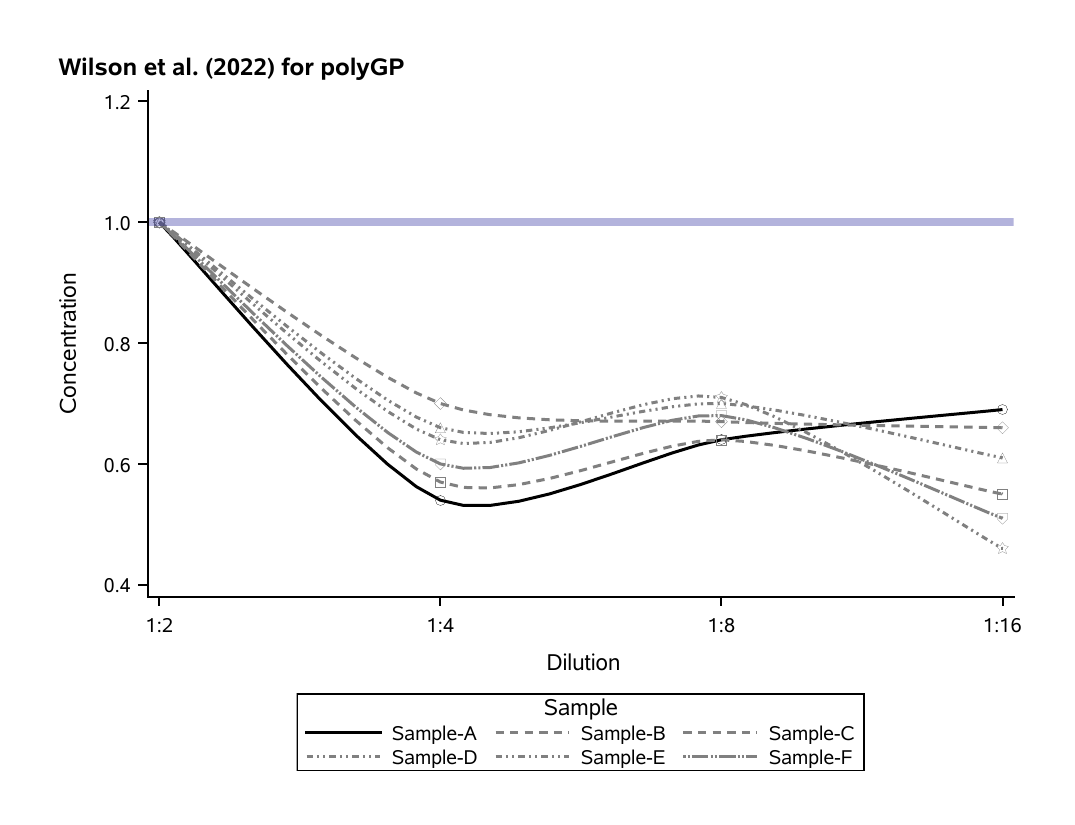}
  \includegraphics[width=0.4\textwidth]{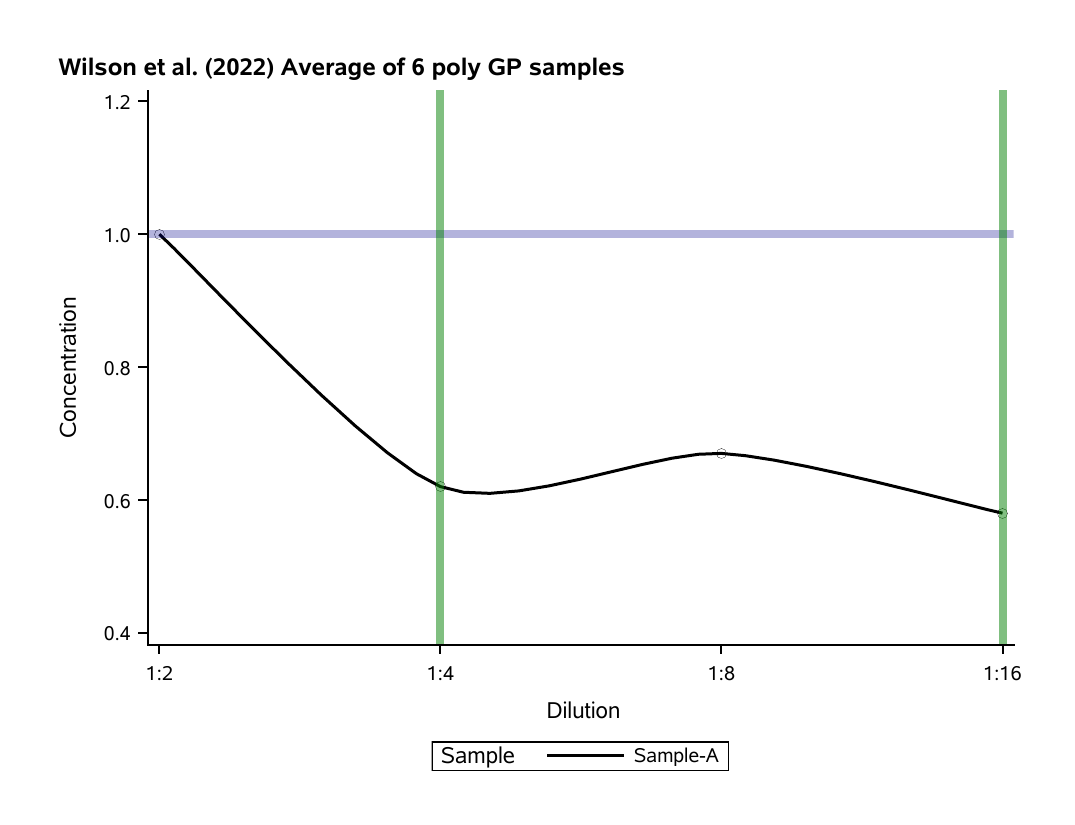}
  \caption{Partial Parallelism plots for quantification of
    poly-glutamin expansions from native plasma samples. The data were
    taken from Figure 3H in reference~\cite{Wil2022_761}. The plot to
    the left shows the individual dilution curves. The graph to the
    right shows the average for 6 plasma samples.  On a group level,
    the plot shows partial parallelism within a dilution range of
    1:4--1:16.}\label{f_Wil2022_761}
\end{figure}

\begin{figure}\centering
  \includegraphics[width=0.4\textwidth]{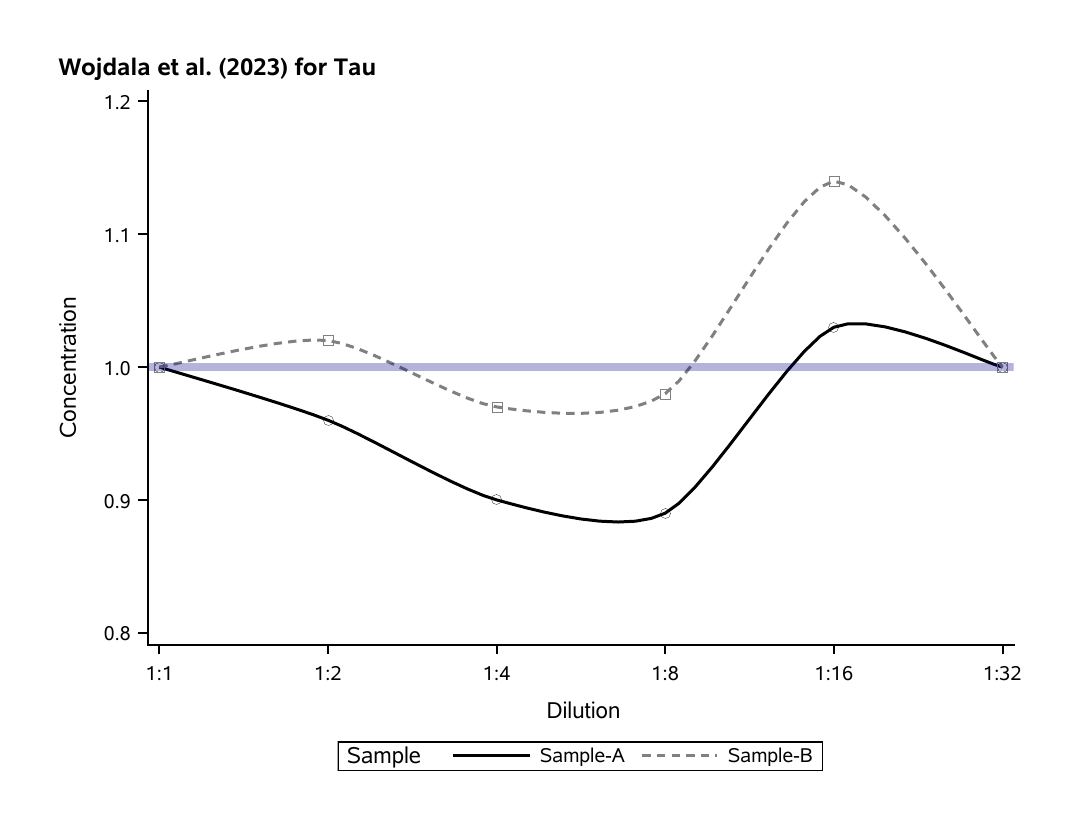}
  \caption{Partial Parallelism plot for quantification of
    phosphorylated (pTau181, pTau231) tau protein from samples spiked
    with the peptide standard. The data were taken from Supplementary
    Table 3 in reference~\cite{Woj2023_}. Partial parallelism is not
    achieved in this plot.}\label{f_Woj2023_}
\end{figure}

\begin{figure*}\centering
  \includegraphics[width=0.4\textwidth]{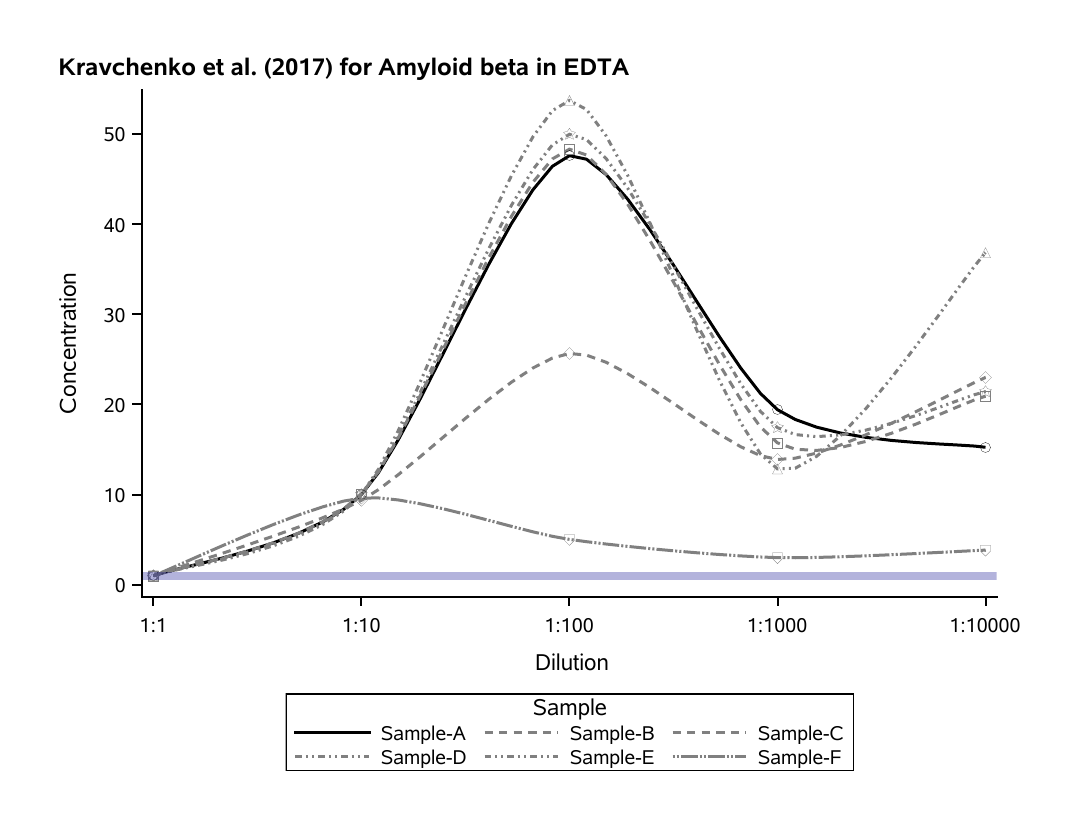}
  \includegraphics[width=0.4\textwidth]{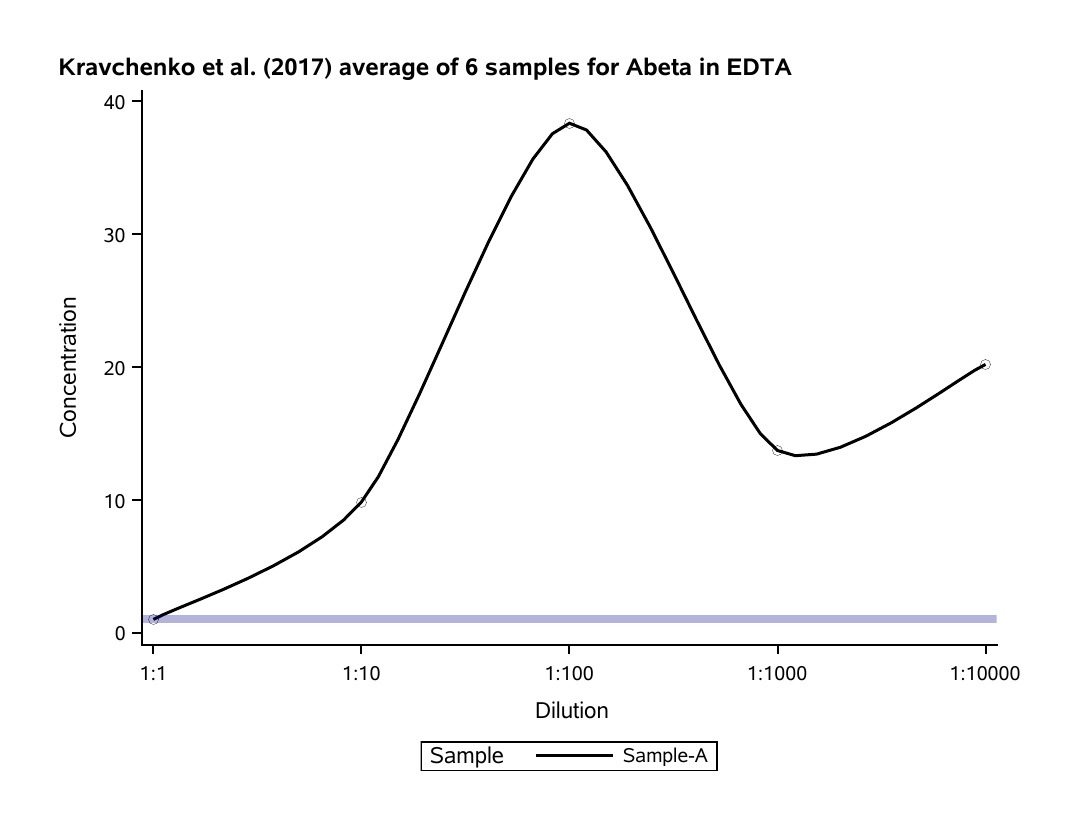}
  \includegraphics[width=0.4\textwidth]{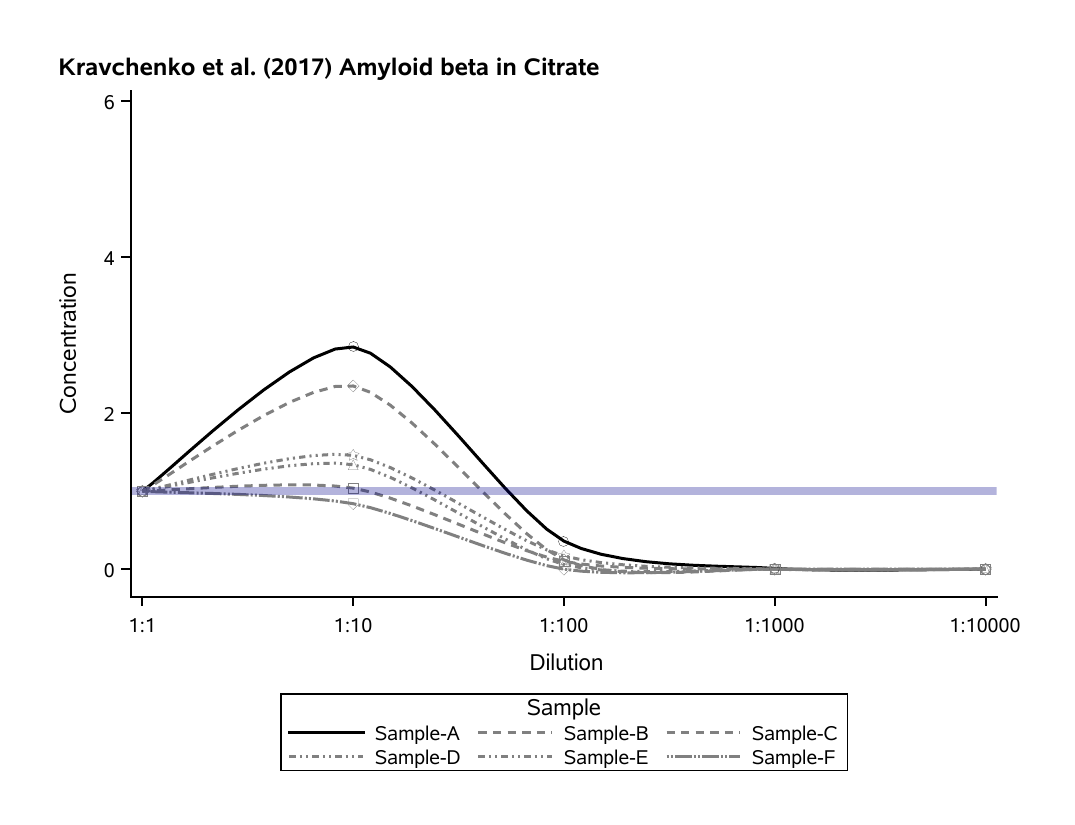}
  \includegraphics[width=0.4\textwidth]{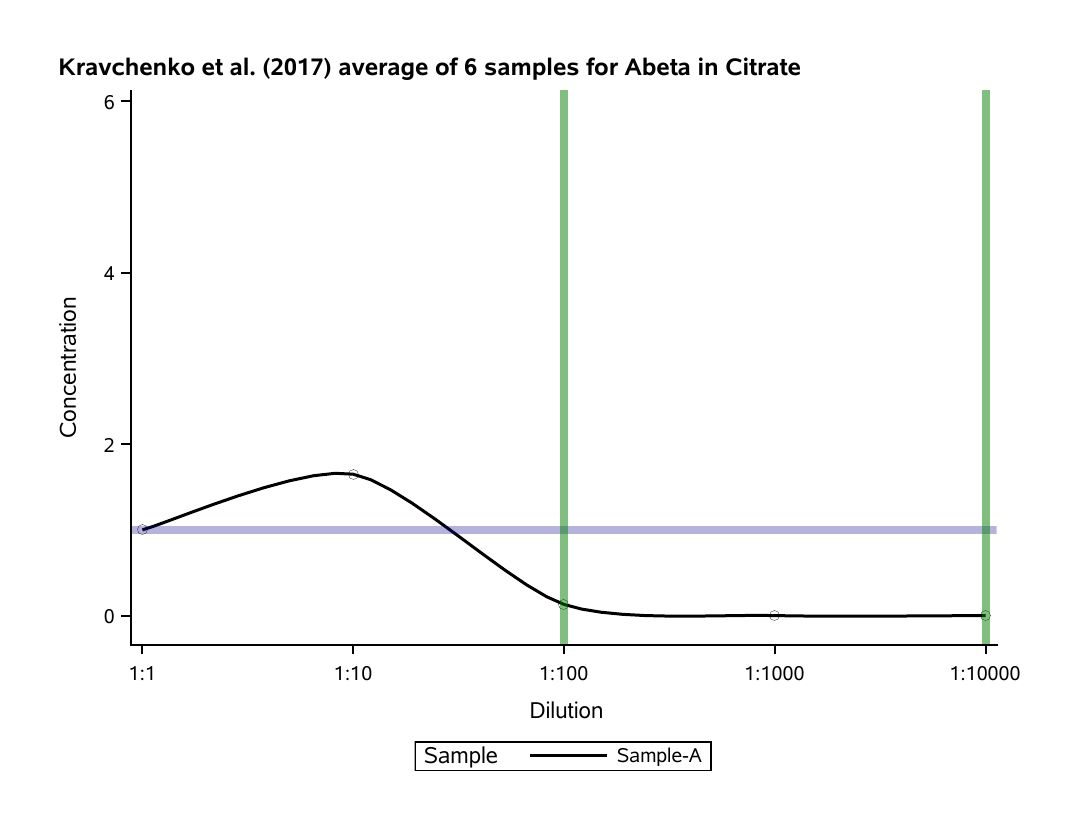}
  \includegraphics[width=0.4\textwidth]{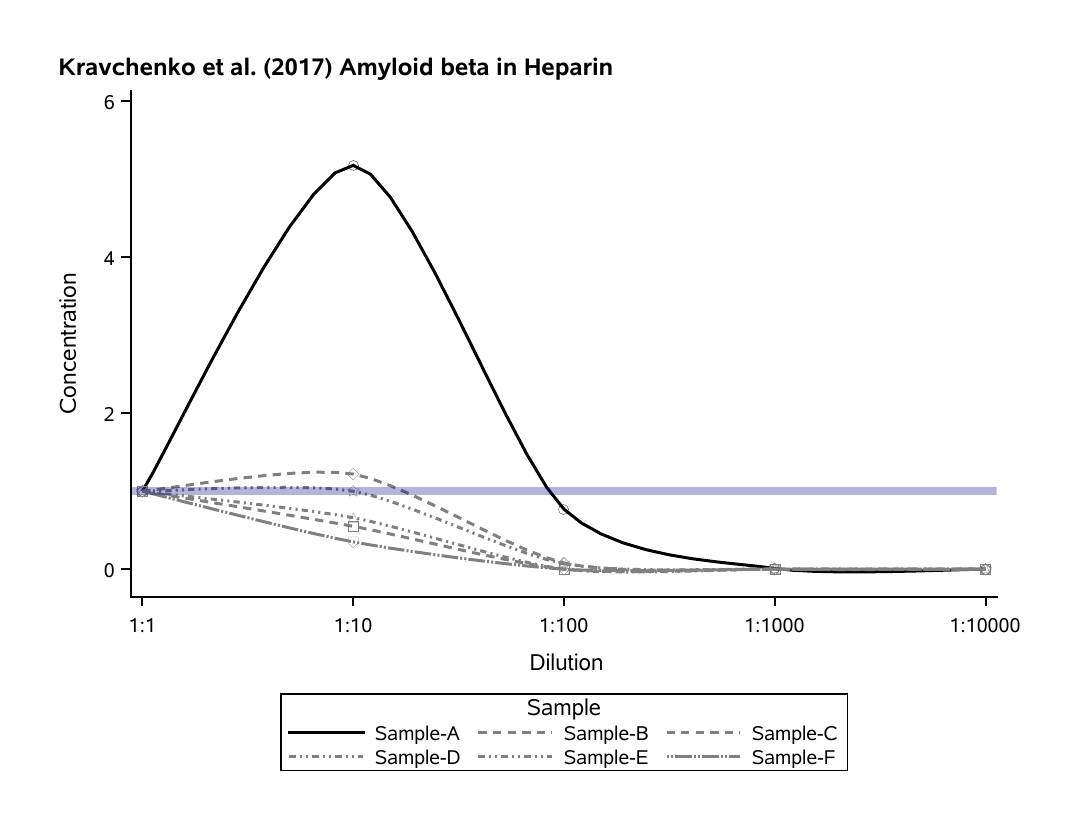}
  \includegraphics[width=0.4\textwidth]{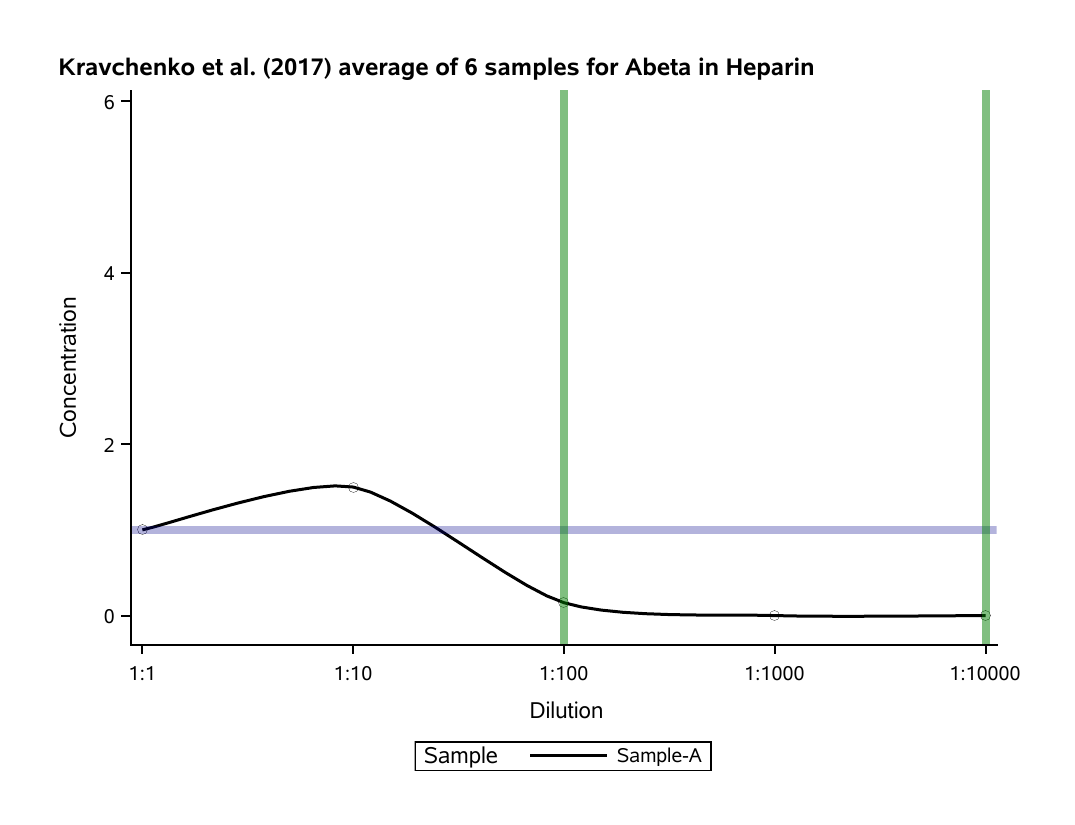}
  \caption{Partial Parallelism Plots for A$\beta$ oligomer
    quantification in different sampling buffers. The figure shows
    partial parallelism plots for A$\beta$ oligomers in EDTA, Citrate,
    and Heparin buffers. Samples A-E are spiked plasma, and Sample F
    is spiked PBS~\cite{Kra2017_465}. The raw data were kindly shared
    by the corresponding author. The plots reveal a lack of
    parallelism for EDTA, while partial parallelism is achieved in
    Citrate and Heparin buffers between dilutions of 1:100 and
    1:10,000 on a group level.}\label{f_Kra2017_465}
\end{figure*}

Partial parallelism plots were generated from 19
studies~\cite{But2021_58,Kru2016_153564,Wil2022_761,Bay2021_,Cul2012_8,Huy2023_734,Kan2014_17,Koe2014_43,Lee2022_,Tri2020_1417,Yam2021_22,Her2017_244,Kir2021_267,Yan2017_9304,Lac2013_897,Lif2019_30,Son2016_58,Woj2023_,Kra2017_465}. Several
studies had analytical results necessitating generation of more than
one partial parallelism plot. In total 49 partial parallelism plots
were generated (Figures~\ref{f_But2021_58}
to~\ref{f_Kra2017_465}). Nine of these plots were on averaged data
from experiments with more than five samples.  The Figures illustrate
that partial parallelism was achieved in only 7 out of 49 experiments
(14\%) (Table~\ref{t_ppp}).

The dilution ranges demonstrating partial parallelism were typically
narrow~\cite{But2021_58,Wil2022_761,Bay2021_,Lee2022_,Lif2019_30,Son2016_58}.
Narrow dilution ranges of about three dilution steps may be acceptable
in laboratory assays investigating parallelism where a narrow range
can be expected~\cite{Guy2023_229}. But for the biomarkers reviewed
here the literature reports ranges over two to three magnitudes of
concentration~\cite{Con2021_2703,And2017_34,For2019_730,Kar2022_1555,Mon2021_1086,Pet2007_94}.

Notably, one study diluted serum pools from healthy controls (low NfL
concentrations) into samples from individuals with multiple sclerosis
(high NfL concentrations) instead of using assay
buffer~\cite{Wil2022_761}. This raises uncertainty about whether these
assays would demonstrate partial parallelism under standard laboratory
conditions~\cite{Pum2019_215}.

These findings underscore limitations in the robustness of assay
performance and variability in partial parallelism across
studies~\cite{Pum2019_215,Pet2024_602,ICH2023_M10}. Further discussion
of these results, including potential sources of bias, is presented in
the subsequent section.

\subsection{Risk of Bias Across Studies}\label{r_bias}
The primary risk of bias across studies pertains to deviations from
established guidelines for testing
parallelism~\cite{ICH2023_M10}. Current recommendations stipulate that
parallelism should be demonstrated using a serially diluted sample
response curve in the same buffer used for the standard. However,
dilution of one pool of samples into another pool, as also
performed~\cite{Wil2022_761}, falls outside these guidelines and risks
overestimating parallelism.

In seven studies, generating reliable partial parallelism plots from
the provided data was not
possible~\cite{Lac2015_42,Liu2022_1186,Thi2021_9736,Waa2016_21,Wil2024_322,Waa2017_310,Wan2024_325}. Several
limitations contributed to this, including hidden biases that were not
pre-specified in the review protocol published on the PROSPERO
Registry. One notable example involved the assessment of parallelism
for Amyloid-$\beta$ immunoassays using spiked calibrators. This
analysis covered a heterogeneous range of dilution curves across six
laboratories employing one or more of seven
assays~\cite{Waa2017_310}. Parallelism was reported as mean
percentages per laboratory, with results ranging from 74\% to 344\%,
indicating significant variability and inconsistency.

Further complicating the assessment, some studies failed to provide
key methodological details, such as sample preparation protocols or
calibration standards, which may have introduced additional biases to
what was reviewed in Table~\ref{t_bias}. Additionally, variations in
experimental design, such as the use of non-standard dilution matrices
or unverified spiking procedures, may have further reduced the
comparability and reproducibility of parallelism testing. This
underscores the need for stricter adherence to standardised guidelines
and more transparent reporting of methodological details to minimise
bias and improve the reliability of partial parallelism
assessments. Our analysis revealed substantial variability in
parallelism assessment. The implications of these findings are
explored in the discussion section.

\section{Discussion}\label{discussion}
This systematic review provides a comprehensive analysis of
parallelism testing in neurodegeneration biomarker assays. The
principal finding is that all current biomarker tests exhibit only
partial~\cite{Pet2024_602}, rather than full,
parallelism. \href{https://discovery.ucl.ac.uk/id/eprint/10185313/1/parallel-v9-accepted.pdf}{Partial
  parallelism plots}, which can be readily generated from existing
data, offer a straightforward visual tool for comparison. However, the
range of partial parallelism is typically narrow, spanning
approximately three doubling dilution steps. This finding has critical
implications for interpreting studies that dilute samples beyond this
range, as such practices can lead to inaccuracies. Biomarker
concentrations may be overestimated when partial parallelism plots
deviate upwards (e.g., Figure~\ref{f_Yam2021_22}) or underestimated
with downward deviations (e.g., Figure~\ref{f_Yan2017_9304}). The
clinical context of such inaccuracies has been discussed for one
example where neurofilaments were quantified at a dilution of 1:400
instead of 1:4. As seen in the present review, other biomarkers used
in the context of neurodegeneration are affected as well.  Notably
however, near perfect partial parallelism~\cite{Pum2019_215} can
sometimes be achieved on a group level (e.g.,
Figure~\ref{f_Lee2022_}).

The distinction between group-level and individual-level partial
parallelism (see Figure~\ref{f_Lee2022_}) highlights two key
points. First, biomarker assays showing group-level partial
parallelism may be suitable for clinical trials, especially as
regulatory agencies like the FDA and EMA increasingly accept
biomarker-based endpoints for rapid drug approvals in
neurodegeneration. Second, the absence of parallelism in individual
samples warrants further investigation. Tests measuring proteins
involved in aggregation-related pathologies may harbor hidden biases,
which could influence results and
interpretations~\cite{Ple2006_750}. The effect of age will be one of
the most obvious demographic factors to be investigated
further~\cite{Mus2025_120014,Her2022_627}. Third, partial agreement
between different assays for quantification of NfL implies presence of
partial parallelism between the two tests~\cite{Pri2025_}. However,
the same method comparison study also showed overestimation of NfL
levels by one assay compared to another at higher
concentrations~\cite{Pri2025_}, indicating a lack of parallelism
outside a narrow dilution range. This is entirely consistent with the
clinical example, NfL in GBS, discussed earlier.

A 20\% failure rate in partial parallelism at the individual sample
level indicates that one in five samples may not yield analytically
comparable results. This degree of non-parallelism, if attributable to
biological phenomena such as protein aggregation inherent to disease
pathology~\cite{Lu2011_143,Adi2018_168}, could introduce significant
bias into quantitative interpretations. It is therefore essential that
deviations from parallelism are not dismissed as technical artefacts
but are instead rigorously interrogated. While matrix effects are
frequently cited, they do not account for all scenarios and are
predominantly a concern in mass spectrometry-based
platforms\cite{Ric2016_2475,Pic2019_2207,Fer2022_66}. Additional
causes include endogenous protein-protein interactions, biochemical
dissimilarity between native analytes and recombinant standards,
analyte instability, suboptimal assay sensitivity, and
supraphysiological biomarker concentrations exceeding the assay's
dynamic
range\cite{Lu2011_143,Adi2018_168,Arn2016_,Ric2016_2475,Pic2019_2207,Fer2022_66}.

Therefore a critical finding in this systematic review is the
potential publication bias, as study abstracts often overstate the
presence of parallelism. Only 14\% of the included studies
demonstrated clear evidence of partial
parallelism~\cite{But2021_58,Wil2022_761,Bay2021_,Lee2022_,Lif2019_30,Son2016_58}.
Interestingly, several studies reported on parallelism in between
different samples instead of in between samples and the standard. This
point could be emphasized stronger in future guidelines on
harmonization of assay validation and implementation of quality
control.

The review encountered several limitations stemming from
inconsistencies in the reporting of methodological details. Dilution
ranges varied widely across studies, with arbitrary steps (e.g., 1:4,
1:5, 1:6 versus 1:1,250, 1:2,500, 1:5,000)~\cite{Lac2013_897}. While
some dilution ranges could be inferred
retrospectively~\cite{Lif2019_30}, this was often based on a single
representative sample, preventing the generation of parallelism plots
for other data. Additionally, some reported values exceeded the assay
measurement range~\cite{Lif2019_30}, indicating inappropriate
extrapolation beyond the standard curve. For the standard curve the
dilution data was frequently not given. Instead, parallelism in
between samples, rather than between samples and the standard curve
was shown.  Table~\ref{t_bias} further highlights that not all samples
were diluted in the assay buffer, a critical limitation when testing
parallelism with a standard curve~\cite{Pum2019_215}.

Overall the number of samples and standard curves was too small for
meaningful statistical modelling~\cite{Got2005_437,Fay2020_721}. There
is need for larger numbers~\cite{Sen2013_1088} of samples, standards
and datapoints (of the dilution ranges) for robust statistical
evidence~\cite{Dur2019_,Sen2013_1088}. For example, one study that was
able to demonstrate partial parallelism, did so for a very narrow
dilution range of 1:1.43--1:2.5 based on one single sample that was
considered to be representative~\cite{Lif2019_30}. The only study with
a very large range of partial parallelism 1:100--1:10,000 reports this
close to the detection limit (16 fM) of the test~\cite{Kra2017_465}.

Moreover, there was concerning use of spiked or artificial samples in
many
studies~\cite{But2021_58,Kru2016_153564,Cul2012_8,Koe2014_43,Lee2022_,Tri2020_1417,Her2017_244,Kir2021_267,Yan2017_9304,Lac2013_897,Son2016_58,Wil2022_761,Woj2023_,Kra2017_465},
instead of the recommended use of native
samples~\cite{And2015_,Bur1994_}. For one study it remained unclear if
samples were spiked or not~\cite{Lif2019_30}. Taken together the use
of spiked samples reduces the generalisability of the findings for
creating representative parallelism plots. Similar issues with have
been reported for other groups of
biomarkers~\cite{Ban2000_131}. Stronger adherence to established
guidelines is recommended~\cite{And2015_}.

Likewise, generalisability is hampered by the lack of sharing data on
the standard curve for the dilution steps
presented~\cite{But2021_58,Cul2012_8,Her2017_244,Huy2023_734,Kan2014_17,Kir2021_267,Kra2017_465,Kru2016_153564,Lac2013_897,Lee2022_,Lif2019_30,Son2016_58,Tri2020_1417,Wil2022_761,Woj2023_,Yam2021_22,Yan2017_9304}. On
revision of the Figures in present systematic review it is not
possible to state with absolute certainty if the standard curve
overlays the (blue) line of unity. It would be desirable to have, in
future studies, data from a representative number of stand curves
available. These data points do not need to be exactly for the same
dilution steps as for samples if the visual representation with
\href{https://discovery.ucl.ac.uk/id/eprint/10185313/1/parallel-v9-accepted.pdf}{partial
  parallelism plots} is chosen. This is another advantage of the
visual approach compared to statistical techniques.

On critical review of the setup for sample dilution it should be
mentioned that for example one of the finally excluded
studies~\cite{Wil2024_322} diluted blood samples from patients with
multiple sclerosis into blood samples of health controls. For
assessment of parallelism between patient samples and the assay's
standard curve it is mandatory to use the the same buffer
solution~\cite{Pum2019_215,ICH2023_M10,Her2022_627}.

Conflicting parallelism results were found for the pTau181 assay. One
study did show partial parallelism~\cite{Lif2019_30}
(Figure~\ref{f_Lif2019_30}B) and one did not~\cite{Bay2021_}
(Figures~\ref{f_Bay2021_}A-F). Clearly, on direct visual comparison of
these partial parallelism plots it is evident that different dilution
ranges were used. The short stretch of partial parallelism
(1:1.43--1:1.25) from one study~\cite{Lif2019_30} is for a dilution
range lower than the one shown for the other (1:4--1:32 and
a:5--1:40)~\cite{Bay2021_}. Because a third study did also not
demonstrate partial parallelism for pTau181~\cite{Woj2023_}, a
balanced interpretation would suggest absence of parallelism for this
biomarker.

Another limitation of this systematic review is the restriction to
studies published between 2010 and 2024, a subset of the 32-years
since the formulation of the amyloid cascade hypothesis
(1992-2024)~\cite{Har1992_184}. This time frame was chosen
intentionally, as concerns regarding the effect of protein aggregation
on dilutional parallelism in immunoassays were first raised in
2010~\cite{Lu2011_143}, and only subsequently acknowledged in
influential white papers~\cite{Ste2013_2903,Her2022_627,Pic2019_2207}
and regulatory guidance
documents~\cite{ICH2023_M10,us2023m10,FDA2025_}.

Taken together, this critical and systematic review emphasises the
importance of adhering rigorously to guidelines for parallelism
testing~\cite{ICH2023_M10,us2023m10,Ste2013_2903,Her2022_627,Pic2019_2207}.
These guidelines are embedded in a well defined framework for
laboratory test validation that is endorsed by regulatory authorities
(Table~\ref{t_validation}).  If observed, the reason for lack of
partial parallelism needs to be explained as it has been done in the
context of protein aggregation~\cite{Lu2011_143}. Authors should
explicitly report the dilution range over which partial parallelism is
achieved and highlight potential biases, such as over- or
underestimation of biomarker concentrations, when dilutions exceed
this range.

For clinical trials, it is crucial to clearly document the range of
partial parallelism, the dilutions performed, and strategies to
minimise biases. Regulatory agencies might consider mandating the
inclusion of data from random individual samples to demonstrate
parallelism at both individual and group levels. Without clear
evidence of partial parallelism, it would be inadvisable to quantify
biomarkers for neurodegeneration at varying dilution steps or across
different time points in clinical trials. In summary, our critical and
systematic review highlights both progress and persistent challenges
in parallelism testing. These insights inform recommendations for
future assay validation and inclusion into the test validation
framework of regulatory authorities (e.g.\ under precision and bias or
under linearity in Table~\ref{t_validation}).

\section{Conclusion}
This systematic review identified
\href{https://discovery.ucl.ac.uk/id/eprint/10185313/1/parallel-v9-accepted.pdf}{partial
  parallelism} in only a small proportion of biomarker tests for
neurodegeneration. A likely biological reason is the presence of
protein aggregates, a key pathological feature in many
neurodegenerative diseases.  Where partial parallelism is absent, the
data suggest specific dilution ranges where it could potentially be
achieved. To advance the field, future research must align with
established guidelines~\cite{ICH2023_M10,FDA2025_}, ensuring
transparent reporting that allows independent researchers and
regulatory bodies to evaluate partial parallelism through standardised
plots.
%

%
%
\subsection{PRISMA 2020 Checklist}\label{prisma_list}
\begin{enumerate}
\item \textbf{Title:} The report is identified as a systematic review in the title\\
\item \textbf{Abstract:} The PRISMA 2020 for Abstracts checklist and the Journal's author guidelines have been followed\\
\item \textbf{Rationale:} The rationale and knowledge for this review have been summarised and referenced, including guidelines from regulatory authorities\\
\item \textbf{Objectives:} The objective has been stated clearly as a review of \textbf{parallelism} in biomarker assays\\
\item \textbf{Eligibility criteria:} Specified in Section~\ref{Eligibility_Criteria}\\
\item \textbf{Information sources:} Specified in Section~\ref{Information_Sources}\\
\item \textbf{Search strategy:} Specified in Section~\ref{Search_Strategy} and in the PROSPERO database\\
\item \textbf{Selection process:} Specified in Section~\ref{Study_Selection}\\
\item \textbf{Data collection process:} Specified in Section~\ref{data_collection}\\
\item \textbf{Data items:} Specified in Section~\ref{data_collection}\\
\item \textbf{Study risk of bias assessment:} Specified in Section~\ref{risk_bias}\\
\item \textbf{Effect measures:} Specified in Section~\ref{summary_measures} as Summary Measures \\
\item \textbf{Synthesis methods:} Specified in Section~\ref{synthesis_results} \\
\item \textbf{Reporting bias assessment:} Specified in Section~\ref{risk_bias_across} \\
\item \textbf{Certainty assessment:} Specified in Section~\ref{risk_bias} and references~\cite{Pum2019_215,Pet2024_602,ICH2023_M10} \\
\item \textbf{Study selection:} Specified in Section~\ref{r_study_selection} \\
\item \textbf{Study characteristics:} Specified in Section~\ref{r_study_character} \\
\item \textbf{Risk of bias in studies:} Specified in Section~\ref{r_risk_bias} \\
\item \textbf{Results of individual studies:} Specified in Section~\ref{r_results_indiv_studies} \\
\item \textbf{Results of syntheses:} Specified in Section~\ref{r_synthesis} \\
\item \textbf{Reporting biases:} Specified in Section~\ref{r_bias} \\
\item \textbf{Certainty of evidence:} Specified in Section~\ref{r_bias} \\
\item \textbf{Discussion:} Specified in Section~\ref{discussion} \\
\item \textbf{Registration and protocol:} specified in Section~\ref{methods}, study number  CRD42024568766\\
\item \textbf{Support:} There was no financial support for this study \\
\item \textbf{Competing interests:} None reported \\
\item \textbf{Availability of data, code and other materials:} Made available for free download from \href{https://www.crd.york.ac.uk/PROSPEROFILES/568766_STRATEGY_20240713.pdf}{PROSPERO} \\
\end{enumerate}
\newpage
\paragraph{Supplementary information}
\paragraph{Acknowledgments} The sharing of the raw data needed to
construct the partial parallelism plots is acknowledged from the
following authors: Oliver Bannach, Dieter Willbold and Marleen
J.A. Koel-Simmelink.
\paragraph{Author contributions} A.P. designed and performed all
experiments, analysed the data, prepared the figures and wrote the
manuscript. D.C. reviewed and co-registered the protocol on PROSPERO
and reviewed the manuscript draft. J.P. reviewed the manuscript draft.
\paragraph{Disclosure of interest} The authors report there are no
competing interests to declare. This study was not funded.
\paragraph{Ethical approval declarations}
This is a systematic review on existing data from studies which had
ethical permission.
\paragraph{Data availability \& Data deposition}
All data have been uploaded to
\href{https://doi.org/10.5522/04/29389949.v1}{Figshare}~\cite{Pet2025_29389949.v1}.
The python code for the search strategy is openly available for
download from the PRESTO registry.

\clearpage\newpage
%
%

\appendix

\begin{appendices}

\section*{Statistical methods for determination of parallelism}
The assessment of parallelism originates in the statistical literature
and is typically framed as a binary decision: parallelism is either
present or absent~\cite{Got2005_437}. This decision is tested through
statistical hypotheses embedded within formal models. Historically,
the first hypothesis documented is Euclid's fifth postulate. It is a
hypothesis that can be tested, and as we learned from history, it can
take centuries to discover that presumed mathematical solutions
eventually turned out to be wrong. The contemporary statistical
approaches to testing parallelism all have in common that they build
on the probability theory~\cite{Dur2019_}. With that comes the law of
large numbers~\cite{Sen2013_1088}. Only with large numbers there is some
guarantee that the averages from random events provide somehow stable
long-term results~\cite{Dur2019_,Sen2013_1088}. That implies that
statistical methods of testing for parallelism on small numbers, are
open to criticism.

One established approach of statistical testing for parallelism is the
extra sum-of-squares analysis of variance (ANOVA) method, which
compares residual sum of squares between nested models
(RSSE$_\text{nonpar}$)~\cite{Bat1988_103,Dra1998_}. For example this
approach forms the basis for using F and $\chi^2$ statistics to test
for parallelism~\cite{Got2005_437}. The authors compare visual and
statistical testing, showcasting the strengths and weaknesses of these
methods. The key message, for presence of nonparallelism, is to
decompose RSSE$_\text{const}$ into the component of nonparallel
origine or RSSE$_\text{nonpar}$ and what can be described by random
variation RSSE$_\text{free}$. Simplified,
RSSE$_\text{const} = $RSSE$_\text{nonpar} + $RSSE$_\text{free}$. Here
the definition of nonparallelism is the extra error that comes because
of lack of similarity between two curves as part of
RSSE$_\text{nonpar}$. The practical calculation of
RSSE$_\text{nonpar}$ depends, and this is very important to realise,
on the assumption of normally distributed data and presence of
parallelism. With this the law of large numbers applies because the
$\chi^2$ test is used, which works on a distributed randome variable:
$df_\text{const} = N_\text{std} + N_\text{uk} - (P + 1)$.

\newpage

To statistically test for this one needs to calculate the
probabilities for the dose (x) for two models.

1. The free model (SSE$_\text{free}$) is described as:
  \small
\begin{align*}
   \text{SSE}_{\text{free}}(\mathbf{p}^{\text{std}},
    \mathbf{p}^{\text{uk}}) = 
    \sum_{i=1}^{N^{\text{std}}}
    w_i^{\text{std}} \left( y_i^{\text{std}} - f(x_i^{\text{std}};
      \mathbf{p}^{\text{std}}) \right)^2 
    + \sum_{i=1}^{N^{\text{uk}}}
    w_i^{\text{uk}} \left( y_i^{\text{uk}} - f(x_i^{\text{uk}};
      \mathbf{p}^{\text{uk}}) \right)^2 
\end{align*}
  \normalsize 

2. The constraint model (SSE$_\text{const}$) is described as: 
  \small
\begin{align*}
   \text{SSE}_{\text{const}}(r, \mathbf{p}) =
    \sum_{i=1}^{N^{\text{std}}} w_i^{\text{std}} \left( y_i^{\text{std}}
      - f(x_i^{\text{std}}; \mathbf{p}) \right)^2 
    +
    \sum_{i=1}^{N^{\text{uk}}} w_i^{\text{uk}} \left( y_i^{\text{uk}} -
      f(r x_i^{\text{uk}}; \mathbf{p}) \right)^2 
\end{align*}
  \normalsize 

Very elegantly the authors elaborate, citing comprehensive reviews,
that highlight the limitations of various statistical factors in this
context~\cite{Got2005_437}.  The application of statistical models for
parallelism assessment is not without limitations. As noted by
Gottschalk \emph{et al.}, \emph{``The existence of similarity between
  two mathematical functions is not difficult to determine. It is less
  straightforward to determine the degree of parallelism between two
  functions that are not exactly similar.''}~\cite{Got2005_437}.

Such demonstration of lack of similarity has been proposed to be
solved by employing Bayesian or frequentist
approaches~\cite{Fay2020_721}. With Bayesian posterior probability one
can test parallel equivalence through:
  \small
\begin{align*}
  p(\gamma, x_L, x_U) = 
  \Pr\left\{ \min_{\rho} \max_{x \in [x_L,
  x_U]} \left| f(\theta_1, x) - f(\theta_2, x + \rho) \right| <
  \gamma \,\middle|\, \text{data} \right\}
\end{align*}
  \normalsize

The auhors give pratical examples, based on simalated data, that
deliver a p-value, using this probabilistic approach. The simulated
data give biomarker concentration ranging from 0.02--125,000 arbitrary
units (Table 1 in reference~\cite{Fay2020_721}).  In this example the
standard error (SE) equals
$\sqrt{\frac{1}{\eta^2} \, \mathrm{Var} \left[ A\left(
      \hat{\boldsymbol{\theta}} \right) \right]}$. Therefore, with
$\alpha = 0.05$ and $\delta = 0.85$, the probability caluclates as
$\Pr\left( T_{2n - p} > \frac{\lambda \left( \hat{\boldsymbol{\theta}}
    \right) - \delta}{SE} \right) = 0.062$. This is not
significant. Consequently, similarity between curves cannot be
declared. This implies that there is no evidence for parallelism in
bespoke example (for visual comparison see Figure 1 in
reference~\cite{Fay2020_721}. This is an extreme example to
demonstrate lack of parallelism, because the two curves cross over
between $\log_2$ and $\log_4$).

As introduced above, the 4PL and 5PL standard curves are now
frequently used for fitting dose-response curves and consequently of
relevance for discussed statistical evaluations of parallelism between
sample and standard
curves~\cite{Fay2020_721,Guy2023_229,Got2005_437,Jon2009_818}. One
final word of caution is warranted here, when employing highly
parameterized non-linear curve models, which can lead to overfitting,
there is a risk to introduce bias into the parallelism metrics that
provide the data fed inot above described statistical models.

As Smith noted, \emph{``The condition [of parallelism] and its importance
are relatively unknown to bioanalytical chemists and many consulting
statisticians''}~\cite{Smi1998_509}. In that work, the authors
illustrated interpretative challenges using simulated data, showing
how deviations in only a portion of the curve can complicate
decision-making. For this reason, our critical review primarily relied
on visual methods to assess the presence or absence of
parallelism~\cite{Pli1994_2441,Gil2002_47,Kle1999_35,Gai2002_35,Pet2024_602}. Visual
inspection offers an intuitive and transparent way to identify
deviations, making the concept accessible to a broader scientific
audience, including laboratory practitioners who may not have advanced
statistical expertise. Importantly, we do not consider visual and
statistical methods to be mutually exclusive. Rather, they are
complementary: visual assessment provides a practical first-line
evaluation, while statistical analyses can serve as confirmatory
tools, or address specific questions in selected situations.

\end{appendices}
\end{document}